# The Distribution of the Elements in the Galactic Disk III. A Reconsideration of Cepheids from l = 30 to 250 Degrees

R. Earle Luck[1] and David L. Lambert[2]

[1]Department of Astronomy, Case Western Reserve University
10900 Euclid Avenue, Cleveland, OH 44106-7215
luck@fafnir.astr.cwru.edu

[2]McDonald Observatory, University of Texas at Austin, Austin, TX 78712
dll@astro.as.utexas.edu

## Abstract

This paper reports on the spectroscopic investigation of 238 Cepheids in the northern sky. Of these stars, about 150 are new to the study of the galactic abundance gradient. These new Cepheids bring the total number of Cepheids involved in abundance distribution studies to over 400. In this work we also consider systematics between various studies and also those which result from the choice of models. We find systematic variations exist at the 0.06 dex level both between studies and model atmospheres. In order to control the systematic effects our final gradients depend only on abundances derived herein. A simple linear fit to the Cepheid data from 398 stars yields a gradient $d[Fe/H]/dR_G = -0.062 \pm 0.002$ dex/kpc which is in good agreement with previously determined values. We have also reexamined the region of the "metallicity island" of Luck et al. (2006). With the doubling of the sample in that region and our internally consistent abundances, we find there is scant evidence for a distinct island. We also find in our sample the first reported Cepheid (V1033 Cyg) with a pronounced Li feature. The Li abundance is consistent with the star being on its red-ward pass towards the first giant branch.

## 1. Introduction

In previous papers of this series (Andrievsky et al. 2002a,b,c, 2004, Luck et al. 2003, 2006, 2011 (collectively Papers I-VII) and Kovtyuhk, Wallerstein, & Andrievsky (2005, (KWA)) characteristic features of the metallicity distribution across galactic disk, as derived from Cepheid variable stars, are described. In the past decade, the gradient has been investigated by variety of technique's, but the derived value of the gradient has not significantly changed from the 2006 value of Luck et al. (see Maciel & Costa 2010 and references therein). In Andrievsky et al. (2004), an apparent step-like character in the iron (and also some other elements) abundance distribution was noted at galactocentric distances of about 10 − 11 kpc. More importantly, in the 2006 work, an ostensible inhomogeneity, a "metallicity island", was detected in the direction of l = 130° at a distance of about 3 kpc from the Sun. The galactocentric radius of the island is about 10 kpc. The step function at 10 kpc is thus intimately related to the metallicity island as without the island, the gradient in that region is rather smooth.

Any structure within the metallicity distribution, such as the alleged "metallicity island", if it is proved to be real, may have important consequences for scenarios of the formation and evolution of our Galaxy. In order to further explore the distribution of elements in the galactic we decided to re-examine in detail the abundances of northern Cepheids paying particular attention to the region of the "metallicity island."

We present here the results of the abundance determinations for 238 galactic Cepheids found in the northern sky, the majority of which have not been investigated spectroscopically before (at least for detailed elemental abundance determination). About 70 of the stars observed are found in Papers I-VI and provide an extensive database for abundance comparison. The list of Cepheids considered can be found in Table 1 along with some basic data.

## 2. Spectroscopic material

High signal-to-noise spectra were obtained in the period 2008 August – November 2010 using the Hobby-Eberly Telescope (HET) and its high-resolution spectrograph (HRS) (Tull 1998). The spectra cover a continuous wavelength range from 440 to 785 nm with a resolving power of about 30000. Typical maximum S/N values (per pixel) for the spectra are in excess of 100. Each night we observed a broad lined B star with a S/N exceeding that of the program stars to enable cancellation of telluric lines where necessary. Additional spectra of several brighter Cepheids and a small sample of Cepheids south of the HET observing limit were obtained with the McDonald Observatory 2.7m telescope and Tull spectrograph (Tull et al. 1995). These spectra have the same wavelength span as the HET spectra but are not continuous. Matching B star spectra were not obtained for a number of these integrations. Table 2 contains details concerning our program Cepheid observations. For the HET stars our observing strategy for the fainter stars was to observe as close to maximum light as possible. This choice is predicated on the desire to maximize signal-to-noise within the time-limit available for a single HET observation – generally about 1 hour. For the brighter stars, generally observed as schedule fill-ins, we accepted any phase. Phase-dependent studies of Cepheids (Luck & Andrievsky 2004, Kovtyukh et al. 2005, Andrievsky, Luck, & Kovtyukh 2005, Luck et al. 2008) have shown that the choice of phase for Cepheid observations is not critical from an abundance determination stand-point.

We used IRAF[1] to perform CCD processing, scattered light subtraction, and echelle order extraction. For these spectra two extractions were done, one uses a zero-order (i.e., the mean) normalization of the flat field which removes the blaze from the extracted spectra. The second uses a high-order polynomial to normalize the flat-field which leaves the blaze function in the extracted spectrum. The latter spectrum reflects more accurately the true counts along the orders. A Windows based graphical package developed by R. Earle Luck (REL) was used to process the spectra. This included Beer's law removal of telluric lines, smoothing with a fast Fourier transform procedure, continuum normalization, and wavelength calibration using template spectra. Echelle orders show significant S/N variations from edge to maximum due to blaze efficiency. To maximize the S/N in the HET spectra we have co-added the order overlap region using as weights the counts from the second data extraction. The co-added spectra were then inspected and the continua sometimes modified by minor amounts in the overlap regions. Equivalent widths from the co-added spectra were then measured using the Gaussian approximation. The line-list is the same as used in our previous studies (Papers I – VII) and derives from Kovtyukh & Andrievsky (1999). The equivalent widths can be obtained from the authors by request.

---

[1] IRAF is distributed by the National Optical Astronomy Observatories, which are operated by the Association of Universities for Research in Astronomy, Inc., under cooperative agreement with the National Science Foundation.

It would be useful to compare equivalent widths from a variety of sources for our program stars but Cepheids do not lend themselves to such efforts. The Cepheids would have to be at the same phase from different authors/measurers and should be from different sources (spectrographs). With respect to the stars in the current paper such data does not exist. As a basic check, we have measured equivalent widths for 15 non-variable supergiants from both the HET/HRS spectrograph and the Sandiford spectrograph (McCarthy et al. 1995) on the McDonald Observatory 2.2m Struve telescope. There are no scale differences between the two datasets and the average fractional difference in equivalent width is of order 7 percent.

### 3. Methods

#### 3.1. Atmospheric Models and Analysis Codes

Atmospheric models were calculated for each Cepheid using the Kurucz ATLAS9 model atmosphere code (Kurucz 1992). The ATLAS9 models are 1D LTE models with opacity distribution function (ODF) line-blanketing and a standard treatment of convection (using the default overshoot scale length). The models used adopt a microturbulent velocity of 4 km s$^{-1}$ for the ODF's. At some phases Cepheids can have microturbulent velocities significantly deviating from this model value; however, our previous test calculations (see Luck et al. 2000) showed that changes of several kilometers per second in the model microturbulent velocity has little effect on the structure of the final model. Thus, a mismatch between the derived microturbulent velocity for a specific star and the microturbulence used in the model computation at a 2 – 3 km s$^{-1}$ level has an insignificant impact on the resulting elemental abundances.

The Cepheids considered here span roughly an order of magnitude in [Fe/H]. Our procedure for dealing with this is to derive initial parameters using solar abundance models. If the derived [Fe/H] was less than -0.4 we re-computed the model atmospheres at [M/H] = -0.5 and re-determined the stellar parameters. For -0.4 ≤ [Fe/H] < -0.15 we utilized [M/H] = -0.3 models, at -0.15 ≤ [M/H] < +0.15 we adopted [M/H] = 0.0 (solar) models, and at [Fe/H] ≥ +0.15 we used [M/H] = +0.3 models. Note that in this work we assume log ε$_{Fe}$ = 7.50.

More sophisticated models than ATLAS9 models are available; e.g. spherically symmetric with opacity sampling (OS) line-blanketing and better element abundance mixes (Gustafsson et al. 2008 and the MARCS website). These models do not consider granulation. We have used these models (which we will refer to as 2D models) with the final line data to derive abundances primarily to determine the effect of differing models on the zero-point of the abundances.

The line analysis code was that of REL which derives from the "LINES" code of Sneden (1973). For syntheses we use a variant of the "MOOG" code of Sneden (1973). The analysis codes have been benchmarked against Kurucz's WIDTH and SYNTHE codes as well the synthesis code used by Andrievsky and Kovtyukh and all codes yield the same results to within expected numerical accuracy and differences due to assumptions (primarily partition functions and damping). For damping in the REL codes we preferentially use the van der Waals coefficients of Barklem, Piskunov, and O'Mara (2000) and otherwise compute the coefficient from the Unsöld approximation (Unsöld 1938). Note that in supergiants van der Waals damping is not especially critical given the low particle density. The basic assumptions of our analysis codes are LTE and 1D geometry. The MARCS models are LTE but assume a 2D geometry. The question of the applicability of analysis codes such as ours when used with the MARCS models has been addressed by Heiter & Eriksson (2005). Their finding is that 1D analysis codes are not a major problem in the parameter range 4000 – 6500 K at (log) gravities of 0.5 to 3.5 relative to a fully consistent 2D treatment. The maximum difference they noted was 0.1 dex in the abundances.

### 3.2. Atmosphere parameters: temperatures, gravities, microturbulent velocities

The effective temperature for each Cepheid was determined using effective temperature relations which originate in the work of Kovtyukh & Gorlova (2000) and Kovtyukh (2007). These relations combine the effective temperature with a set of spectral line depth ratios. The internal accuracy of the effective temperature determined in this way is rather high in the temperature range 5000 K to 6500 K: typically 150 K or less (standard deviation or a standard error ($\sigma/\sqrt{N}$) of 10 to 20 K). Note that this method uses multiple measures (ratios) each obtained from a single observation. In Table 3 we give the standard deviation of the mean temperature and number of ratios used at each phase (observation). An important advantage of this method (or any spectroscopic method) is that it produces reddening-free $T_{eff}$ estimates.

The method used for gravity and microturbulent velocity determination in supergiant stars such as Cepheids is described in detail by Kovtyukh & Andrievsky (1999). This method determines the microturbulent velocity using Fe II lines: the dominant ionization species of iron and hence less susceptible to any non-LTE effects which might be in play in supergiant atmospheres. The gravity value is found by enforcing the ionization balance condition; i.e., the mean iron abundance from Fe II lines equals the iron abundance which results from the Fe I – EW relation extrapolated to zero equivalent width.

The uncertainty in the microturbulent velocity and the gravity is more difficult to assess than the statistical error in the effective temperature. For the microturbulence, a variation of ±0.5 km s$^{-1}$ from the adopted velocity causes a significant slope in the relation between Fe II line abundance and equivalent width. We therefore adopt ±0.5 km s$^{-1}$ as the uncertainty in the microturbulence. For *log g* we adopt ±0.1 dex as the formal uncertainty based on the numerical result that a change in gravity at that level will result in a difference of 0.05 dex between the total iron abundance as computed from the Fe I and Fe II lines. Since we have forced an ionization balance we do not allow a spread larger than 0.05 dex in the total abundance of iron as derived from the two ions and thus our uncertainty estimate. Iron abundance details are also given in Table 3 along with the stellar parameters.

The final results of the determinations of $T_{eff}$, $log\ g$ and $V_t$ are given in Table 3. Note that Table 3 (and Table 4) contains the results of a re-analysis of all southern Cepheids from Luck et al. (2011). The rationale for the re-analysis will be discussed in §4.4.1 which deals with abundance comparisons.

### 3.3. Atomic Data

The oscillator strengths for $Z > 10$ elements used in this (and all preceding Cepheid analyses of this series) come from Kovtyukh & Andrievsky (1999). The values are based on an inverted solar analysis which used the solar abundances of Grevesse et al. (1996). The solar model used was taken from the Kurucz (1992) model atmosphere grid.

For lithium, carbon, nitrogen, and oxygen we have performed spectrum syntheses for the features of interest. In these cases we have preferred to utilize laboratory oscillator strengths where available. For the lithium feature we have used all components of $^7Li$ (using the data presented by Andersen, Gustafsson, & Lambert (1984)) in the 670.7 nm hyperfine doublet to match the observed profiles. In all cases except one, the syntheses yield only upper limits. Per phase lithium abundances are given in Table 3.

Carbon abundances have been derived from C I lines at 505.2 nm, 538.0 nm, and 711.5 nm. For these lines we have utilized the oscillator strengths of Biémont et al. (1993) or Hibbert et al. (1993). These oscillator strengths have been used in recent determinations of the solar carbon abundance (Asplund et. al. 2005). To form the carbon abundance we combine the individual features as follows: for $T_{eff} > 5500K$, 505.2, and 538.0 each have weight 1 while 711.5 has weight 2. At T < 5500 K, 505.2 has weight 1 while 538.0 and 711.5 both have weight 3. The weights are based on relative strength and blending. Typical spreads in abundance for the features are 0.15 dex. For the purpose of abundances with respect to solar values, we adopt $log\ \varepsilon_C = 8.45$, very close to the Asplund et al. (2009) recommended solar carbon abundance of 8.43. Per phase average carbon abundances can be found in Table 3.

Nitrogen abundances are derived from the N I lines at 744.2 and 746.8 nm. Oscillator strengths for these lines were taken from the inverted solar analysis of Kovtyukh and Andrievsky (1999). The nitrogen abundance is formed by averaging the best fit as determined from a χ-square minimization to two lines together using a single abundance with the best fit values for the two lines individually. This process minimizes blending and noise problems. For better quality spectra with little blending the total spread in abundance is usually about 0.03 dex. To determine nitrogen abundances relative to the Sun we use $log\ \varepsilon_C = 7.99$ – the Grevesse et al. (1996) solar nitrogen abundance used to compute the oscillator strengths. The per phase abundances are given in Table 3.

Oxygen abundance indicators in the available spectral range are rather limited: the O I triplet at 615.6 nm and the [O I] lines at 630.0 and 636.3 nm. The O I triplet at 777.5 nm is heavily affected by non-LTE effects and thus not usable in a standard LTE analysis. The O I 615.6 lines are problematic in abundance analyses with only the 615.8 nm line being retained in solar oxygen analyses (Asplund et al. 2004). We have synthesized the O I lines using the NIST atomic parameters (Ralchenko et al. 2010) which were also used by Asplund et al. For the forbidden oxygen lines only 630.0 nm is usable as 636.3 nm is weak, heavily blended, and complicated by the presence of the Ca I autoionization feature. For our syntheses of 630.0 nm we have used the line data presented by Allende Prieto, Lambert, & Asplund (2001) except that we have used the experimental oscillator strength for the blending Ni I line (Johansson et al. 2003). In our syntheses we have assumed [Ni/Fe] = 0. To form a final oxygen abundance we have averaged our data in the following manner: for $T_{eff}$ > 6000 K O I has weight 2 and [O I] weight 1, for 5500 < $T_{eff}$ < 6000 K both O I and [O I] have weight 1, and for $T_{eff}$ < 5500 K O I has weight 1 and [O I] has weight 2. Oxygen abundance from the individual analyses can be found in Table 3.

### 3.4. Line Selection

One of the primary problems in spectroscopic abundance determination is line selection. It would be preferable that each star in an analysis be analyzed using exactly the same line list. However, this is not feasible in this analysis given the temperature, gravity, and abundance variations exhibited by these Cepheids. For the Fe I and Fe II data we have examined the data for each individual dataset (star and phase) with an interactive editing program. The editing program allows one to check for discrepant points in the data, determine if there are untoward trends in items such as abundance versus wavelength (indicative of a continuum problem if present), and check how elimination of lines in one relation (such as abundance versus excitation potential) affects another (abundance versus equivalent width). We do eliminate grossly discrepant lines, but try to keep as many lines as possible.

For the Z > 10 elements we examined a subset of about 85 Cepheids which span the stellar parameter range. These stars were treated on an individual basis to identify and remove discrepant lines using the same editing process as used for Fe I and Fe II. This data was then used to determine the line list to be used for the final abundances: to be retained as a "good" line a line had to be retained in at least 50% of the stars in the preliminary analysis. The dominant number of retained lines appeared in 80% or more of the stars of the subset considered. This list of lines was then used to determine Z > 10 abundances for all stars/phases.

This procedure was used for all species *not* examined individually – specifically Fe I and Fe II – or by detailed spectrum synthesis; i.e., lithium, carbon, nitrogen, and oxygen. A caveat to the line list must be stated: our selection procedure yields a consistent set of lines for analysis but does not insure that each line is actually present in each star; it only guarantees that no other lines than these will be considered. Shifting parameters and vagaries in the observational data can remove lines from consideration but none can be added.

## 3.5. Possible Non-LTE Effects

The consideration of the effects of the failure of local thermodynamic equilibrium have not been explicitly considered in this analysis. However, some discussion is in order relative to the iron and light element abundances. We wish to make three general points: 1) the stellar parameter range of Cepheids has not been adequately examined for non-LTE effects, 2) non-LTE effects decrease with decreasing temperature and increase with decreasing gravity with neutral species being more affected than first-ionized, and 3) it appears that as the atomic models become more complete that non-LTE effects become smaller.

Non-LTE effects in iron in cooler stars have been studied starting with Tanaka (1971) and have proceeded to the recent work of Mashonkina et al. (2011) and Mashonkina (2011). The primary emphasis in these studies has been on dwarfs. In the work of Mashonkina et al. we find that in the temperature range of Cepheids (though at higher gravities) that non-LTE calculations yield abundances that are not substantially different than LTE values. At lower gravity and temperature in the metal-poor star HD 122563, the differences can be substantial – 0.3 dex, but depend strongly on the poorly known strength of inelastic hydrogen collisions.

We can expect non-LTE effects to be both gravity and temperature dependent. Cepheids undergo systematic changes in both parameters but *not* in abundance. If there were strong non-LTE effects in iron in Cepheids, we should see them in phase dependent abundances. Phase dependent abundance studies (Luck & Andrievsky 2004, Kovtyukh et al. 2005, Andrievsky, Luck, & Kovtyukh 2005, Luck et al. 2008) have shown no such effects. Examination of the multi-phase data within this study shows no untoward trends relative to differences in temperature or gravity in the iron abundance data.

Non-LTE effects in CNO have been considered by Fabbian et al. (2006) for carbon, by Lyubmikov et al. (2011) for nitrogen, and by Takeda & Takada-Hidai (1998) and Fabbian et al. for oxygen. The Fabbian et al. study is of F type dwarfs while Takeda & Takada-Hidai and Lyubmikov et al. considered A and F supergiants. All overlap with the higher temperature range of Cepheids but not necessarily with the gravity range. For carbon, the Fabbian et al. results yield averaged non-LTE corrections (in dwarfs) of −0.2 to −0.3 (in the sense LTE being higher). How this would scale to supergiants is unknown but a preliminary calculation for ℓ Car ($T_{eff}$ = 5265, log g = 1.12) indicates only minor non-LTE effects for our C I lines (Sergei Andrievsky – private communication). This is consistent with the Fabbian et al. statement that C I 505.2 and 538.0 nm are relatively unaffected by non-LTE.

In the temperature range 5800 – 6400 K, Lyubmikov et al. (2011) give the non-LTE correction for N I as +0.2 to +0.3 dex in the sense that LTE abundances are larger. Our gf values for N I 744.2 and 746.8 nm are about 0.2 dex higher than assumed by Lyubmikov et al. This means our LTE values would be at -0.2 dex relative to theirs, thus when the non-LTE correction is added our nitrogen abundances should be on the same scale as theirs. Whether this is actually true, we cannot say.

Oxygen as determined from the O I triplet at 615.6 nm has been considered by Takeda & Takada-Hidai (1998) in A and F supergiants. Takeda & Takada-Hidai found that in the F supergiant range that the non-LTE correction for the 615.6 triplet is ≲ 0.1 dex. Our oxygen abundances are based on a combination of [O I] 630.0 and the triplet. The forbidden line is not affected by non-LTE processes, so if non-LTE effects are present, then they could manifest as a difference between the two indicators which depends on effective temperature and/or gravity. We find no such dependences through the scatter is relatively large. The mean difference in oxygen abundance between the two indicators is 0.04 dex with a standard deviation of 0.18.

While this discussion does not prove one way or the other the importance of non-LTE effects, it does show that there is a need for further work in this area. In the discussion to follow we must keep in mind that there may be effects in the abundances that will ultimately vitiate the conclusions reached here.

### 3.6. Distances

For the determination of the Cepheid galactocentric distances the following standard formula was used:

$$R_G = [R_{G,\odot}^2 + (d\cos b)^2 - 2R_{G,\odot} d\cos b\cos l]^{1/2}$$

where $R_{G,\odot}$ is the galactocentric distance of the Sun (7.9 kpc, McNamara et al. 2000), $d$ is the heliocentric distance of the Cepheid, and $l$ and $b$ are the galactic longitude and latitude respectively.

To determine the heliocentric distance we need the absolute V magnitude, the mean apparent V magnitude, and the reddening. To estimate the absolute magnitude we used "absolute magnitude – pulsational period" relation of Fouqué et al. (2007) for Johnson V magnitudes. The formal uncertainly of this relation is about ±0.2 mag which leads to a distance uncertainty of 9%. Periods (and epochs) are from Berdnikov (2006 – private communication to S. M. Andrievsky) or the GCVS. For first overtone pulsators (DCEPS), we assume $P_0 = P_1 / 0.71$ where $P_1$ is the observed period. The mean visual (V) magnitudes are from Fernie et al. (1995) or as noted in Table 1. Reddenings are from Fouqué et al. (2007), Fernie et al. (1995) with the systematic correction of Fouqué et al. applied (i.e., E(B-V) = 0.952 E(B-V)$_{Fernie}$), or from the source given in Table 1. These reddening are on the E(B-V) system of Laney & Caldwell (2007). We follow Fouqué et al. and adopt $R_V = A_V / E(B-V) = 3.23$. Galactocentric and heliocentric distances for program stars are listed in Table 1 along with the adopted E(B-V) and <V>. Note that 18 of the program stars (all from Wils & Greaves (2004)) lack distances as they do not have Johnson B or V magnitude data available.

Adopting a distance modulus combined uncertainty of ±0.3 mag the distances to our Cepheids are imprecise at the 13% level. Propagating this uncertainty into the galactocentric distances we find possible distance errors that range from 10's of parsecs for Cepheids at the solar galactocentric radius to more than a kiloparsec for stars well outside the solar circle. We indicate in the lower panel of Figure 1 two representative distance uncertainties.

# 4. Results

## 4.1. Elemental abundances

The elemental abundances in our program stars are in Table 4. Iron abundance details for our Cepheids are given in Table 3 along with the stellar parameters. The mean abundances of Tables 4 have been averaged using equal weighting for each spectra/observation of each star.

### 4.1.1. Comparison with other Abundance Studies

Abundance reliability in Cepheids as a function of phase has been addressed in a recent series of papers (Luck & Andrievsky 2004, Kovtyukh et al. 2005, Andrievsky, Luck, & Kovtyukh 2005, and Luck et al. 2008). Our finding is that accurate parameters and [Fe/H] ratios can be derived at any phase in a Cepheid independent of the period of the Cepheid. Based on this we would not expect any untoward effects in the abundances due to phase (stellar) parameter problems. This is borne out by inspection of the per phase data give in Table 3. Typical differences in iron abundance from phase-to-phase are of order 0.1 dex or less.

We have re-observed and derived new abundances for 81 stars of Papers I-VI. A comparison of the iron data for the common objects shows an offset of the iron abundance of 0.07 dex ($\sigma = 0.09$) with this work having the higher abundance. This difference was noted in Luck et al. 2006 (Paper VI) but was based on small number of stars. While 0.07 dex is a relatively small difference given the complexity of Cepheid abundance analyses, we would still like to know the source of the variance. We could reanalyze the previous spectra from Papers I-VI, but a better alternative exists. In Luck et al. (2011 – Paper VII), a large sample of southern Cepheids were analyzed using ESO spectra. These spectra were processed by REL to a final state including B star division, order co-addition, and continua and wavelength normalization; that is, they were reduced in exactly the same way as the northern Cepheid spectra of this study were. The southern Cepheid spectra were then transferred to Sergei Andrievsky and Valery Kovtyukh for the determination of equivalent widths and abundances. We can use the southern spectra to investigate the source of the abundance offset. To do this we have finished processing the southern spectra to equivalent widths, line depths, and finally abundances in the same manner as used for the northern Cepheids. The parameters derived for these stars are also given in Table 3 and their mean abundances in Table 4. Comparison of the iron abundances derived here with those of Paper VII shows the same offset found here in the northern Cepheids – our iron abundances are 0.07 dex larger than the previous values.

What could be the origin of the difference? We can eliminate oscillator strengths and atmospheric models immediately as both analyses are equivalent in these areas. Another possibility is differences in equivalent widths. We compared the values used for Paper VII with our values and find excellent agreement. There are variations due to different smoothing and procedures for handling blends, but the scales are the same and the absolute median percentage variation in equivalent width is of order two percent over the 42621 common measures. To make sure that our analysis code returns the same abundance as the code used for Paper VII, we have used the GU Nor iron data from Paper VII with an ATLAS9 stellar model having the same parameters as used in Paper VII in our analysis code. The mean difference in abundance with respect to the Paper VII per line abundance is +0.009 dex with a maximum difference of about 0.03 dex. There is no problem in the analysis software. Another possibility is stellar parameters. The effective temperatures differ in the mean by 5 K, the gravities by +0.11 with ours being the larger, and the microturbulent velocities differ in the mean by 0.01 km s$^{-1}$. While these mean variations are small, the key to the abundance variance is in the difference in the gravities.

As pointed out in §3.2, a change in gravity (at constant temperature) of +0.1 (logarithmic) results in a change in the iron abundance of +0.05 dex. Since the effective temperatures used here and in Paper VII are essentially the same, we would expect that since our gravities are systematically higher, that our [Fe/H] values would follow. This is exactly as observed and the scaling is as expected. The question now becomes, why are the gravities higher?

Gravities and microturbulent velocities are simultaneously determined by the process briefly described in §3.2. One calculates the per line abundance for a series of microturbulences and gravities, selects the microturbulence that gives no dependence of iron abundance on equivalent width as determined from Fe II lines, and then finds the gravity that yields the same total iron abundance from both Fe I and Fe II using that microturbulence. What is not said but which influences the derived values, is that as this process proceeds a subjective line editing is performed to remove outliers from the data. While Fe I has many lines and thus is not too sensitive to this process, Fe II has fewer lines and thus the editing or lack thereof, can affect the relations greatly. Thus a personal equation enters into the parameter determination. Two individuals editing the same data will not necessarily remove the same lines, and hence, the final answers will vary. It appears that our process differs somewhat from that used in Papers I-VII. As to the question of which process is correct, that is impossible to say. We can only note that both produce internally consistent but somewhat different answers.

Other than our previous work there are two primary Cepheid abundance analyses to consider. The larger number of analyses are from Romaniello et al. (2008) – 25 stars. The [Fe/H] ratios of Romaniello et al. are based on a solar iron abundance of 7.51, very similar to our assumed value of 7.50. The raw mean difference between this work (including the southern stars) is +0.11 ($\sigma$ = 0.11). This difference while larger than found in Paper VII (+0.03 dex) is as expected due to the offset in the abundance zero-point. Another source of comparison is Yong et al. (2006) with whom we have 15 stars in common. Yong et al. assume a solar iron abundance of 7.54. The mean difference found here is that our [Fe/H] ratios average 0.3 dex higher with an indication that larger differences are found for the more metal-poor stars. The latter conclusion was also reached by Lemasle et al. (2008) in regard to their abundance determinations.

### 4.2. Model Atmosphere Abundance Dependence

Another issue that affects the question of the zero-point of abundances is the choice of model atmospheres. Our primary results here are derived using ATLAS9 models (Kurucz 1992) in an attempt to maintain the greatest degree of consistency with previous analyses. Newer 2D, opacity sampling (OS) models are available (Gustafsson et al. 2008). We have used these new models with the parameters given in Table 3 to investigate the dependence of the iron abundance on the choice of models. To do this we have utilized 2, 5, and 10 $M_\odot$ mass models with Doppler broadening velocities (DBV) of 2 and 5 km s$^{-1}$. The models have moderate CNO processing abundances and [α/Fe] = 0. Inspection of the abundances shows that the mass of the models has little effect on the abundances at constant DBV – the total range from 2 to 10 $M_\odot$ is typically 0.015 dex or less with the lower mass models having the lower abundance. At constant mass, changing from a 2 to 5 km s$^{-1}$ DBV model increases the abundance by about 0.03 dex. In order to compare the MARCS models to ATLAS models we use 5 $M_\odot$ mass models with a DBV of 5 km s$^{-1}$.

We have interpolated a model from the MARCS grid at the stellar parameters (including metallicity) for each Cepheid. The mean difference in the iron content in the sense ATLAS – MARCS is +0.06 dex with a marked dependence on temperature and gravity. At $T_{eff}$ around 5000 K, the difference in iron abundance between the ATLAS and MARCS models is about 0.02 dex increasing to 0.08 to 0.12 dex at 6500 K. In the MARCS models Fe I and Fe II vary in near lock-step so that difference between the two species remains about 0. We also note that the slopes in the Fe I per line abundance versus lower excitation potential relations from the MARCS models are near identical with those found in the ATLAS models. Taken together, this means that the MARCS models fit the stellar parameters as well as the ATLAS models. The iron zero-point of the MARCS models is lower than that found from the ATLAS models. This further means that the overall uncertainty in current epoch metallicity must be at least ±0.1 dex based solely upon the differences related to the models.

### 4.3. Spatial Abundance Distributions

Using the distance and abundance information from Tables 1, 3, and 4 we have constructed radial (1D) distributions of elemental abundances. The radial gradients are found in Table 5.

### 4.3.1. [Fe/H] Distribution

In Figure 1 we show the radial gradient for iron discriminating between the older data, the re-analyzed stars of Luck el al. (2011), and the data from the northern Cepheids of this study. The iron data from Papers I-VI have been scaled in accord with the findings of §4.1.1. While the gradient is evident in the data, there are a number of discrepant stars. After dealing briefly with the problem of discriminating between Type I and Type II Cepheids, we shall deal with each of the discrepant stars individually.

Type I or classical Cepheids, are young evolved intermediate mass stars of the disk, while Type II Cepheids (aka W Vir stars), are low mass evolved giants of the halo. While this distinction seems simple, it is difficult to separate the two types observationally. They have comparable periods, range over the same spectral types, and show similar light curves. As pointed out by Harris (1985), perhaps a better way to sort them is by distance from the galactic plane as classical Cepheids are part of the youngest disk population. This method, while reasonable, is not fool-proof.

The first star to consider is QQ Per which is not shown in Figure 1. If it is a classical Cepheid, then it would lie at a galactocentric radius of 37.5 kpc. Both the GCVS and SIMBAD indicate that it is a CEP variable. Wallerstein, Kovtyukh, & Andrievsky (2008) conclude that QQ Per is most likely a Type II Cepheid and thus the distance is in error as the classical Cepheid PL relation does not apply. We also note that our [Fe/H] ratio of -0.67 for QQ Per agrees very well with their value of -0.6.

BC Aql, EK Del, GP Per, and FQ Lac are all of type CEP. From the point of view of an abundance analysis the primary problem with them is that their parameters and hence, their abundances are very uncertain. This is easily seen by noting that the standard deviation of the effective temperature determination is in excess of 400 K for each while typical standard deviations for the effective temperature are less than 150 K. While the spectra for these stars have S/N ratios towards the lower end for this dataset, other Cepheids with similar S/N have normal temperature uncertainties. The problem is that these spectra are much more blended than normal Cepheids due to high line density (molecules?) and what appears to be substantial line broadening. This makes continuum and wavelength determinations difficult. Looking at Figure 3, we note that all except GP Per are at a substantial distance from the galactic plane. In fact, Harris(1985) classifies EK Del as a Type II Cepheid and BC Aql as a probable Type II. It appears that Berdnikov (1985) considers FQ Lac a W Vir star. We eliminate them from further discussion involving abundances and the gradient.

HK Cas is a DCEP variable with a typical effective temperature uncertainty and reasonable stellar parameters for a Cepheid: $T_{eff}$ = 6058 K and log g = 2.23. Its primary spectral peculiarity is strong Hα emission. However, more problematic is the result that it is a carbon star and the fact that it is far from the galactic plane. The formal C/O ratio is 2.7. All C and O indicators are in accord. This result is unprecedented and unexpected for a classical Cepheid. However, Harris (1985) classifies it as Type II Cepheid based on its position. Significant fractions of Type II Cepheids are known to be carbon-stars (Maas et al. 2007). We thus eliminate HK Cas from our gradient discussion.

Eliminating the above stars allows us to hone in on the gradient and we show in Figure 1 (bottom panel) the iron gradient from the retained stars only. Fitting a simple linear model to the entire sample of 398 stars we find that [Fe/H] = −0.062 (±0.002) * $R_G$ + 0.605 (±0.021). This is somewhat steeper than the value given in Paper VI: slope = −0.055, but is in overall agreement with the values of Romaniello et al. (2008) and Pedicelli et al. (2009). If one limits the sample to the 313 stars with analyses from this paper the gradient changes very little: [Fe/H] = −0.061 (±0.002) * $R_G$ + 0.609 (±0.024). This is not especially surprising as the stars of this study (including the reanlyzed Carina stars) dominate the sample at distances greater than 1 kpc from the solar galactocentric radius.

If we divide the whole range into three parts (zone I: $R_G$ < 6.6 kpc, zone II 6.6 < $R_G$ < 10.6 kpc, and zone III $R_G$ > 10.6 kpc), then the statistics for each zone and the total sample for iron are the following:

Zone I:  [Fe/H] = −0.103 (±0.040) * $R_G$ + 0.894 (±0.246)
 <[Fe/H]> = +0.271 (s.d. = 0.126, n = 34)

Zone II:    [Fe/H] = −0.042 (0.005) * $R_G$ + 0.443 (± 0.042)
            <[Fe/H]> = +0.098 (s.d. = 0.098, n = 292)

Zone III :  [Fe/H] = −0.068 (± 0.009) * $R_G$ + 0.686 (± 0.119)
            <[Fe/H]> = −0.172 (s.d. = 0.169, n = 72)

There is little difference in the slopes and means for the various regions as computed from the total sample (398 stars) versus the values determined from the objects analyzed here (313 stars). The largest difference is in Zone II where using the new abundances only yields a gradient of −0.047 dex kpc$^{-1}$. In agreement with other studies (Romaniello et al. 2008, Pedicelli et al. 2009), as well as our previous work, we find the gradient in the inner zone shows a steepening towards the galactic center (see bottom panel of Figure 1).

The zones given above are those used in Andrievsky et al. (2004 – Paper V) and roughly correspond to points of inflection in the LOWESS (Locally Weighted Scatter Plot Smoothing) fit in Figure 1 (see especially the lower panel). There appears to be some flattening of the gradient in Zone II, but the difference in the LOWESS and linear fit at the point of maximum variance at $R_G \approx 9.4$ kpc is only 0.04 dex in [Fe/H]. The behavior of the gradient in the 9 – 10 kpc region relates to the contention that the imputed change in slope at about 9 kpc is due to dynamical effects at the co-rotation radius (see Andrievsky et al. 2004 – Paper V, Cruz et al. 2011). While there may be dynamical effects which can cause a break in an abundance gradient, the existing evidence from Cepheids on such a discontinuity is scant and appears to be disappearing with better abundances and more stars.

Maciel & Costa (2010) emphasize the flattening of the gradient outside of galactocentric radius 10 kpc. Cepheid abundances whether based on the total sample or the current sample indicate that the outer zone gradient is much the same as the overall gradient; that is, there is no flattening of the gradient in the outer portion of the galactic radius sampled here.

Abundances at larger galactocentric radii preferentially come from this study. Of the 43 stars with galactocentric radius greater than 12 kpc, only 3 have abundances from older studies. The [Fe/H] ratio appears to have more scatter at larger galactocentric radius ($R_G > 12$ kpc) than in the solar region. A typical range in abundance at a particular $R_G$ is approximately ±0.3 dex in Zone II. For the more distant Cepheids, the total range in [Fe/H] at fixed $R_G$ is easily 0.6 dex. While this range is not different than seen in inner regions, what is different it that there appears to be little concentration about a mean value. That is, the distribution in [Fe/H] appears to resemble more a uniform rather than a normal distribution. The range in $R_G$ for these stars is about 5 kpc (all Cepheids exterior to $R_G = 12$ kpc) but the physical separation spans about 15 kpc (see Figure 2). Could these stars be indicating that the outer regions of the Galaxy show much more significant variations in abundance than do the regions closer to the solar circle? This would not be especially surprising given the overall lower gas densities and star formation rates which could be significantly influenced by local events. This also means that gradient values can be influenced in the outer region quite strongly by candidate selection.

In Figure 2 we show the spatial distribution of our total sample relative to the galactic center. As expected, the density of observed stars decreases with galactocentric distance and distance from the Sun. The reason is twofold: one, stars far from the solar region require the investment of significant observing time; and two, as one approaches the edge of the Galaxy ($R_G \approx 16$ kpc or d ≈ 8 kpc from the Sun towards $l = 180°$) the stellar density decreases. The galactic stellar scale length is of order 3 kpc for the Milky Way (Zheng et al. 2001, López-Corredoira et al. 2002) which implies a stellar disk of about 12 kpc based on external spirals (de Grijs, Kregel & Wesson 2001). López-Corredoira et al. however find no indication of a cut-off in the galactic stellar disk at radii $R_G < 15$ kpc. Additionally, there is evidence of spiral structure from molecular cloud and H I mapping at $R_G$ from 15 to 24 kpc (McClure-Griffiths et al. 2004, Nakagawa et al. 2005, Strasser et al. 2007, Dickey et al. 2009). Thus, it is possible that the Cepheids thus far sampled do not approach the full radius of the Milky Way. Unfortunately, Cepheids will not be able to say much more about the gradient and the extent of the disk as there are few known distant Cepheids that remain without abundances.

While Cepheids are primarily located near the galactic plane, this sample is large enough to investigate the Z (distance from the plane) distribution in the stars themselves. Figure 3 shows the Z distance as a function of galactocentric distance. A number of things are readily apparent: 1) there are a number of Cepheids at large distance from the galactic plane, 2) there is a tendency for those that deviate the most to lie at larger galactocentric radii, and 3) the mean Z height of the Cepheids is −30 to −40 pc which reflects the solar position above the galactic plane. One wonders if the Cepheids lying away from the plane exert an undue influence on the abundance gradient. If we eliminate all stars with |Z| > 300 pc the value of the gradient changes very little: the slope in [Fe/H] decreases from −0.062 to −0.058. We conclude that they do not influence the gradient, but that they do warrant further consideration especially in terms of kinematics. The question is, if they are truly young massive stars, how did they get to distances approaching a kiloparsec from the plane?

In Figure 4, we show several azimuthal cuts of the data at different galactocentric radius. Each cut has a depth of 1 kpc. The angle shown is as seen from the galactic center with the solar region lying in the center at 0°. The Carina region lies to the left of center and the Cepheus/Cygnus region lies to the right. To put the azimuthal variation into better perspective, note that the 80° arc in the 7 – 8 kpc cut spans about 11 kpc. As was found in Paper VII, there is no believable evidence in the data for any variation in abundance as a function of azimuthal angle.

One of the principal objectives of this study was to investigate the "metallicity island" found by Luck el al. (2006 – Paper VI). In galactocentric radius the island lies at about 9 to 10 kpc. We have added numerous stars in the region – the older sample had 27 stars while we currently have 56 stars of which 46 have new abundance determinations. The mean [Fe/H] ratio in the 9 – 10 kpc region is +0.07 dex which is higher the value predicted by the overall gradient (+0.03 dex) and the Zone II gradient (+0.04 dex) but not dramatically so. Our current view of the "metallicity island" is that there is little evidence for a pronounced region of enhanced abundances. This view is supported by Figure 5 which presents a contour map of position (X,Y relative to the Sun) versus [Fe/H]. What is most obvious about the behavior of iron is the simple linear behavior independent of azimuth. What is seen is a monotonic gradient with an increase in slope starting at about $R_G$ about 6.6 kpc (X = -1.3) and proceeding inwards. The outer slope appears overall to be well behaved, but it is possible, if not probable, that the outer disk shows a larger dispersion in abundance than does the region at 12 kpc and inwards. The "metallicity island" in the Cygnus region (X =4 kpc, Y = 1 kpc) has for all practical purposes disappeared.

The intrinsic metallicity dispersion of the sample is also of interest. The two most heavily populated bins are the 1 kpc annuli centered at 7.5 and 8.5 kiloparsecs. Both have about 125 stars in them. Each has a standard deviation about the mean [Fe/H] abundance of 0.09 dex with a range about the mean of ±0.3 dex . Much of the dispersion can be attributed to observational scatter. The multiphase observations done in this study yield a mean [Fe/H] difference of 0.02 dex (average over unique phase-to-phase [Fe/H] difference for a single Cepheid) with a standard deviation of 0.07. Assuming the statistical scatter in the multiphase data represents the uncertainty in the [Fe/H] value, then the correspondence between the two standard deviations means a large fraction of the star-to-star scatter comes from the observations. This then means that the intrinsic scatter in the [Fe/H] values at these galactocentric radii is small. The intrinsic largest scatter (Gaussian) that can be accommodated by the [Fe/H] data has a standard deviation of about 0.07 with a total range of ±0.2.

As a last item concerning the [Fe/H] gradient we have computed the gradient using the iron abundance data derived from the MARCS models discussed in §4.2. To do this we have simply averaged the total iron abundance as derived from Fe I and Fe II on a per phase basis, and for the stars with multiple phases the final iron content is a simple average over phase. For the MARCS models we obtain $d[Fe/H]/dR_G$ = -0.060 dex kpc$^{-1}$. Though the abundance zero-point is different for the MARCS models, the gradient is essentially the same as obtained using ATLAS models.

### 4.3.2. Lithium and V1033 Cyg

In an evolved intermediate-mass star one expects the lithium abundance to be severely diluted due to the combined effects of mass-loss on the Main Sequence and the subsequent first dredge-up. Standard stellar evolution (without mass loss) predicts a dilution about a factor of 60 relative to the initial value (Iben 1966). The sensitivity to mass-loss stems from the fact that in B stars (the progenitors of Cepheids) Li remains in only the outer 2% of the star at the end of the Main Sequence. Assuming an initial lithium content of log $\varepsilon_{Li}$ = 3.0 dex, we would expect Cepheids to show lithium abundances of log $\varepsilon_{Li}$ < 1.2 dex. Inspection of the upper limits for lithium in Table 3 bears out this expectation except for one extraordinary case: V1033 Cyg which has a strong lithium line and an LTE lithium abundance of log $\varepsilon_{Li}$ ≈ 3.2 dex. Non-LTE and LTE abundance calculations at the stellar parameters of V1033 Cyg yield near identical results (Lind et al. 2009).

How is it that V1033 Cyg has lithium in its photosphere? The simplest answer is that V1033 Cyg is crossing the HR diagram towards the first giant branch. This would mean that the photospheric composition has not yet been altered by the dredge-up and that we are seeing unaltered abundances. For this to be the case, the CNO content should be in its original state; and indeed, this appears to be the case. The [C/Fe] and [N/Fe] ratios are +0.1 and +0.3 respectively while the C/O ratio is 0.55. The [N/Fe] ratio is a bit high but could be overestimated by up to 0.2-0.3 dex due to non-LTE effects (Lyubmikov et al. 2011). The [C/Fe] and C/O ratios are significantly larger than the typical ratios of -0.31 and 0.25 respectively found in Cepheids. The various CNO abundances and ratios found in V1033 Cyg are typical of those found in young, unevolved stars.

Another way to potentially ascertain the evolutionary status of a Cepheid is to look for systematic period changes over time. This has been pursued by Berdnikov & Turner (2010 and references therein). They have examined eight Cepheids finding period changes in five. However, none of the period changes detected have the characteristics that would indicate a first-crossing Cepheid. Thus, while other Cepheids have been forwarded as pre-dredge-up (SV Vul for example – Luck et al. 2001); we have in V1033 Cyg the best case yet for a star evolving red-ward for the first time in the HR diagram.

### 4.3.3. Light Element Abundances and Gradients

While the primary result of these studies is the [Fe/H] gradient, we have information on other elements. The light elements carbon, nitrogen, and oxygen are especially important as they are abundant species and the product of primary energy generation nucleosynthesis. Additionally, they are the species that are most often investigated in gradient studies involving B stars or H II regions. The difficulty with these species in Cepheids is that they are affected by nucleosynthesis and subsequent mixing events during the first dredge-up which alter the surface composition of carbon and nitrogen. While standard evolution leaves the surface oxygen content unaltered, abundance studies starting with Luck (1978) have shown that oxygen in intermediate-mass stars is subsolar (Luck & Lambert 1981, 1985, 1992, Luck 1994, Luck et al. 2006). While the current best solar oxygen value (log $\varepsilon_O \approx 8.75$ – Asplund et al. 2009) alleviates this discrepancy somewhat, the problem has persisted (Luck et al. 2011 – Paper VII). Note that the newest solar results for oxygen discussed in Asplund et al. (2009) reflect a near convergence of 3D models and modern 1D models for the oxygen content. Another element that is affected by nucleosynthesis and mixing is Na. In this case the agent is Ne-Na cycle which brings sodium enriched material to the stellar surface (Sasselov 1986; Luck 1994; Denissenkov 1993a, 1993b, 1994). Before considering gradients in these species derived from Cepheids, we must try to ascertain if evolutionary processes have left differential effects in these stars that would vitiate any gradient result. To examine this question we will look at our abundance data for these species as a function of period.

In Figure 6, we show [Fe/H], [C/Fe], [O/Fe], and [Na/Fe] as a function of log (Period). Since the most reliable CNO abundances come from this analysis we will limit the sample to those stars analyzed here. The Cepheids span a period range of about 2 to 70 days and thus have a significant mass range. We show [Fe/H] to emphasize that there is no meaningful dependence on period in a species not affected by the first dredge-up (if there were a dependence, it could mean there is a problem in gravity determinations). The remaining species are normalized to iron to remove the overall metallicity from affecting the result. There is scant evidence in Figure 6 for there being period (mass) dependent mixing affecting these stars.

Perhaps the most striking aspect of Figure 6 is that a few of these Cepheids are extremely carbon deficient; and thus, may be related to the weak G-band stars. The stars that are very carbon-poor are FN Aql and NT Pup, both of which have [C/Fe] = -1.3. FN Aql was first pointed out as carbon-poor by Usenko el al. (2001) and Hurley et al. (2008) show that it has a time-variable dust extinction which is probably localized in a circumstellar shell. The difficulty is that weak G-band stars are typically G/K giants of 1 – 1.5 M$_\odot$ and sometimes show strong Li features (Sneden et al. 1978, Lambert & Sawyer 1984, Gratton 1991). These carbon-poor Cepheids are more massive and show no Li, but do show evidence for deep-mixing in their extreme carbon deficiencies. This phenomenon needs further explication as do the weak G-band stars.

To return to the oxygen problem briefly, Figure 6 illustrates that the [O/Fe] ratios are very close to 0. In fact, the mean [O/Fe] value for the total sample is +0.05. This is different than the oxygen result of Luck et al. (2011) who found <[O/Fe]> = −0.09. They noted that because their oxygen abundances were based on solar gf values that the subsolar oxygen abundance problem persisted in intermediate mass stars. They noted a need for an oxygen analysis of intermediate mass stars based on absolute oscillator strengths identical to those used for the solar abundance. This is precisely what this analysis presents. Our [O/H] ratios assume a solar oxygen abundance of log $\varepsilon_O$ = 8.69 dex based on the analysis of Asplund et al. (2004). If one prefers the Asplund et al. (2009) oxygen abundance of 8.75, then the [O/H] and [O/Fe] ratios of the Cepheids would decrease by 0.06 yielding a mean [O/Fe] of -0.01. Progress in solar abundances as well in the analysis of Cepheids has removed the oxygen deficiency in intermediate-mass stars.

Another classification related question that needs some explication is the status of V526 Aql (see Figures 7 and 8). We previously noted that HK Cas is formally a carbon star. This is also true of V526 Aql. Both stars have very similar stellar parameters (at the individual phase of observation). While HK Cas is classified a Type II Cepheid by Harris (1985), there is no such classification in the literature for V526 Aql. The GCVS does list it as a CEP variable, but it is very close to the galactic plane. Perhaps it is a Type II Cepheid. There are undoubtedly other Type II Cepheids in this study but they do not, in general, make themselves obvious. A possible exception is DR Cep, a 19 day DCEP 1.6 kpc above the galactic plane. However, there is nothing peculiar about its spectrum or abundances. Note that our gradient calculations include V526 Aql as eliminating it from the gradients has little effect on the values.

In Figures 7 through 11 (and Table 5) we present gradient information for carbon, nitrogen, and oxygen. The carbon gradient (Figure 7) is well defined presenting a gradient d[C/H]/dR = −0.08 dex/kpc. This is somewhat steeper than found for iron and very much like that found by Luck et al. (2011). There is a gradient determination for carbon from B stars by Daflon & Cunha (2004). They find d[C/H]dR = −0.037 dex/kpc which is significantly lower than the value determined here. Could evolution in Cepheids be altering the carbon abundance in a non-homogeneous manner? From Figure 7 it is also apparent that there is a large increase in the dispersion of the carbon abundance at $R_G$ > 12 kpc. In lower panel of Figure 7 we show [C/Fe] against galactocentric distance. The slope for this data is d[C/Fe]/dR$_G$ = −0.02 dex/kpc reflecting the difference in the gradients of the two species. If one looks at the confidence interval on the gradient in d[C/Fe]/dR, one notes that d[C/Fe]/dR = 0 is not excluded. The outer galaxy shows a higher fraction of low [C/Fe] values than does the inner galaxy. Since in evolved stars carbon and nitrogen are linked abundances, we next turn to the nitrogen data.

Nitrogen ratios are given as a function of galactocentric radius in Figure 8. Nitrogen shows a fair amount of scatter, but has a definite gradient in [N/H]. The nitrogen gradient is d[N/H]/dR$_G$ =

−0.049 dex/kpc. This gradient is in excellent agreement with the Daflon & Cunha (2004) value for B stars of −0.046 dex/kpc. In some ways the nitrogen data is a mirror image of the carbon data in that in the outer galaxy the [C/Fe] ratio are low, while the [N/Fe] ratios are high. In standard stellar evolution the sum of C+N remains constant which leads us to examine the behavior of the summed species as shown in Figure 9. [(C+N)/H] shows the about the same gradient as iron leading to no net gradient in [(C+N)/Fe]. However, it appears that the outer galaxy shows more scatter than the inner galaxy in C+N and that beyond about 15 kpc the stars are enhanced in C+N relative to iron. However, this result must be interpreted with caution as it depends on very few stars. Perhaps oxygen will tell us more.

Studies of H II regions, B stars, and planetary nebula most often quote gradients in terms of d[O/H]/dR$_G$. Maciel & Costa (2010) in their review of gradients note that H II regions yield gradients in the range of −0.044 dex/kpc (Esteban et al. 2005) to −0.060 dex/kpc (Rudolph et al. 2006). The more common determination for the oxygen gradient from H II regions is about −0.04 dex/kpc. Our information for [O/H] is presented in Figure 10 (and Table 5). The gradient d[O/H]/dR$_G$ found here is −0.055 dex/kpc which is in accord with the determinations from H II regions. Since Cepheids are young stars freshly struck from H II regions and GMCs, it is not surprising they give much the same gradient. B stars are the direct predecessors of Cepheids and one expects that gradients determined from them should be very similar to those found in Cepheids. Daflon & Cunha (2004) have examined a number of OB associations and derive gradients for a mix of elements. For oxygen they derive d[O/H]/dR = −0.031 dex/kpc. This is in fair agreement with our value.

Gradients from studies of planetary nebulae suffer from the lack of homogeneity in age of planetary nebulae. In fact, the ages of planetary nebulae range from about 1 Gyr to 8 Gyr (Maciel & Costa 2010). This means that gradients derived from planetary nebula are not strictly comparable to those determined from a young object such as a Cepheid. Nonetheless, the gradient in d[O/H]/dR$_G$ for planetary nebulae is about −0.04 to −0.09 dex/kpc (Maciel & Costa 2010). They also note that a number of studies indicate very flat to non-existent gradients. The latter result brings to mind the effects of galactic "churning" (Sellwood & Binney 2002, Schönrich & Binney 2008, Casagrande et al. 2011) which erases gradients over time due to orbit migration.

Examination of the [O/Fe] data reveals a modest increase in the ratio in the outer regions of the Galaxy as well as an increase in the scatter. Since [Fe/H] also depends on galactocentric radius in Cepheids, we can expect an increase in [O/Fe] with decreasing [Fe/H] which is found in the data. This is somewhat reminiscent of the relation found in metal-poor stars. In the metal-poor stars the dependence is a result of age-dependent contributions from Type I versus Type II supernovae. In Cepheids, the composition is the sum of all previous processing up to/through the current epoch. The dependence of [O/Fe] on [Fe/H] in Cepheids is thus a result of the dependence of each species on galactocentric radius and differential nucleosynthesis and processing as a function of galactocentric radius, not age. However, this does not imply that each region in the galaxy is merely a scaled version of others. If one looks at the C+N+O trends in Figure 11 we see a coherent gradient in [(C+N+O)/H] with the same slope as found in iron. The outer regions of the Galaxy may show more scatter in [(C+N+O)/Fe] than do the inner regions potentially implying that the outer regions are more susceptible to local effects than are the inner region. What is obviously needed are more abundances from the outer regions of the Galaxy.

### 4.3.4. Alpha and Heavy Element Gradients

Figure 12 shows an array of radial gradients. Si and Ca are typical α species while Y, Nd, and Eu are s/r process elements. Table 5 gives the gradients for all of the species in Figure 12 along with all other elements examined in this analysis. These figures are similar to those found in Luck et al. (2011). What is different is that with this larger self-consistent sample gradients are now certain in all examined species including the lanthanides. Of interest is the Si gradient of Daflon & Cunha (2004) derived from O/B stars. They find d[Si/H]/dR = −0.040 dex/kpc which compares very well with our value of −0.048 dex/kpc. Daflon & Cunha also derive gradients for Mg and Al of −0.052 and −0.048 dex/kpc respectively while we find for the same elements gradients of −0.048 and −0.049. The agreement is excellent.

### 5. Summary

With the newly analyzed galactic Cepheids we have enlarged our homogeneous sample of stars used for galactic abundance gradient investigation to about 310 objects. Including the stars of Papers I-VI the total Cepheid sample approaches 400 objects. The primary results obtained in present study are:

- The abundance distribution in galactic disk shows what is on the whole a linear structure. The galactic abundance gradient in the present epoch is approximately d[Fe/H]/dR$_G$ = −0.06 dex / kpc.
- There is little or no evidence in our data (especially [Fe/H]) for a flattening of the gradient beyond 10 kpc from the galactic center. However, that region shows an increased dispersion in abundance which given inadequate sampling could lead to an under- or over- estimate of the gradient. We do not claim that we have definitive evidence one way or the other, only that our results are more consistent with a large variation in abundance.
- The CNO data shows gradients consistent with that found from the iron data. There is an indication that at larger radii that CNO, like iron, shows increased scatter.
- We have found the first Cepheid with a pre- first dredge-up lithium content.
- The existence of very carbon-poor Cepheids reveals that intermediate-mass stars can undergo an as yet theoretically unpredicted deep-mixing event.

What is next in the study of the distribution of the elements in the Galaxy? Extending these studies to larger distances will be challenging given the paucity of candidates. From the point of view of abundances and understanding of Cepheids themselves, the next target should be the peculiar Cepheids identified here. This means not only abundance studies, but also theoretical evolutionary studies to identify the stage where a W Vir star can present itself as a carbon star. Also of use would be a re-analysis of the Paper I-VI stars not re-observed as part of this study. A general non-LTE study of CNO in supergiants would greatly improve the reliability of those abundances. A starting point would be a non-LTE treatment of the O I triplet at 777 nm. There are a number of Cepheids here that do not have distances due to a lack of B and V magnitude data. That should be rectified. As a last point, better distances by way of GAIA will certainly reduce the uncertainties in the gradient.

We would like to thank Bengt Edvardsson and Ulrike Heiter for providing additional models to supplement the standard MARCS atmospheric model grid. DLL thanks the Robert A. Welch Foundation of Houston, Texas for support through grant F-634. We would like to thank the Resident Astronomers of the Hobby-Eberly Telescope (HET) for their able help in obtaining these observations. The HET is a joint project of the University of Texas at Austin, the Pennsylvania State University, Stanford University, Ludwig-Maximilians-Universität München, and Georg-August-Universität Göttingen. The HET is named in honor of its principal benefactors, William P. Hobby and Robert E. Eberly.

**Table 1**
Program Stars

| Cepheid | Type | Period (days) | Mv | <V> | E(B-V) | d (kpc) | Rg (kpc) |
|---:|:---:|---:|---:|---:|---:|---:|---:|
| U Aql | DCEP | 7.02 | -3.54 | 6.45 | 0.36 | 0.58 | 7.42 |
| TT Aql | DCEP | 13.75 | -4.32 | 7.14 | 0.44 | 1.02 | 7.10 |
| BC Aql | CEP | 2.91 | -2.52 | 13.15 | 0.37 | 7.82 | 3.74 |
| EV Aql | CEP | 38.71 | -5.53 | 11.90 | 0.75 | 10.04 | 7.54 |
| FM Aql | DCEP | 6.11 | -3.38 | 8.27 | 0.59 | 0.89 | 7.29 |
| FN Aql | DCEPS | 9.48 | -4.29 | 8.38 | 0.48 | 1.67 | 6.68 |
| KL Aql | DCEP | 6.11 | -3.38 | 10.17 | 0.23 | 3.62 | 6.54 |
| V336 Aql | DCEP | 7.30 | -3.59 | 9.85 | 0.61 | 1.95 | 6.38 |
| V493 Aql | DCEP | 2.99 | -2.55 | 11.08 | 0.65 | 2.02 | 6.30 |
| V526 Aql | CEP | 4.21 | -2.95 | 12.71 | 0.94 | 3.36 | 5.82 |
| V916 Aql | DCEP | 13.44 | -4.30 | 10.79 | 1.04 | 2.22 | 6.56 |
| V1344 Aql | DCEP | 7.48 | -3.61 | 7.77 | 0.55 | 0.84 | 7.27 |
| V1359 Aql | DCEPS: | 3.73 | -2.81 | 9.06 | 0.66 | 0.88 | 7.26 |
| Y Aur | DCEP | 3.86 | -2.85 | 9.61 | 0.38 | 1.77 | 9.63 |
| RT Aur | DCEP | 3.73 | -2.81 | 5.45 | 0.06 | 0.41 | 8.30 |
| RX Aur | DCEP | 11.62 | -4.13 | 7.66 | 0.26 | 1.54 | 9.40 |
| SY Aur | DCEP | 10.14 | -3.97 | 9.07 | 0.43 | 2.14 | 9.97 |
| YZ Aur | DCEP | 18.19 | -4.65 | 10.33 | 0.54 | 4.45 | 12.28 |
| AN Aur | DCEP | 10.29 | -3.99 | 10.46 | 0.56 | 3.34 | 11.16 |
| AO Aur | DCEP | 6.76 | -3.50 | 10.86 | 0.43 | 3.92 | 11.81 |
| AS Aur | DCEP | 3.18 | -2.62 | 11.94 | 0.41 | 4.45 | 12.34 |
| AX Aur | DCEP | 3.05 | -2.57 | 12.41 | 0.57 | 4.25 | 12.15 |
| BK Aur | DCEP | 8.00 | -3.69 | 9.43 | 0.42 | 2.24 | 10.01 |
| CO Aur | CEP(B) | 1.78 | -1.95 | 7.71 | 0.22 | 0.62 | 8.51 |
| CY Aur | DCEP | 13.85 | -4.33 | 11.85 | 0.77 | 5.50 | 13.21 |
| ER Aur | DCEP | 15.69 | -4.48 | 11.52 | 0.49 | 7.59 | 15.36 |
| EW Aur | DCEP | 2.66 | -2.41 | 13.51 | 0.60 | 6.24 | 14.02 |
| FF Aur | DCEP | 2.12 | -2.15 | 13.70 | 0.59 | 6.13 | 13.91 |
| GT Aur | DCEP | 4.40 | -3.00 | 12.19 | 0.36 | 6.39 | 14.13 |
| GV Aur | DCEP | 5.26 | -3.21 | 12.08 | 0.55 | 5.00 | 12.86 |
| IN Aur | DCEP | 4.91 | -3.13 | 13.81 | 0.90 | 6.38 | 14.22 |
| V335 Aur | DCEP | 3.41 | -2.70 | 12.46 | 0.63 | 4.25 | 12.11 |
| RW Cam | DCEP | 16.41 | -4.53 | 8.69 | 0.63 | 1.72 | 9.35 |
| RX Cam | DCEP | 7.91 | -3.68 | 7.68 | 0.53 | 0.85 | 8.61 |
| TV Cam | DCEP | 5.29 | -3.21 | 11.66 | 0.61 | 3.79 | 11.20 |
| AB Cam | DCEP | 5.79 | -3.32 | 11.85 | 0.66 | 4.07 | 11.44 |
| AC Cam | DCEP | 4.16 | -2.93 | 12.60 | 0.92 | 3.27 | 10.65 |
| AD Cam | DCEP | 11.26 | -4.09 | 12.56 | 0.86 | 5.92 | 13.05 |
| AM Cam | CEP | 4.00 | -2.89 | 13.50 | 0.98 | 4.39 | 11.62 |
| CK Cam | DCEP | 3.29 | -2.66 | 7.54 | 0.43 | 0.58 | 8.42 |
| LO Cam | DCEP | 12.64 | -4.23 | 11.05 | 0.55 | 4.98 | 12.32 |

**Table 1**
Program Stars

| Cepheid | Type | Period (days) | Mv | <V> | E(B-V) | d (kpc) | Rg (kpc) |
|---|---|---|---|---|---|---|---|
| MN Cam | DCEP | 8.20 | -3.72 | … | … | … | … |
| MQ Cam | DCEP | 6.60 | -3.47 | … | … | … | … |
| SS CMa | DCEP | 12.36 | -4.20 | 9.92 | 0.55 | 2.92 | 9.72 |
| TV CMa | DCEP | 4.67 | -3.07 | 10.58 | 0.56 | 2.35 | 9.65 |
| RS Cas | DCEP | 6.30 | -3.41 | 9.93 | 0.83 | 1.35 | 8.55 |
| SZ Cas | DCEPS | 13.62 | -4.71 | 9.85 | 0.79 | 2.51 | 9.83 |
| UZ Cas | DCEP | 4.26 | -2.96 | 11.34 | 0.50 | 3.47 | 10.31 |
| VV Cas | DCEP | 6.21 | -3.40 | 10.72 | 0.53 | 3.05 | 10.14 |
| VW Cas | DCEP | 5.99 | -3.36 | 10.70 | 0.45 | 3.30 | 10.15 |
| XY Cas | DCEP | 4.50 | -3.02 | 9.94 | 0.53 | 1.77 | 8.98 |
| AP Cas | DCEP | 6.85 | -3.51 | 11.54 | 0.79 | 3.15 | 9.89 |
| AS Cas | CEP(B) | 3.02 | -2.56 | 12.25 | 0.83 | 2.68 | 9.53 |
| AW Cas | DCEP | 4.28 | -2.97 | 12.07 | 0.87 | 2.78 | 9.79 |
| AY Cas | DCEP | 2.87 | -2.50 | 11.54 | 0.74 | 2.14 | 9.37 |
| BF Cas | DCEP | 3.63 | -2.77 | 12.47 | 0.74 | 3.72 | 10.21 |
| BP Cas | DCEP | 6.27 | -3.41 | 10.92 | 0.90 | 1.92 | 9.14 |
| BV Cas | DCEP | 5.40 | -3.24 | 12.40 | 0.99 | 3.07 | 10.06 |
| BY Cas | DCEPS | 3.22 | -3.03 | 10.37 | 0.76 | 1.55 | 8.97 |
| CD Cas | DCEP | 7.80 | -3.66 | 10.74 | 0.78 | 2.38 | 9.18 |
| CE Cas A | DCEP | 5.14 | -3.18 | 10.92 | 0.56 | 2.89 | 9.55 |
| CE Cas B | DCEP | 4.48 | -3.02 | 11.06 | 0.53 | 2.99 | 9.62 |
| CF Cas | DCEP | 4.88 | -3.12 | 11.14 | 0.55 | 3.11 | 9.70 |
| CG Cas | DCEP | 4.37 | -2.99 | 11.34 | 0.66 | 2.75 | 9.47 |
| CT Cas | DCEP | 3.81 | -2.83 | 12.26 | 0.75 | 3.39 | 10.01 |
| CZ Cas | DCEP | 5.66 | -3.29 | 11.71 | 0.82 | 2.96 | 9.52 |
| DW Cas | DCEP | 5.00 | -3.15 | 11.11 | 0.84 | 2.03 | 8.92 |
| EX Cas | DCEP | 4.31 | -2.97 | 12.80 | 0.75 | 4.69 | 10.87 |
| FO Cas | DCEP | 6.80 | -3.50 | 14.30 | 0.76 | 11.65 | 16.93 |
| FW Cas | DCEP | 6.24 | -3.40 | 12.44 | 0.87 | 4.06 | 10.59 |
| GL Cas | DCEP | 4.01 | -2.89 | 12.74 | 0.71 | 4.69 | 11.61 |
| GM Cas | DCEP | 7.47 | -3.61 | 11.88 | 1.15 | 2.27 | 9.63 |
| GO Cas | DCEP | 3.24 | -2.64 | 13.22 | 1.05 | 3.10 | 10.35 |
| HK Cas | DCEP | 2.50 | -2.34 | 14.30 | 0.38 | 12.11 | 17.66 |
| IO Cas | DCEP | 5.60 | -3.28 | 13.67 | 0.58 | 10.30 | 16.51 |
| KK Cas | DCEP | 8.19 | -3.72 | 11.92 | 0.94 | 3.30 | 9.77 |
| LT Cas | DCEP | 5.90 | -3.34 | 12.53 | 0.69 | 5.38 | 12.32 |
| NP Cas | DCEP | 6.17 | -3.39 | 13.58 | 1.12 | 4.67 | 11.00 |
| NY Cas | DCEPS | 2.82 | -2.88 | 13.33 | 0.33 | 10.65 | 16.22 |
| OP Cas | DCEP | 5.51 | -3.26 | 11.91 | 0.97 | 2.57 | 9.53 |
| OZ Cas | DCEP | 5.08 | -3.17 | 13.47 | 1.55 | 2.12 | 9.29 |

**Table 1**
Program Stars

| Cepheid | Type | Period (days) | Mv | <V> | E(B-V) | d (kpc) | Rg (kpc) |
|---:|:---:|---:|---:|---:|---:|---:|---:|
| PW Cas | DCEP | 4.00 | -2.89 | 13.08 | 0.69 | 5.59 | 11.31 |
| V342 Cas | DCEP | 3.92 | -2.86 | 12.03 | 0.69 | 3.42 | 9.56 |
| V395 Cas | DCEP | 4.04 | -2.90 | 10.72 | 0.57 | 2.26 | 9.55 |
| V407 Cas | DCEP | 4.58 | -3.05 | 12.00 | 0.72 | 3.48 | 9.95 |
| V556 Cas | DCEP | 6.03 | -3.36 | … | … | … | … |
| V1017 Cas | DCEP | 4.64 | -3.06 | … | … | … | … |
| V1019 Cas | DCEPS | 3.62 | -3.17 | … | … | … | … |
| V1020 Cas | DCEP | 4.74 | -3.08 | … | … | … | … |
| Delta Cep | DCEP | 5.37 | -3.23 | 3.95 | 0.08 | 0.24 | 7.97 |
| AK Cep | DCEP | 7.23 | -3.58 | 11.18 | 0.67 | 3.30 | 9.32 |
| CN Cep | DCEP | 9.50 | -3.89 | 12.38 | 1.11 | 3.42 | 9.80 |
| DR Cep | DCEP | 19.08 | -4.70 | 12.85 | 0.78 | 10.18 | 13.38 |
| IY Cep | CEP | 5.66 | -3.29 | 12.84 | 0.86 | 4.65 | 9.91 |
| MU Cep | CEP | 3.77 | -2.82 | 12.24 | 0.79 | 3.16 | 9.20 |
| CD Cyg | DCEP | 17.07 | -4.58 | 8.95 | 0.49 | 2.43 | 7.47 |
| EP Cyg | DCEP | 4.29 | -2.97 | 12.74 | 0.65 | 5.24 | 7.55 |
| EU Cyg | CEP | 14.99 | -4.42 | 13.91 | 1.06 | 9.59 | 9.92 |
| EX Cyg | DCEP | 4.85 | -3.11 | 12.98 | 1.03 | 3.59 | 7.32 |
| EZ Cyg | DCEP | 11.66 | -4.13 | 11.09 | 0.78 | 3.44 | 7.28 |
| GH Cyg | DCEP | 7.82 | -3.67 | 9.92 | 0.63 | 2.05 | 7.33 |
| GI Cyg | DCEP | 5.78 | -3.32 | 11.75 | 0.74 | 3.41 | 7.47 |
| GL Cyg | CEP | 3.37 | -2.69 | 13.72 | 0.76 | 6.18 | 8.69 |
| IY Cyg | CEP | 21.76 | -4.86 | 13.10 | 1.14 | 7.15 | 8.45 |
| KX Cyg | DCEP | 20.05 | -4.76 | 11.94 | 1.68 | 1.80 | 7.75 |
| V347 Cyg | DCEP | 8.74 | -3.80 | 12.48 | 1.05 | 3.79 | 8.52 |
| V356 Cyg | DCEP | 5.06 | -3.16 | 12.44 | 0.74 | 4.39 | 8.92 |
| V396 Cyg | DCEP | 33.24 | -5.35 | 11.42 | 1.09 | 4.44 | 8.29 |
| V459 Cyg | DCEP | 7.25 | -3.58 | 10.60 | 0.76 | 2.21 | 8.22 |
| V438 Cyg | DCEP | 11.21 | -4.09 | 10.94 | 1.08 | 2.04 | 7.72 |
| V492 Cyg | DCEP | 7.58 | -3.63 | 12.39 | 1.01 | 3.54 | 7.65 |
| V495 Cyg | DCEP | 6.72 | -3.49 | 10.62 | 0.98 | 1.55 | 7.60 |
| V514 Cyg | DCEP | 5.10 | -3.17 | 11.43 | 1.08 | 1.66 | 7.93 |
| V520 Cyg | DCEP | 4.05 | -2.90 | 10.85 | 0.76 | 1.81 | 8.03 |
| V538 Cyg | DCEP | 6.12 | -3.38 | 10.46 | 0.64 | 2.25 | 8.41 |
| V547 Cyg | DCEP | 6.23 | -3.40 | 13.39 | 0.95 | 5.57 | 8.10 |
| V609 Cyg | DCEP | 31.08 | -5.27 | 11.00 | 1.26 | 2.77 | 8.64 |
| V621 Cyg | DCEP | 5.86 | -3.33 | 11.75 | 0.53 | 4.72 | 9.41 |
| V1020 Cyg | DCEP | 4.92 | -3.13 | 13.71 | 1.27 | 3.55 | 7.43 |
| V1025 Cyg | DCEP | 6.96 | -3.53 | 12.98 | 1.06 | 4.16 | 7.42 |
| V1033 Cyg | CEP | 4.95 | -3.13 | 13.03 | 1.02 | 3.77 | 7.50 |
| V1046 Cyg | CEP | 4.95 | -3.13 | 12.36 | 1.09 | 2.48 | 7.62 |

**Table 1**
Program Stars

| Cepheid | Type | Period (days) | Mv | <V> | E(B-V) | d (kpc) | Rg (kpc) |
|---:|:---:|---:|---:|---:|---:|---:|---:|
| V1364 Cyg | DCEP | 12.97 | -4.26 | 13.26 | 1.24 | 5.05 | 7.99 |
| V1397 Cyg | DCEP | 4.64 | -3.06 | … | … | … | … |
| EK Del | CEP | 2.05 | -2.11 | 12.31 | 0.12 | 6.40 | 6.72 |
| Zeta Gem | DCEP | 10.15 | -3.97 | 3.92 | 0.01 | 0.37 | 8.25 |
| W Gem | DCEP | 7.91 | -3.68 | 6.95 | 0.26 | 0.92 | 8.78 |
| RZ Gem | DCEP | 5.53 | -3.26 | 10.01 | 0.56 | 1.95 | 9.84 |
| AA Gem | DCEP | 11.30 | -4.10 | 9.72 | 0.31 | 3.66 | 11.55 |
| AD Gem | DCEP | 3.79 | -2.82 | 9.86 | 0.17 | 2.66 | 10.48 |
| BB Gem | DCEP | 2.31 | -2.25 | 11.43 | 0.43 | 2.86 | 10.64 |
| BW Gem | DCEP | 2.64 | -2.40 | 12.00 | 0.56 | 3.30 | 11.17 |
| DX Gem | DCEPS | 3.14 | -3.00 | 10.75 | 0.43 | 2.96 | 10.75 |
| BB Her | DCEP | 7.51 | -3.62 | 10.09 | 0.39 | 3.08 | 6.05 |
| V Lac | DCEP | 4.98 | -3.14 | 8.94 | 0.34 | 1.58 | 8.48 |
| X Lac | DCEPS | 5.44 | -3.64 | 8.41 | 0.34 | 1.56 | 8.48 |
| Y Lac | DCEP | 4.32 | -2.98 | 9.15 | 0.21 | 1.95 | 8.42 |
| Z Lac | DCEP | 10.89 | -4.05 | 8.42 | 0.37 | 1.80 | 8.56 |
| BG Lac | DCEP | 5.33 | -3.22 | 8.88 | 0.30 | 1.69 | 8.16 |
| DF Lac | DCEP | 4.48 | -3.02 | 11.87 | 0.56 | 4.17 | 9.68 |
| RR Lac | DCEP | 6.42 | -3.44 | 8.85 | 0.32 | 1.78 | 8.55 |
| FQ Lac | CEP: | 11.27 | -4.09 | 13.74 | 0.55 | 16.16 | 19.83 |
| V411 Lac | DCEPS | 2.91 | -2.91 | 7.95 | 0.15 | 1.18 | 8.21 |
| V473 Lyr | DCEPS: | 1.49 | -1.74 | 6.18 | 0.02 | 0.37 | 7.73 |
| T Mon | DCEP | 27.02 | -5.11 | 6.12 | 0.18 | 1.35 | 9.15 |
| TW Mon | CEP | 7.10 | -3.55 | 12.58 | 0.66 | 6.28 | 13.61 |
| TX Mon | DCEP | 8.70 | -3.79 | 10.96 | 0.49 | 4.33 | 11.74 |
| TY Mon | DCEP | 4.02 | -2.89 | 11.74 | 0.54 | 3.76 | 11.23 |
| TZ Mon | DCEP | 7.43 | -3.61 | 10.76 | 0.42 | 4.00 | 11.44 |
| UY Mon | DCEPS | 2.40 | -2.69 | 9.39 | 0.06 | 2.37 | 10.09 |
| VZ Mon | DCEP | 5.09 | -3.17 | 13.59 | 0.99 | 5.14 | 12.85 |
| WW Mon | DCEP | 4.66 | -3.07 | 12.51 | 0.60 | 5.29 | 12.94 |
| XX Mon | DCEP | 5.46 | -3.25 | 11.90 | 0.57 | 4.60 | 11.95 |
| YY Mon | DCEP | 3.45 | -2.72 | 13.76 | 0.64 | 7.65 | 14.60 |
| AA Mon | DCEP | 3.94 | -2.87 | 12.71 | 0.79 | 4.01 | 11.36 |
| AC Mon | DCEP | 8.01 | -3.70 | 10.07 | 0.48 | 2.75 | 10.12 |
| BE Mon | DCEP | 2.71 | -2.43 | 10.58 | 0.59 | 1.66 | 9.43 |
| BV Mon | DCEP | 3.01 | -2.56 | 11.43 | 0.58 | 2.64 | 10.28 |
| CS Mon | CEP | 6.73 | -3.49 | 11.00 | 0.51 | 3.73 | 11.40 |
| CU Mon | DCEP | 4.71 | -3.08 | 13.61 | 0.75 | 7.10 | 14.45 |
| CV Mon | DCEP | 5.38 | -3.23 | 10.30 | 0.72 | 1.74 | 9.46 |
| EE Mon | DCEP | 4.81 | -3.10 | 12.94 | 0.46 | 8.10 | 15.02 |

**Table 1**
Program Stars

| Cepheid | Type | Period (days) | Mv | <V> | E(B-V) | d (kpc) | Rg (kpc) |
|---|---|---|---|---|---|---|---|
| EK Mon | DCEP | 3.96 | -2.88 | 11.05 | 0.56 | 2.66 | 10.19 |
| FG Mon | DCEP | 4.50 | -3.02 | 13.31 | 0.65 | 7.02 | 13.94 |
| FI Mon | DCEP | 3.29 | -2.66 | 12.92 | 0.51 | 6.10 | 13.11 |
| FT Mon | DCEP | 4.83 | -3.11 | 12.73 | 0.28 | 9.71 | 17.14 |
| V446 Mon | CEP | 1.92 | -2.03 | 14.54 | 0.84 | 5.93 | 13.34 |
| V447 Mon | CEP | 2.48 | -2.33 | 14.11 | 0.92 | 4.94 | 12.47 |
| V465 Mon | DCEP | 2.71 | -2.44 | 10.38 | 0.24 | 2.54 | 10.09 |
| V484 Mon | CEP | 3.14 | -2.60 | 13.73 | 0.68 | 6.75 | 14.04 |
| V495 Mon | DCEP | 4.10 | -2.92 | 12.43 | 0.61 | 4.73 | 12.11 |
| V504 Mon | CEP | 2.77 | -2.46 | 11.81 | 0.54 | 3.22 | 10.68 |
| V508 Mon | DCEP | 4.13 | -2.93 | 10.52 | 0.31 | 3.09 | 10.71 |
| V510 Mon | DCEP | 7.46 | -3.61 | 12.68 | 0.80 | 5.50 | 12.96 |
| V526 Mon | DCEPS | 2.68 | -2.82 | 8.60 | 0.09 | 1.68 | 9.32 |
| CR Ori | DCEP | 4.91 | -3.13 | 12.29 | 0.54 | 5.44 | 13.20 |
| DF Ori | DCEP | 3.18 | -2.62 | 13.52 | 0.74 | 5.62 | 13.33 |
| AU Peg | WVir | 2.40 | -2.29 | 9.23 | 0.10 | 1.74 | 7.48 |
| QQ Per | WVir | 11.17 | -4.08 | 13.71 | 0.10 | 31.22 | 37.52 |
| SV Per | DCEP | 11.13 | -4.08 | 9.02 | 0.41 | 2.27 | 10.09 |
| SX Per | DCEP | 4.29 | -2.97 | 11.16 | 0.47 | 3.34 | 11.06 |
| UX Per | DCEP | 4.57 | -3.04 | 11.66 | 0.51 | 4.08 | 11.10 |
| UY Per | DCEP | 5.37 | -3.23 | 11.34 | 0.87 | 2.24 | 9.64 |
| VY Per | DCEP | 5.53 | -3.26 | 11.26 | 0.95 | 1.97 | 9.40 |
| VX Per | DCEP | 10.89 | -4.05 | 9.31 | 0.48 | 2.32 | 9.63 |
| AS Per | DCEP | 4.97 | -3.14 | 9.72 | 0.67 | 1.37 | 9.15 |
| AW Per | DCEP | 6.46 | -3.45 | 7.49 | 0.49 | 0.74 | 8.62 |
| BM Per | DCEP | 22.96 | -4.92 | 10.39 | 0.87 | 3.15 | 10.85 |
| CI Per | CEP: | 3.30 | -2.66 | 12.68 | 0.26 | 7.90 | 14.46 |
| DW Per | DCEP | 3.65 | -2.78 | 11.56 | 0.63 | 2.90 | 10.20 |
| GP Per | CEP | 2.04 | -2.11 | 14.09 | 0.70 | 6.09 | 13.72 |
| HQ Per | CEP | 8.64 | -3.78 | 11.60 | 0.56 | 5.15 | 12.90 |
| HZ Per | DCEP | 11.28 | -4.09 | 13.77 | 1.21 | 6.14 | 13.77 |
| MM Per | DCEP | 4.12 | -2.92 | 10.80 | 0.49 | 2.68 | 10.30 |
| OT Per | DCEP | 26.09 | -5.07 | 13.48 | 1.37 | 6.66 | 14.27 |
| X Pup | DCEP | 25.97 | -5.06 | 8.46 | 0.40 | 2.79 | 9.73 |
| RS Pup | DCEP | 41.39 | -5.60 | 6.95 | 0.46 | 1.64 | 8.54 |
| VX Pup | CEP(B) | 3.01 | -2.56 | 8.33 | 0.13 | 1.24 | 8.64 |
| VZ Pup | DCEP | 23.17 | -4.93 | 9.62 | 0.46 | 4.11 | 10.40 |
| WX Pup | DCEP | 8.94 | -3.82 | 9.06 | 0.30 | 2.41 | 9.30 |
| AQ Pup | DCEP | 30.08 | -5.23 | 8.79 | 0.52 | 2.95 | 9.49 |
| AD Pup | DCEP | 13.60 | -4.31 | 9.86 | 0.31 | 4.28 | 10.61 |
| BN Pup | DCEP | 13.67 | -4.32 | 9.88 | 0.42 | 3.72 | 9.92 |

**Table 1**
Program Stars

| Cepheid | Type | Period (days) | Mv | <V> | E(B-V) | d (kpc) | Rg (kpc) |
|---|---|---|---|---|---|---|---|
| V335 Pup | DCEPS | 4.86 | -3.51 | 8.72 | 0.15 | 2.22 | 9.19 |
| Y Sct | DCEP | 10.34 | -3.99 | 9.63 | 0.76 | 1.72 | 6.37 |
| Z Sct | DCEP | 12.90 | -4.25 | 9.60 | 0.52 | 2.73 | 5.60 |
| RU Sct | DCEP | 19.70 | -4.74 | 9.47 | 0.92 | 1.76 | 6.40 |
| SS Sct | DCEP | 3.67 | -2.79 | 8.21 | 0.33 | 0.98 | 7.03 |
| TY Sct | DCEP | 11.05 | -4.07 | 10.83 | 0.97 | 2.27 | 5.99 |
| BX Sct | DCEP | 6.41 | -3.44 | 12.24 | 1.25 | 2.11 | 6.14 |
| CK Sct | DCEP | 7.42 | -3.61 | 10.59 | 0.76 | 2.24 | 5.98 |
| CM Sct | DCEP | 3.92 | -2.86 | 11.11 | 0.72 | 2.13 | 6.08 |
| CN Sct | DCEP | 9.99 | -3.95 | 12.48 | 1.21 | 3.21 | 5.29 |
| EV Sct | DCEPS | 3.09 | -2.99 | 10.14 | 0.66 | 1.59 | 6.48 |
| AA Ser | DCEP | 17.14 | -4.58 | 12.23 | 1.53 | 2.38 | 5.99 |
| DV Ser | DCEP | 23.06 | -4.92 | 13.65 | 2.08 | 2.35 | 5.97 |
| DG Sge | DCEP | 4.44 | -3.01 | 13.22 | 1.36 | 2.33 | 6.86 |
| GX Sge | DCEP | 12.90 | -4.25 | 12.46 | 1.25 | 3.44 | 6.53 |
| GY Sge | DCEP | 51.56 | -5.86 | 10.15 | 1.19 | 2.73 | 6.72 |
| ST Tau | DCEP | 4.03 | -2.90 | 8.22 | 0.37 | 0.97 | 8.83 |
| AE Tau | CEP | 3.90 | -2.86 | 11.68 | 0.58 | 3.43 | 11.33 |
| AV Tau | DCEP | 3.62 | -2.77 | 12.34 | 0.85 | 2.97 | 10.87 |
| S Vul | DCEP | 68.04 | -6.18 | 8.96 | 0.73 | 3.63 | 7.07 |
| U Vul | DCEP | 7.99 | -3.69 | 7.13 | 0.60 | 0.59 | 7.58 |
| AS Vul | DCEP | 12.23 | -4.19 | 12.25 | 1.19 | 3.33 | 7.10 |
| DG Vul | CEP | 13.61 | -4.31 | 11.37 | 1.20 | 2.29 | 7.23 |
| GSC 0754-1993 | DCEP | 4.98 | -3.14 | … | … | … | … |
| GSC 2418-1443 | DCEP | 7.85 | -3.67 | … | … | … | … |
| GSC 3706-0233 | DCEP | 3.22 | -2.64 | … | … | … | … |
| GSC 3725-0174 | DCEP | 3.09 | -2.59 | … | … | … | … |
| GSC 3726-0565 | DCEP | 5.12 | -3.17 | … | … | … | … |
| GSC 3729-1127 | DCEP | 5.07 | -3.16 | … | … | … | … |
| GSC 3732-0183 | DCEP | 6.60 | -3.47 | … | … | … | … |
| GSC 4009-0024 | DCEP | 4.74 | -3.08 | … | … | … | … |
| GSC 4038-1585 | DCEP | 4.71 | -3.08 | … | … | … | … |
| GSC 4040-1803 | DCEP | 2.11 | -2.15 | … | … | … | … |
| GSC 4265-0193 | DCEP | 4.28 | -2.97 | … | … | … | … |
| GSC 4265-0569 | DCEP | 9.00 | -3.83 | … | … | … | … |

Notes:

Type: Variability type from the General Catalog of Variable Stars (Samus , N. N. et al. 2007-2009) or Wils & Greaves (2004). Stars from Wils & Greaves are identifiable as they lack <V> etc.
Period: Period in days from Berdnikov (2006– private communication to S. Andrievsky) , GCVS, or Wils & Greaves(2004).
$M_V$: Absolute V magnitude computed assuming listed period is the fundamental except for DCEPS where $P_0 = P/0.71$. Period-Luminosity relation and R (=Av/E(B-V) = 3.23) from Fouqué et al. (2007).
<V>: Intensity mean apparent V magnitude from Fernie (1995). Exceptions are CK Cam and V411 Lac from van Leeuven et al. (2007) and LO Cam from Garcia-Melendo (2001).
E(B-V): From Fouqué et al. (2007) or Fernie (1995) as modified by Fouqué et al.
Exceptions are:
BC Aql and BW Gem - Diethelm (1990).
EV Aql, KL Aql, and V1344 Aql – Fernie (1990)
AS Aur, GT Aur, GM Cas, HK Cas, FT Mon, DF Ori, and DV Ser – computed using the method of Kovtyukh et al (2010)
DR Cep, EX Cyg, GL Cyg, EK Del, BB Gem, BB Her, and AU Peg – Harris (1985)
CK Cam and V411 Lac - van Leeuven et al. (2007)
LO Cam – Garcia-Melendo (2001)
d and Rg: Distance in kiloparsecs from the Sun and the galactic center respectively. The Sun is assumed to be 7.9 kpc from the galactic center (McNamara et al. 2000).

**Table 2**
Observing Log

| Cepheid | Telescope | UT Date | UT Time | Exposure (s) | S/N | Phase |
|---:|---:|---:|---:|---:|---:|---:|
| U Aql | H | 2010-05-28 | 10:13:29.93 | 60 | 270 | 0.534 |
| U Aql | H | 2010-05-29 | 09:59:11.33 | 60 | 281 | 0.675 |
| U Aql | H | 2010-07-07 | 07:16:09.13 | 80 | 317 | 0.211 |
| U Aql | H | 2010-08-26 | 04:06:41.33 | 60 | 207 | 0.311 |
| U Aql | H | 2010-09-05 | 03:22:50.66 | 100 | 390 | 0.730 |
| TT Aql | H | 2010-09-27 | 03:10:38.23 | 300 | 201 | 0.394 |
| TT Aql | H | 2010-10-11 | 02:13:39.76 | 150 | 514 | 0.409 |
| BC Aql | H | 2010-06-12 | 08:49:18.14 | 2400 | 76 | 0.039 |
| EV Aql | H | 2010-07-06 | 04:45:14.54 | 1800 | 149 | 0.158 |
| FM Aql | H | 2010-09-19 | 04:10:14.87 | 200 | 270 | 0.279 |
| FM Aql | H | 2010-09-27 | 03:41:27.03 | 720 | 159 | 0.585 |
| FN Aql | H | 2010-04-06 | 11:31:16.32 | 540 | 332 | 0.942 |
| FN Aql | H | 2010-06-02 | 10:48:43.37 | 180 | 278 | 0.951 |
| KL Aql | H | 2009-10-24 | 03:03:58.72 | 600 | 182 | 0.252 |
| V336 Aql | H | 2009-10-10 | 01:43:04.94 | 300 | 188 | 0.040 |
| V336 Aql | H | 2010-04-12 | 11:12:29.33 | 800 | 357 | 0.287 |
| V493 Aql | H | 2010-04-16 | 11:03:50.13 | 900 | 202 | 0.803 |
| V526 Aql | H | 2010-09-04 | 05:02:19.83 | 3000 | 75 | 0.283 |
| V916 Aql | H | 2009-09-23 | 03:50:06.34 | 1200 | 145 | 0.293 |
| V1344 Aql | H | 2009-10-23 | 01:26:55.77 | 100 | 180 | 0.828 |
| V1359 Aql | H | 2010-05-28 | 10:04:25.57 | 300 | 247 | ... |
| V1359 Aql | H | 2010-05-29 | 09:49:56.95 | 300 | 245 | ... |
| V1359 Aql | H | 2010-06-02 | 10:34:37.57 | 250 | 247 | ... |
| Y Aur | H | 2010-11-11 | 11:43:08.17 | 630 | 158 | 0.733 |
| Y Aur | H | 2010-11-22 | 10:51:40.73 | 630 | 196 | 0.574 |
| Y Aur | H | 2010-11-29 | 04:49:55.56 | 630 | 155 | 0.323 |
| RT Aur | H | 2010-11-14 | 07:02:29.56 | 60 | 412 | 0.111 |
| RT Aur | H | 2010-11-22 | 06:42:40.91 | 75 | 489 | 0.253 |
| RX Aur | H | 2010-11-11 | 11:20:59.55 | 150 | 253 | 0.857 |
| SY Aur | H | 2010-11-11 | 11:27:06.72 | 440 | 194 | 0.251 |
| YZ Aur | H | 2010-11-05 | 06:13:33.56 | 1200 | 277 | 0.499 |
| AN Aur | H | 2010-11-08 | 06:00:19.16 | 1200 | 228 | 0.835 |
| AO Aur | H | 2009-02-19 | 05:45:09.34 | 750 | 87 | 0.101 |
| AO Aur | H | 2009-03-27 | 03:27:26.57 | 800 | 96 | 0.410 |
| AO Aur | H | 2010-01-04 | 08:16:27.73 | 900 | 144 | 0.286 |
| AS Aur | H | 2009-01-22 | 07:22:30.33 | 1500 | 99 | 0.346 |
| AX Aur | H | 2009-01-17 | 07:27:18.14 | 2040 | 126 | 0.201 |
| AX Aur | H | 2010-11-19 | 05:53:23.56 | 1200 | 146 | 0.421 |
| BK Aur | H | 2010-11-15 | 05:24:56.36 | 600 | 222 | 0.804 |
| CO Aur | M | 2008-11-15 | 07:45:15.02 | 1800 | 195 | 0.001 |
| CO Aur | M | 2008-11-17 | 08:56:00.64 | 1200 | 312 | 0.148 |
| CO Aur | H | 2009-02-01 | 06:52:40.55 | 180 | 218 | 0.721 |
| CY Aur | H | 2010-11-14 | 05:23:27.55 | 3600 | 226 | 0.975 |
| ER Aur | H | 2010-11-09 | 06:02:05.34 | 3000 | 178 | 0.379 |

**Table 2**
Observing Log

| Cepheid | Telescope | UT Date | UT Time | Exposure (s) | S/N | Phase |
|---:|:---:|:---:|:---:|---:|---:|---:|
| EW Aur | H | 2010-03-22 | 02:36:55.14 | 3300 | 164 | 0.892 |
| FF Aur | H | 2010-09-04 | 09:56:05.23 | 4260 | 34 | 0.381 |
| GT Aur | H | 2009-01-18 | 01:37:48.74 | 2000 | 88 | 0.514 |
| GV Aur | H | 2009-01-10 | 02:41:32.34 | 1500 | 72 | 0.838 |
| GV Aur | H | 2009-01-15 | 07:26:17.75 | 1500 | 108 | 0.826 |
| GV Aur | H | 2009-11-28 | 05:07:24.57 | 2100 | 93 | 0.077 |
| IN Aur | H | 2009-03-01 | 04:07:44.12 | 4140 | 64 | 0.390 |
| V335 Aur | H | 2009-01-19 | 01:41:49.15 | 2100 | 99 | 0.679 |
| RW Cam | H | 2010-11-05 | 05:46:59.36 | 240 | 275 | 0.678 |
| RX Cam | H | 2010-11-11 | 09:21:19.95 | 200 | 300 | 0.879 |
| TV Cam | H | 2009-01-04 | 05:36:23.13 | 1200 | 97 | 0.500 |
| TV Cam | H | 2009-11-27 | 07:59:02.16 | 1667 | 52 | 0.276 |
| TV Cam | H | 2009-11-28 | 08:16:30.57 | 1400 | 95 | 0.467 |
| AB Cam | H | 2010-11-06 | 09:41:49.56 | 2100 | 147 | 0.411 |
| AC Cam | H | 2009-01-31 | 03:00:13.32 | 2400 | 119 | 0.985 |
| AD Cam | H | 2010-11-05 | 04:53:51.93 | 1066 | 88 | 0.656 |
| AD Cam | H | 2010-11-06 | 05:11:16.96 | 3600 | 133 | 0.746 |
| AM Cam | H | 2009-01-03 | 04:19:44.74 | 1860 | 48 | 0.717 |
| AM Cam | H | 2009-01-17 | 04:05:57.34 | 5720 | 128 | 0.217 |
| CK Cam | H | 2009-03-22 | 02:29:29.36 | 180 | 313 | 0.317 |
| CK Cam | H | 2009-11-16 | 06:00:27.70 | 180 | 266 | 0.903 |
| LO Cam | H | 2009-12-17 | 07:10:08.12 | 1200 | 127 | ... |
| MN Cam | H | 2010-01-05 | 06:02:12.95 | 1200 | 114 | 0.412 |
| MQ Cam | H | 2009-12-14 | 02:49:21.16 | 1900 | 50 | 0.929 |
| MQ Cam | H | 2009-12-23 | 02:01:14.34 | 1800 | 94 | 0.288 |
| SS Cma | M | 2008-11-16 | 10:05:45.35 | 1800 | 218 | 0.524 |
| TV Cma | M | 2008-11-17 | 09:53:15.79 | 5400 | 226 | 0.746 |
| RS Cas | M | 2008-08-08 | 07:49:31.02 | 1800 | 145 | 0.214 |
| SZ Cas | H | 2009-10-31 | 09:03:44.90 | 600 | 186 | 0.447 |
| SZ Cas | H | 2009-11-08 | 04:13:54.32 | 700 | 153 | 0.020 |
| SZ Cas | H | 2009-12-12 | 01:35:33.95 | 600 | 210 | 0.508 |
| UZ Cas | H | 2008-09-28 | 09:43:57.72 | 1200 | 130 | 0.522 |
| VV Cas | M | 2008-11-17 | 03:37:38.28 | 3600 | 173 | 0.360 |
| VW Cas | M | 2008-11-16 | 03:04:02.18 | 3600 | 163 | 0.377 |
| XY Cas | H | 2010-11-20 | 05:34:33.96 | 900 | 331 | 0.969 |
| AP Cas | H | 2008-09-05 | 06:42:00.99 | 1200 | 118 | 0.556 |
| AS Cas | H | 2009-09-29 | 05:51:46.32 | 2100 | 112 | 0.947 |
| AW Cas | H | 2009-11-01 | 03:36:15.74 | 1800 | 195 | 0.987 |
| AW Cas | H | 2009-12-07 | 04:22:05.73 | 2000 | 194 | 0.409 |
| AY Cas | H | 2009-11-07 | 03:50:35.94 | 1800 | 131 | 0.711 |
| BF Cas | H | 2008-09-24 | 08:31:38.12 | 2250 | 130 | 0.220 |
| BP Cas | M | 2008-11-17 | 04:42:03.74 | 3600 | 192 | 0.104 |
| BP Cas | H | 2010-11-14 | 04:14:18.33 | 2700 | 308 | 0.997 |
| BV Cas | H | 2009-11-09 | 07:30:51.32 | 2700 | 142 | 0.327 |
| BY Cas | M | 2008-11-16 | 05:46:13.54 | 3600 | 203 | 0.664 |

**Table 2**
Observing Log

| Cepheid | Telescope | UT Date | UT Time | Exposure (s) | S/N | Phase |
|---:|:---:|:---:|:---:|---:|---:|---:|
| CD Cas | M | 2008-11-16 | 01:58:22.59 | 3600 | 201 | 0.086 |
| CE Cas A | H | 2010-08-05 | 08:41:53.92 | 1800 | 213 | 0.699 |
| CE Cas B | H | 2010-08-10 | 10:57:45.57 | 1800 | 253 | 0.199 |
| CF Cas | H | 2010-08-13 | 07:01:24.57 | 1800 | 268 | 0.109 |
| CG Cas | H | 2009-11-08 | 05:27:04.33 | 1500 | 134 | 0.340 |
| CG Cas | H | 2010-01-03 | 02:24:33.38 | 1200 | 197 | 0.138 |
| CT Cas | H | 2008-09-29 | 08:14:18.71 | 1800 | 130 | 0.329 |
| CT Cas | H | 2008-10-10 | 04:07:27.75 | 1800 | 133 | 0.171 |
| CZ Cas | H | 2009-10-31 | 04:40:25.11 | 1620 | 99 | 0.440 |
| CZ Cas | H | 2009-12-21 | 01:52:17.59 | 2400 | 170 | 0.423 |
| CZ Cas | H | 2010-10-15 | 05:32:29.35 | 2100 | 243 | 0.060 |
| DW Cas | H | 2009-10-23 | 02:17:52.54 | 900 | 130 | 0.027 |
| DW Cas | H | 2010-10-07 | 07:29:08.29 | 1200 | 258 | 0.901 |
| EX Cas | H | 2008-08-09 | 08:16:59.62 | 2940 | 93 | 0.837 |
| FO Cas | H | 2009-11-07 | 05:42:27.71 | 7500 | 75 | 0.390 |
| FW Cas | H | 2008-08-25 | 09:48:04.00 | 2100 | 52 | 0.319 |
| FW Cas | H | 2008-09-21 | 05:58:12.51 | 2100 | 117 | 0.623 |
| GL Cas | H | 2008-09-21 | 07:08:52.72 | 2760 | 98 | 0.206 |
| GM Cas | H | 2009-10-24 | 08:27:51.92 | 1800 | 152 | 0.135 |
| GM Cas | H | 2009-12-21 | 01:22:12.17 | 1500 | 156 | 0.860 |
| GO Cas | H | 2008-12-07 | 06:32:47.94 | 3579 | 151 | 0.200 |
| HK Cas | H | 2010-01-07 | 03:03:04.35 | 4447 | 91 | 0.713 |
| IO Cas | H | 2009-11-13 | 02:59:00.94 | 6900 | 106 | 0.165 |
| KK Cas | H | 2008-09-20 | 07:57:15.74 | 1400 | 139 | 0.342 |
| LT Cas | H | 2008-09-18 | 10:14:12.31 | 2220 | 112 | 0.253 |
| NP Cas | H | 2009-11-26 | 04:22:32.93 | 6000 | 158 | 0.735 |
| NY Cas | H | 2009-11-06 | 06:08:56.14 | 2100 | 92 | 0.863 |
| OP Cas | H | 2009-10-15 | 07:27:22.37 | 2100 | 126 | 0.168 |
| OZ Cas | H | 2009-11-01 | 07:07:41.12 | 5400 | 150 | 0.722 |
| PW Cas | H | 2009-10-10 | 06:10:58.56 | 4200 | 98 | 0.187 |
| V342 Cas | H | 2008-08-06 | 10:36:56.61 | 1500 | 126 | 0.876 |
| V395 Cas | H | 2008-12-12 | 02:25:32.34 | 500 | 102 | 0.747 |
| V395 Cas | M | 2008-11-17 | 05:48:11.70 | 3600 | 168 | 0.595 |
| V407 Cas | H | 2008-09-30 | 07:59:49.94 | 1800 | 129 | 0.079 |
| V556 Cas | H | 2010-08-06 | 10:14:55.17 | 2400 | 157 | 0.458 |
| V1017 Cas | H | 2010-01-10 | 02:11:47.93 | 1800 | 113 | 0.591 |
| V1019 Cas | H | 2009-10-24 | 03:21:18.76 | 750 | 82 | 0.223 |
| V1019 Cas | H | 2010-08-03 | 07:38:17.13 | 1500 | 199 | 0.450 |
| V1020 Cas | H | 2009-11-06 | 04:50:29.35 | 2100 | 84 | 0.491 |
| Delta Cep | H | 2010-11-05 | 04:45:52.33 | 20 | 567 | 0.764 |
| AK Cep | H | 2008-08-06 | 10:18:16.39 | 750 | 158 | 0.028 |
| CN Cep | H | 2008-08-25 | 09:09:14.59 | 1920 | 109 | 0.926 |
| DR Cep | H | 2009-10-11 | 04:27:49.36 | 3030 | 78 | 0.696 |
| IY Cep | H | 2008-09-04 | 08:28:32.40 | 3000 | 106 | 0.281 |

**Table 2**
Observing Log

| Cepheid | Telescope | UT Date | UT Time | Exposure (s) | S/N | Phase |
|---:|:---:|:---:|:---:|---:|---:|---:|
| MU Cep | H | 2008-08-04 | 06:41:39.59 | 1740 | 102 | 0.973 |
| CD Cyg | H | 2009-10-15 | 03:54:46.14 | 180 | 180 | 0.146 |
| EP Cyg | H | 2008-10-13 | 03:54:44.54 | 2469 | 98 | 0.300 |
| EU Cyg | H | 2010-05-06 | 09:04:10.94 | 3900 | 97 | 0.243 |
| EX Cyg | H | 2008-11-10 | 01:42:30.13 | 3400 | 97 | 0.746 |
| EZ Cyg | H | 2008-07-29 | 03:54:50.49 | 750 | 106 | 0.391 |
| GH Cyg | M | 2008-11-16 | 00:51:25.50 | 3600 | 138 | 0.403 |
| GI Cyg | H | 2008-08-05 | 02:58:58.79 | 1200 | 106 | 0.394 |
| GL Cyg | H | 2010-11-11 | 02:20:31.13 | 4100 | 93 | 0.856 |
| IY Cyg | H | 2010-04-10 | 10:45:06.94 | 3750 | 75 | 0.964 |
| KX Cyg | H | 2009-11-06 | 03:10:23.34 | 2100 | 196 | 0.157 |
| V347 Cyg | H | 2008-10-08 | 05:11:03.71 | 2160 | 113 | 0.098 |
| V356 Cyg | H | 2008-09-26 | 06:04:43.15 | 2100 | 94 | 0.697 |
| V396 Cyg | H | 2008-08-05 | 03:51:22.63 | 900 | 100 | 0.892 |
| V396 Cyg | H | 2008-10-19 | 03:58:45.13 | 900 | 170 | 0.148 |
| V438 Cyg | H | 2009-11-09 | 02:48:04.74 | 780 | 182 | 0.961 |
| V438 Cyg | H | 2010-05-19 | 08:49:58.32 | 900 | 244 | 0.020 |
| V438 Cyg | H | 2010-11-12 | 02:29:22.58 | 1500 | 265 | 0.785 |
| V459 Cyg | M | 2008-08-09 | 08:32:16.94 | 3600 | 166 | 0.634 |
| V492 Cyg | H | 2008-10-08 | 04:31:00.11 | 2100 | 128 | 0.929 |
| V492 Cyg | H | 2008-10-18 | 03:40:34.52 | 2100 | 142 | 0.243 |
| V495 Cyg | M | 2008-11-17 | 01:20:11.84 | 3600 | 246 | 0.010 |
| V514 Cyg | H | 2009-11-24 | 02:34:51.74 | 1482 | 226 | 0.182 |
| V514 Cyg | H | 2009-11-28 | 01:57:48.53 | 1500 | 124 | 0.961 |
| V514 Cyg | H | 2010-09-04 | 07:30:20.65 | 1800 | 106 | 0.921 |
| V520 Cyg | M | 2008-11-17 | 02:29:00.00 | 3600 | 175 | 0.526 |
| V538 Cyg | M | 2008-08-09 | 07:27:23.77 | 3600 | 180 | 0.191 |
| V547 Cyg | H | 2010-11-03 | 02:47:11.33 | 3947 | 137 | 0.273 |
| V609 Cyg | H | 2009-10-31 | 03:43:13.92 | 1000 | 194 | 0.238 |
| V621 Cyg | H | 2008-08-05 | 04:36:07.98 | 1200 | 107 | 0.832 |
| V1020 Cyg | H | 2010-11-09 | 02:30:17.53 | 3488 | 51 | 0.976 |
| V1025 Cyg | H | 2008-09-26 | 04:59:49.54 | 3400 | 106 | 0.237 |
| V1033 Cyg | H | 2010-10-15 | 04:05:20.35 | 3800 | 166 | 0.982 |
| V1046 Cyg | H | 2010-10-29 | 03:14:59.92 | 2800 | 148 | 0.274 |
| V1364 Cyg | H | 2010-08-06 | 08:45:44.53 | 3975 | 152 | 0.346 |
| V1397 Cyg | H | 2009-10-15 | 05:07:56.57 | 810 | 58 | 0.030 |
| V1397 Cyg | H | 2010-09-19 | 02:29:18.46 | 2000 | 135 | 0.069 |
| EK Del | H | 2009-10-10 | 04:35:02.75 | 1830 | 66 | 0.274 |
| Zeta Gem | H | 2010-11-15 | 07:52:52.95 | 30 | 697 | 0.687 |
| RZ Gem | M | 2008-11-16 | 06:50:43.39 | 3600 | 233 | 0.316 |
| RZ Gem | H | 2009-01-30 | 07:04:10.52 | 500 | 151 | 0.879 |
| RZ Gem | H | 2010-02-06 | 01:27:38.13 | 750 | 232 | 0.117 |
| RZ Gem | H | 2010-03-14 | 03:53:15.54 | 1200 | 315 | 0.646 |
| W Gem | H | 2010-11-11 | 12:14:05.13 | 70 | 271 | 0.034 |
| W Gem | H | 2010-11-22 | 07:22:01.95 | 100 | 264 | 0.399 |

**Table 2**
Observing Log

| Cepheid | Telescope | UT Date | UT Time | Exposure (s) | S/N | Phase |
|---|---|---|---|---|---|---|
| AA Gem | H | 2010-11-08 | 07:08:14.14 | 750 | 265 | 0.422 |
| AA Gem | H | 2010-11-22 | 06:25:29.34 | 750 | 202 | 0.658 |
| AD Gem | H | 2010-11-18 | 07:13:52.11 | 900 | 250 | 0.412 |
| AD Gem | H | 2010-11-22 | 06:48:08.55 | 900 | 229 | 0.463 |
| BB Gem | H | 2010-11-12 | 08:00:17.35 | 2400 | 152 | 0.709 |
| BW Gem | H | 2009-01-29 | 06:51:13.35 | 1500 | 108 | 0.219 |
| BW Gem | H | 2009-11-26 | 05:58:43.15 | 1800 | 124 | 0.437 |
| DX Gem | H | 2010-11-11 | 07:38:33.74 | 1800 | 241 | 0.151 |
| BB Her | H | 2010-06-14 | 10:21:39.34 | 600 | 195 | 0.221 |
| BB Her | H | 2010-09-19 | 03:46:37.26 | 600 | 200 | 0.104 |
| BB Her | H | 2010-10-11 | 02:20:14.15 | 600 | 283 | 0.026 |
| V Lac | H | 2010-11-11 | 04:44:29.16 | 360 | 322 | 0.016 |
| X Lac | H | 2010-11-11 | 04:53:49.57 | 250 | 303 | 0.133 |
| Y Lac | H | 2010-11-13 | 04:00:47.14 | 420 | 244 | 0.006 |
| Z Lac | H | 2010-11-11 | 04:26:36.74 | 250 | 322 | 0.260 |
| RR Lac | H | 2010-11-11 | 04:34:09.16 | 360 | 254 | 0.808 |
| BG Lac | H | 2010-11-13 | 04:11:21.95 | 330 | 205 | 0.210 |
| DF Lac | H | 2008-09-27 | 02:19:59.15 | 2700 | 155 | 0.944 |
| DF Lac | H | 2009-11-05 | 04:41:38.73 | 2100 | 179 | 0.157 |
| FQ Lac | H | 2009-08-21 | 10:08:19.55 | 5175 | 37 | 0.868 |
| V411 Lac | H | 2009-08-23 | 09:55:04.15 | 180 | 248 | ... |
| V473 Lyr | H | 2009-11-14 | 01:25:10.19 | 150 | 154 | 0.697 |
| V473 Lyr | H | 2010-04-25 | 09:11:47.57 | 40 | 315 | 0.581 |
| V473 Lyr | H | 2010-06-26 | 10:22:10.94 | 30 | 209 | 0.204 |
| V473 Lyr | H | 2010-07-07 | 04:27:19.13 | 300 | 442 | 0.417 |
| V473 Lyr | H | 2010-07-07 | 04:32:42.76 | 90 | 108 | 0.421 |
| T Mon | H | 2010-11-11 | 12:08:22.92 | 105 | 324 | 0.102 |
| T Mon | H | 2010-11-22 | 11:07:59.35 | 150 | 515 | 0.507 |
| TW Mon | H | 2009-12-26 | 06:10:37.71 | 3150 | 94 | 0.710 |
| TX Mon | H | 2009-02-01 | 05:46:19.53 | 900 | 134 | 0.258 |
| TX Mon | H | 2009-11-28 | 08:16:30.57 | 1200 | 154 | 0.746 |
| TY Mon | H | 2008-12-29 | 08:25:20.34 | 1080 | 107 | 0.809 |
| TZ Mon | H | 2009-01-29 | 06:35:57.54 | 600 | 122 | 0.149 |
| TZ Mon | H | 2009-11-17 | 08:38:50.73 | 750 | 171 | 0.470 |
| UY Mon | H | 2009-01-30 | 07:16:53.94 | 450 | 191 | 0.980 |
| UY Mon | H | 2009-11-26 | 07:29:57.57 | 350 | 204 | 0.074 |
| UY Mon | H | 2010-03-22 | 03:50:52.35 | 900 | 422 | 0.379 |
| VZ Mon | H | 2008-12-08 | 10:04:31.91 | 3489 | 61 | 0.518 |
| WW Mon | H | 2009-01-19 | 03:18:40.31 | 2220 | 90 | 0.841 |
| WW Mon | H | 2010-01-18 | 03:38:55.96 | 2550 | 115 | 0.918 |
| WW Mon | H | 2010-01-21 | 03:35:24.75 | 2662 | 85 | 0.561 |
| XX Mon | H | 2009-02-25 | 03:57:32.92 | 1400 | 89 | 0.061 |
| XX Mon | H | 2009-11-25 | 08:25:10.33 | 1800 | 136 | 0.128 |
| YY Mon | H | 2009-01-26 | 05:13:11.31 | 3480 | 63 | 0.375 |

**Table 2**
Observing Log

| Cepheid | Telescope | UT Date | UT Time | Exposure (s) | S/N | Phase |
|---:|:---:|:---:|:---:|---:|---:|---:|
| AA Mon | H | 2009-01-14 | 06:23:09.31 | 2760 | 79 | 0.841 |
| AC Mon | M | 2008-11-16 | 10:39:06.43 | 3600 | 254 | 0.007 |
| AC Mon | H | 2009-03-01 | 03:17:53.51 | 500 | 148 | 0.067 |
| AC Mon | H | 2009-11-28 | 09:31:02.73 | 600 | 140 | 0.037 |
| AC Mon | H | 2010-03-22 | 01:56:57.36 | 900 | 336 | 0.221 |
| BE Mon | M | 2008-11-16 | 09:00:18.37 | 3600 | 187 | 0.468 |
| BE Mon | H | 2008-11-17 | 07:46:58.95 | 800 | 166 | 0.812 |
| BE Mon | H | 2009-02-20 | 05:30:31.11 | 750 | 140 | 0.890 |
| BE Mon | H | 2009-11-26 | 07:08:32.55 | 900 | 249 | 0.038 |
| BV Mon | H | 2009-01-14 | 04:35:29.31 | 800 | 94 | 0.159 |
| BV Mon | H | 2009-11-26 | 07:40:51.71 | 900 | 174 | 0.049 |
| CS Mon | H | 2009-01-04 | 08:10:58.51 | 660 | 105 | 0.674 |
| CS Mon | H | 2009-11-25 | 07:04:57.53 | 750 | 155 | 0.945 |
| CU Mon | H | 2009-01-01 | 05:11:40.14 | 2026 | 43 | 0.558 |
| CU Mon | H | 2009-02-19 | 04:19:10.11 | 3100 | 87 | 0.959 |
| CV Mon | M | 2008-11-16 | 07:56:24.85 | 3600 | 227 | 0.589 |
| CV Mon | H | 2008-11-17 | 08:04:49.94 | 660 | 180 | 0.773 |
| CV Mon | H | 2009-02-19 | 05:27:08.74 | 750 | 142 | 0.229 |
| CV Mon | H | 2009-11-27 | 07:38:36.57 | 900 | 71 | 0.490 |
| EE Mon | H | 2009-01-15 | 05:41:28.71 | 2640 | 118 | 0.094 |
| EK Mon | H | 2009-01-26 | 06:15:25.94 | 660 | 131 | 0.377 |
| EK Mon | H | 2010-01-25 | 04:19:22.15 | 1100 | 136 | 0.324 |
| FG Mon | H | 2009-01-30 | 05:17:40.71 | 2800 | 81 | 0.433 |
| FI Mon | H | 2009-01-01 | 07:03:34.51 | 2900 | 82 | 0.083 |
| FT Mon | H | 2009-12-28 | 05:11:59.38 | 3150 | 145 | 0.967 |
| V446 Mon | H | 2010-01-05 | 07:58:33.93 | 3150 | 47 | 0.441 |
| V447 Mon | H | 2010-03-11 | 03:56:33.56 | 3150 | 80 | 0.328 |
| V447 Mon | H | 2010-01-25 | 03:30:24.76 | 3150 | 53 | 0.205 |
| V465 Mon | H | 2009-01-30 | 06:34:55.15 | 500 | 121 | 0.092 |
| V465 Mon | H | 2010-02-16 | 03:19:12.75 | 750 | 117 | 0.820 |
| V465 Mon | H | 2010-03-19 | 03:16:12.94 | 1200 | 299 | 0.244 |
| V484 Mon | H | 2009-01-04 | 07:02:54.72 | 3000 | 60 | 0.497 |
| V495 Mon | H | 2009-12-23 | 06:38:46.96 | 3150 | 135 | 0.021 |
| V504 Mon | H | 2009-11-26 | 09:24:45.35 | 1600 | 117 | 0.864 |
| V504 Mon | H | 2010-01-07 | 06:47:56.94 | 1500 | 102 | 0.963 |
| V508 Mon | H | 2009-11-28 | 07:26:59.32 | 750 | 106 | 0.374 |
| V508 Mon | H | 2009-12-10 | 09:50:26.74 | 900 | 89 | 0.301 |
| V510 Mon | H | 2009-03-22 | 03:09:44.93 | 2580 | 101 | 0.295 |
| V526 Mon | H | 2009-02-19 | 05:16:28.94 | 200 | 127 | 0.971 |
| V526 Mon | H | 2010-01-10 | 05:40:34.72 | 250 | 213 | 0.467 |
| V526 Mon | H | 2010-03-25 | 03:11:21.94 | 900 | 334 | 0.091 |
| V526 Mon | M | 2008-11-15 | 08:45:33.11 | 3600 | 191 | 0.146 |
| CR Ori | H | 2009-01-14 | 07:18:12.92 | 1800 | 83 | 0.458 |
| CR Ori | H | 2010-01-25 | 02:35:23.92 | 2400 | 103 | 0.990 |
| CR Ori | H | 2010-01-27 | 02:28:10.12 | 2400 | 110 | 0.397 |

**Table 2**
Observing Log

| Cepheid | Telescope | UT Date | UT Time | Exposure (s) | S/N | Phase |
|---:|:---:|:---:|:---:|---:|---:|---:|
| DF Ori | H | 2009-12-07 | 10:17:06.16 | 3240 | 118 | 0.256 |
| AU Peg | H | 2009-10-31 | 04:04:26.53 | 300 | 187 | 0.353 |
| QQ Per | H | 2010-02-06 | 03:18:43.15 | 4448 | 82 | 0.268 |
| SV Per | H | 2010-11-12 | 05:03:48.94 | 400 | 229 | 0.889 |
| SX Per | H | 2009-01-23 | 05:45:00.90 | 900 | 116 | 0.904 |
| SX Per | H | 2009-11-17 | 09:57:10.54 | 750 | 123 | 0.410 |
| UX Per | H | 2010-11-11 | 03:42:13.54 | 3600 | 255 | 0.359 |
| UY Per | H | 2008-12-26 | 01:41:18.54 | 900 | 82 | 0.666 |
| UY Per | H | 2009-01-20 | 03:34:44.35 | 900 | 126 | 0.340 |
| VX Per | H | 2010-11-12 | 03:16:18.92 | 500 | 284 | 0.538 |
| VX Per | H | 2010-11-22 | 07:08:18.36 | 480 | 257 | 0.471 |
| VY Per | H | 2009-01-20 | 03:17:57.51 | 750 | 112 | 0.659 |
| AS Per | H | 2010-11-12 | 04:48:08.96 | 720 | 336 | 0.979 |
| AW Per | H | 2010-11-05 | 05:55:29.72 | 180 | 350 | 0.812 |
| BM Per | H | 2010-11-08 | 05:32:25.73 | 1200 | 276 | 0.225 |
| CI Per | H | 2008-09-21 | 10:55:00.51 | 2520 | 122 | 0.328 |
| DW Per | H | 2009-01-14 | 03:59:48.55 | 1050 | 115 | 0.086 |
| GP Per | H | 2009-12-12 | 08:25:37.54 | 4650 | 73 | 0.086 |
| HQ Per | H | 2009-03-22 | 02:44:19.14 | 1000 | 109 | 0.456 |
| HQ Per | H | 2009-11-27 | 09:48:33.96 | 1400 | 96 | 0.430 |
| HQ Per | H | 2010-11-11 | 05:23:15.74 | 3600 | 214 | 0.808 |
| HZ Per | H | 2009-01-17 | 06:01:41.93 | 5050 | 101 | 0.867 |
| MM Per | H | 2009-12-07 | 07:56:21.56 | 1100 | 236 | 0.129 |
| OT Per | H | 2008-12-27 | 02:03:29.13 | 1650 | 55 | 0.032 |
| OT Per | H | 2009-01-19 | 06:01:00.91 | 4950 | 107 | 0.920 |
| X Pup | M | 2009-11-23 | 09:58:56.87 | 1800 | 215 | 0.127 |
| RS Pup | M | 2009-12-27 | 08:23:36.11 | 1200 | 402 | 0.151 |
| VX Pup | M | 2009-11-23 | 09:25:02.61 | 1800 | 211 | 0.233 |
| VZ Pup | M | 2009-11-23 | 10:35:49.84 | 3600 | 187 | 0.767 |
| WX Pup | M | 2009-12-23 | 07:08:51.65 | 3600 | 241 | 0.253 |
| AD Pup | M | 2009-12-26 | 07:13:53.26 | 3600 | 225 | 0.550 |
| AQ Pup | M | 2009-12-23 | 08:49:49.53 | 2400 | 301 | 0.533 |
| BN Pup | M | 2009-12-26 | 08:22:10.56 | 3600 | 80 | 0.113 |
| V335 Pup | M | 2009-12-23 | 08:14:23.91 | 1800 | 204 | 0.064 |
| Y Sct | H | 2010-05-02 | 11:03:02.35 | 300 | 156 | 0.920 |
| Y Sct | H | 2010-06-24 | 07:19:16.13 | 600 | 319 | 0.030 |
| Z Sct | H | 2010-06-11 | 08:15:43.59 | 350 | 161 | 0.347 |
| RU Sct | H | 2010-05-02 | 11:13:04.35 | 400 | 237 | 0.407 |
| SS Sct | H | 2010-05-19 | 09:31:06.35 | 150 | 242 | 0.167 |
| SS Sct | H | 2010-05-31 | 09:08:06.53 | 150 | 195 | 0.431 |
| SS Sct | H | 2010-06-11 | 08:08:33.75 | 150 | 152 | 0.416 |
| TY Sct | H | 2010-06-02 | 08:46:52.96 | 900 | 207 | 0.963 |
| BX Sct | H | 2010-06-09 | 08:48:08.57 | 2400 | 192 | 0.306 |
| CK Sct | H | 2010-04-29 | 10:57:35.92 | 600 | 153 | 0.952 |

**Table 2**
Observing Log

| Cepheid | Telescope | UT Date | UT Time | Exposure (s) | S/N | Phase |
|---:|:---:|:---:|:---:|---:|---:|---:|
| CM Sct | H | 2010-05-18 | 10:04:31.33 | 900 | 190 | 0.048 |
| CN Sct | H | 2010-06-23 | 07:17:49.37 | 712 | 80 | 0.105 |
| CN Sct | H | 2010-09-01 | 02:56:54.51 | 2325 | 146 | 0.092 |
| CN Sct | H | 2010-09-06 | 02:30:00.71 | 2850 | 141 | 0.591 |
| EV Sct | H | 2010-04-26 | 11:14:19.94 | 500 | 177 | 0.923 |
| EV Sct | H | 2010-07-06 | 04:45:14.54 | 660 | 181 | 0.809 |
| AA Ser | H | 2010-04-19 | 10:36:17.35 | 3600 | 207 | 0.342 |
| DV Ser | H | 2010-05-07 | 09:40:48.75 | 3000 | 143 | 0.814 |
| DG Sge | H | 2010-04-16 | 10:22:19.35 | 3300 | 176 | 0.069 |
| GX Sge | H | 2010-06-11 | 06:39:04.17 | 3000 | 191 | 0.196 |
| GY Sge | H | 2010-04-06 | 10:55:47.13 | 1800 | 108 | 0.258 |
| GY Sge | H | 2010-04-12 | 10:26:44.15 | 2400 | 659 | 0.374 |
| GY Sge | H | 2010-06-23 | 05:52:47.57 | 1200 | 466 | 0.767 |
| ST Tau | H | 2010-11-14 | 06:53:10.38 | 280 | 305 | 0.994 |
| ST Tau | H | 2010-11-20 | 06:33:14.71 | 350 | 328 | 0.478 |
| AE Tau | H | 2010-11-15 | 06:00:56.54 | 3600 | 213 | 0.248 |
| AV Tau | H | 2009-01-26 | 06:32:05.74 | 1860 | 97 | 0.658 |
| S Vul | H | 2010-04-25 | 09:38:54.96 | 1200 | 634 | 0.144 |
| S Vul | H | 2010-08-13 | 02:21:41.98 | 400 | 305 | 0.756 |
| U Vul | H | 2009-11-02 | 02:15:10.95 | 90 | 425 | 0.233 |
| AS Vul | H | 2008-11-13 | 01:44:50.91 | 1800 | 108 | 0.281 |
| DG Vul | H | 2010-07-21 | 09:37:31.57 | 1500 | 238 | 0.969 |
| GSC 0754-1993 | H | 2010-01-12 | 04:02:04.73 | 1560 | 65 | 0.125 |
| GSC 2418-1443 | H | 2009-12-12 | 09:55:47.57 | 1200 | 103 | 0.971 |
| GSC 3706-0233 | H | 2009-12-21 | 06:06:38.75 | 1800 | 89 | 0.425 |
| GSC 3725-0174 | H | 2009-12-12 | 07:12:54.74 | 1800 | 106 | 0.817 |
| GSC 3726-0565 | H | 2010-01-03 | 06:09:44.54 | 1800 | 127 | 0.334 |
| GSC 3729-1127 | H | 2009-11-14 | 08:51:31.93 | 1500 | 221 | 0.065 |
| GSC 3732-0183 | H | 2010-03-05 | 02:54:51.35 | 3600 | 107 | 0.900 |
| GSC 4009-0024 | H | 2009-11-01 | 05:48:26.93 | 1260 | 105 | 0.551 |
| GSC 4038-1585 | H | 2010-09-07 | 08:02:48.16 | 2400 | 170 | 0.160 |
| GSC 4040-1803 | H | 2009-12-14 | 02:27:30.95 | 1000 | 113 | 0.202 |
| GSC 4265-0193 | H | 2009-09-09 | 07:32:12.14 | 1140 | 99 | 0.655 |
| GSC 4265-0569 | H | 2009-10-15 | 04:51:43.73 | 660 | 44 | 0.168 |

Telescope:  H = Hobby-Eberly Telescope
            M = McDonald Observatory 2.7m
S/N:        Signal to noise (per pixel) near 600 nm at order center.
Phase:      Periods and epochs from Berdnikov (2006– private communication to S. Andrievsky) , GCVS, or Wils & Greaves(2004).

Table 3
Stellar Parameters and Abundances

| Object | Phase | T (K) | σ | N | Log(g) (cm/s^2) | $V_t$ (km/s) | $V_M$ (km/s) | Fe I | σ | N | Fe II | Sigma | N | Li | C | N | O |
|---|---|---|---|---|---|---|---|---|---|---|---|---|---|---|---|---|---|
| U Aql | 0.534 | 5434 | 33 | 66 | 1.75 | 4.02 | 7.5 | 7.63 | 0.12 | 319 | 7.63 | 0.06 | 27 | 1.3 | 8.32 | 8.14 | 8.89 |
| U Aql | 0.675 | 5326 | 30 | 64 | 1.75 | 4.59 | 8.3 | 7.67 | 0.12 | 270 | 7.67 | 0.08 | 24 | 1.3 | 8.30 | 8.38 | 8.99 |
| U Aql | 0.211 | 5883 | 39 | 64 | 2.06 | 4.25 | 8.1 | 7.71 | 0.11 | 273 | 7.71 | 0.10 | 33 | 1.4 | 8.30 | 8.37 | 8.91 |
| U Aql | 0.311 | 5782 | 49 | 68 | 2.04 | 3.88 | 7.6 | 7.71 | 0.11 | 309 | 7.71 | 0.09 | 34 | 1.4 | 8.36 | 8.30 | 8.94 |
| U Aql | 0.730 | 5338 | 42 | 67 | 1.95 | 5.16 | 10.5 | 7.62 | 0.14 | 264 | 7.62 | 0.18 | 23 | 1.3 | 8.39 | 8.50 | 8.97 |
| TT Aql | 0.394 | 5256 | 70 | 69 | 1.29 | 3.54 | 5.0 | 7.72 | 0.11 | 313 | 7.73 | 0.13 | 40 | 1.0 | 8.47 | 8.36 | 9.03 |
| TT Aql | 0.409 | 5224 | 67 | 69 | 1.35 | 3.61 | 5.0 | 7.71 | 0.11 | 340 | 7.71 | 0.10 | 29 | 0.9 | 8.49 | 8.26 | 9.05 |
| BC Aql | 0.039 | 6035 | 406 | 69 | 1.82 | 3.91 | 7.3 | 7.22 | 0.16 | 291 | 7.22 | 0.13 | 39 | 1.5 | 8.14 | 7.98 | 8.84 |
| EV Aql | 0.158 | 5404 | 69 | 64 | 1.36 | 5.80 | 7.7 | 7.56 | 0.14 | 296 | 7.57 | 0.12 | 28 | 1.0 | 8.34 | 8.35 | 9.00 |
| FM Aql | 0.279 | 5808 | 71 | 64 | 2.38 | 5.03 | 14.0 | 7.75 | 0.18 | 243 | 7.75 | 0.12 | 26 | 1.2 | 8.36 | 8.35 | 8.97 |
| FM Aql | 0.585 | 5494 | 40 | 56 | 1.99 | 5.04 | 14.2 | 7.72 | 0.14 | 218 | 7.72 | 0.11 | 19 | 1.2 | 8.29 | 8.19 | 8.99 |
| FN Aql | 0.942 | 5850 | 71 | 65 | 1.79 | 4.87 | 8.1 | 7.41 | 0.10 | 300 | 7.41 | 0.11 | 38 | 1.2 | 7.09 | 8.38 | 8.75 |
| FN Aql | 0.951 | 5886 | 59 | 62 | 2.02 | 5.11 | 7.7 | 7.47 | 0.11 | 309 | 7.47 | 0.09 | 37 | 1.2 | 7.07 | 8.46 | 8.79 |
| KL Aql | 0.252 | 5672 | 91 | 71 | 2.15 | 3.96 | 6.7 | 7.83 | 0.10 | 319 | 7.83 | 0.10 | 32 | 1.4 | 8.57 | 8.15 | 9.11 |
| V336 Aql | 0.040 | 6218 | 40 | 55 | 2.35 | 5.42 | 8.7 | 7.72 | 0.12 | 270 | 7.72 | 0.11 | 29 | 1.4 | 8.34 | 8.37 | 8.91 |
| V336 Aql | 0.287 | 5743 | 32 | 61 | 2.03 | 4.03 | 6.1 | 7.63 | 0.11 | 330 | 7.63 | 0.09 | 32 | 1.0 | 8.35 | 8.21 | 8.93 |
| V493 Aql | 0.803 | 6186 | 134 | 53 | 2.55 | 5.18 | 14.5 | 7.53 | 0.11 | 207 | 7.53 | 0.10 | 27 | 1.9 | 8.13 | 8.64 | 8.70 |
| V526 Aql | 0.283 | 5986 | 105 | 62 | 2.36 | 5.03 | 8.1 | 8.00 | 0.16 | 292 | 8.00 | 0.11 | 29 | 1.4 | 9.35 | 8.23 | 9.18 |
| V916 Aql | 0.293 | 5299 | 101 | 71 | 1.44 | 3.58 | 6.9 | 7.89 | 0.15 | 285 | 7.89 | 0.09 | 23 | 1.0 | 8.47 | 8.40 | 9.00 |
| V1344 Aql | 0.828 | 5582 | 42 | 63 | 2.08 | 6.23 | 9.0 | 7.65 | 0.09 | 266 | 7.65 | 0.10 | 24 | 1.0 | 8.27 | 8.53 | 8.95 |
| V1359 Aql | … | 4605 | 108 | 52 | 3.08 | 2.80 | 5.3 | 7.77 | 0.16 | 388 | 7.76 | 0.15 | 19 | 1.3 | 8.70 | 8.07 | 8.97 |
| V1359 Aql | … | 4597 | 110 | 54 | 3.17 | 3.03 | 5.3 | 7.80 | 0.15 | 388 | 7.80 | 0.16 | 23 | 1.2 | 8.81 | 9.01 | 9.01 |
| V1359 Aql | … | 4636 | 114 | 53 | 3.14 | 2.63 | 5.3 | 7.79 | 0.13 | 349 | 7.79 | 0.09 | 17 | 1.3 | 8.78 | 8.33 | 8.94 |
| Y Aur | 0.733 | 5706 | 77 | 61 | 2.08 | 4.07 | 11.5 | 7.49 | 0.10 | 221 | 7.49 | 0.08 | 24 | 1.8 | 8.09 | 8.22 | 8.79 |
| Y Aur | 0.574 | 5745 | 51 | 61 | 2.19 | 3.74 | 9.3 | 7.51 | 0.12 | 299 | 7.51 | 0.10 | 39 | 1.5 | 8.15 | 8.35 | 8.70 |
| Y Aur | 0.323 | 6074 | 52 | 60 | 2.38 | 3.64 | 8.7 | 7.60 | 0.10 | 260 | 7.60 | 0.10 | 33 | 1.7 | 8.23 | 8.30 | 8.50 |
| RT Aur | 0.111 | 6595 | 38 | 47 | 2.54 | 3.86 | 8.9 | 7.63 | 0.10 | 263 | 7.63 | 0.07 | 39 | 1.2 | 8.40 | 8.52 | 9.00 |
| RT Aur | 0.253 | 6210 | 31 | 56 | 2.40 | 3.48 | 6.8 | 7.64 | 0.09 | 316 | 7.64 | 0.08 | 43 | 1.5 | 8.42 | 8.34 | 8.98 |
| RX Aur | 0.857 | 6331 | 61 | 45 | 2.33 | 4.55 | 13.6 | 7.60 | 0.15 | 230 | 7.60 | 0.07 | 25 | 1.9 | 8.25 | 8.25 | 8.83 |
| SY Aur | 0.251 | 5911 | 47 | 59 | 2.13 | 4.18 | 5.9 | 7.54 | 0.10 | 300 | 7.54 | 0.07 | 34 | 1.3 | 8.16 | 8.17 | 8.87 |
| YZ Aur | 0.499 | 5004 | 151 | 55 | 0.82 | 3.81 | 12.6 | 7.20 | 0.16 | 222 | 7.20 | 0.14 | 18 | 1.2 | 8.12 | … | 8.74 |
| AN Aur | 0.835 | 6026 | 89 | 62 | 2.13 | 5.06 | 7.1 | 7.40 | 0.14 | 324 | 7.40 | 0.11 | 39 | 1.4 | 8.06 | 8.05 | 8.72 |

Table 3
Stellar Parameters and Abundances

| Object | Phase | T (K) | σ | N | Log(g) (cm/s^2) | $V_t$ (km/s) | $V_M$ (km/s) | Fe I | σ | N | Fe II | Sigma | N | Li | C | N | O |
|---|---|---|---|---|---|---|---|---|---|---|---|---|---|---|---|---|---|
| AO Aur | 0.101 | 5903 | 106 | 61 | 1.83 | 3.82 | 7.6 | 7.27 | 0.10 | 239 | 7.27 | 0.13 | 41 | 1.3 | 7.96 | 8.03 | 8.55 |
| AO Aur | 0.410 | 5494 | 138 | 67 | 1.46 | 4.08 | 8.0 | 7.15 | 0.12 | 279 | 7.15 | 0.14 | 38 | 1.1 | 7.87 | 7.83 | 8.48 |
| AO Aur | 0.286 | 5733 | 127 | 68 | 2.12 | 4.06 | 6.1 | 7.28 | 0.14 | 355 | 7.28 | 0.12 | 41 | 1.7 | 7.99 | 8.01 | 8.58 |
| AS Aur | 0.346 | 5844 | 117 | 68 | 2.02 | 3.41 | 6.4 | 7.30 | 0.14 | 365 | 7.30 | 0.14 | 45 | 1.0 | 7.91 | 8.15 | 8.73 |
| AX Aur | 0.201 | 6192 | 154 | 61 | 2.22 | 3.77 | 11.2 | 7.41 | 0.16 | 295 | 7.42 | 0.11 | 37 | 1.3 | 7.96 | 8.23 | 8.65 |
| AX Aur | 0.421 | 5962 | 103 | 59 | 2.37 | 4.35 | 8.4 | 7.52 | 0.25 | 384 | 7.52 | 0.28 | 51 | 1.3 | 8.03 | 8.34 | 8.72 |
| BK Aur | 0.804 | 5658 | 79 | 54 | 2.31 | 4.29 | 16.0 | 7.47 | 0.22 | 213 | 7.47 | 0.15 | 24 | 1.3 | 8.07 | 8.40 | 8.97 |
| CO Aur | 0.001 | 6593 | 59 | 48 | 2.73 | 3.18 | 7.4 | 7.50 | 0.14 | 349 | 7.50 | 0.09 | 26 | 1.2 | 8.20 | 8.35 | 8.77 |
| CO Aur | 0.148 | 6361 | 65 | 62 | 2.56 | 3.40 | 5.3 | 7.46 | 0.10 | 339 | 7.46 | 0.11 | 35 | 1.6 | 8.28 | 8.47 | 8.80 |
| CO Aur | 0.721 | 6328 | 87 | 60 | 2.52 | 3.08 | 8.0 | 7.53 | 0.10 | 278 | 7.53 | 0.13 | 39 | 1.0 | 8.22 | 8.39 | 8.86 |
| CY Aur | 0.975 | 6019 | 109 | 59 | 2.05 | 4.57 | 12.2 | 7.38 | 0.17 | 257 | 7.38 | 0.12 | 34 | 1.2 | 7.96 | 8.04 | 8.69 |
| ER Aur | 0.379 | 6385 | 136 | 48 | 2.12 | 4.38 | 6.3 | 7.23 | 0.12 | 294 | 7.23 | 0.15 | 52 | 1.2 | 7.95 | 8.06 | 8.59 |
| EW Aur | 0.892 | 6555 | 175 | 42 | 2.03 | 3.67 | 13.3 | 6.96 | 0.16 | 160 | 6.96 | 0.13 | 42 | 1.7 | 7.76 | 7.93 | 8.37 |
| FF Aur | 0.381 | 5927 | 142 | 30 | 2.09 | 3.34 | 12.4 | 6.99 | 0.41 | 224 | 6.99 | 0.26 | 32 | 1.5 | 7.68 | … | 8.32 |
| GT Aur | 0.514 | 6039 | 93 | 64 | 1.96 | 3.19 | 7.9 | 7.48 | 0.14 | 312 | 7.48 | 0.12 | 39 | 1.0 | 8.14 | 8.22 | 8.55 |
| GV Aur | 0.838 | 5502 | 156 | 70 | 1.36 | 3.22 | 11.2 | 7.25 | 0.20 | 303 | 7.25 | 0.18 | 34 | 1.2 | 7.84 | 8.15 | 8.29 |
| GV Aur | 0.826 | 5484 | 119 | 64 | 1.56 | 3.25 | 10.4 | 7.32 | 0.17 | 312 | 7.32 | 0.13 | 26 | 1.1 | 7.94 | 8.15 | 8.43 |
| GV Aur | 0.077 | 6362 | 141 | 48 | 2.40 | 5.23 | 12.4 | 7.30 | 0.14 | 204 | 7.30 | 0.19 | 30 | 1.8 | 7.97 | 8.13 | 8.68 |
| IN Aur | 0.390 | 5707 | 117 | 65 | 1.42 | 2.98 | 9.1 | 7.22 | 0.16 | 292 | 7.22 | 0.20 | 41 | 1.3 | 7.80 | 7.93 | 8.44 |
| V335 Aur | 0.679 | 6410 | 99 | 41 | 1.91 | 3.39 | 13.2 | 7.20 | 0.22 | 232 | 7.20 | 0.20 | 40 | 1.6 | 7.86 | 8.46 | 8.62 |
| RW Cam | 0.678 | 4838 | 64 | 55 | 1.32 | 6.40 | 10.4 | 7.61 | 0.19 | 151 | 7.61 | 0.15 | 14 | 0.8 | 8.09 | … | 8.76 |
| RX Cam | 0.879 | 5872 | 92 | 62 | 2.22 | 5.78 | 13.0 | 7.61 | 0.13 | 221 | 7.62 | 0.08 | 21 | 1.2 | 8.14 | 8.17 | 8.74 |
| TV Cam | 0.500 | 5715 | 50 | 68 | 1.65 | 3.34 | 6.9 | 7.45 | 0.11 | 316 | 7.46 | 0.13 | 48 | 1.0 | 8.07 | … | 8.68 |
| TV Cam | 0.276 | 6017 | 112 | 54 | 2.65 | 3.89 | 6.1 | 7.68 | 0.23 | 330 | 7.68 | 0.17 | 45 | 1.2 | 8.18 | 8.29 | 8.93 |
| TV Cam | 0.467 | 5753 | 65 | 68 | 1.96 | 3.12 | 6.0 | 7.49 | 0.12 | 328 | 7.48 | 0.09 | 42 | 1.2 | 8.04 | 8.51 | 8.70 |
| AB Cam | 0.411 | 5678 | 61 | 63 | 1.98 | 3.66 | 7.3 | 7.43 | 0.14 | 380 | 7.43 | 0.11 | 41 | 1.0 | 8.06 | 8.07 | 8.58 |
| AC Cam | 0.985 | 6091 | 95 | 62 | 2.00 | 4.40 | 11.0 | 7.37 | 0.15 | 251 | 7.37 | 0.15 | 41 | 1.0 | 7.94 | 8.25 | 8.50 |
| AD Cam | 0.656 | 5238 | 92 | 62 | 0.92 | 3.17 | 7.7 | 7.27 | 0.16 | 336 | 7.27 | 0.16 | 37 | 0.8 | 7.51 | 7.94 | 8.57 |
| AD Cam | 0.746 | 5208 | 110 | 61 | 1.29 | 4.36 | 10.9 | 7.24 | 0.17 | 291 | 7.24 | 0.12 | 26 | 0.9 | 7.42 | 8.01 | 8.60 |
| AM Cam | 0.717 | 5589 | 174 | 60 | 1.52 | 3.71 | 15.1 | 7.44 | 0.20 | 217 | 7.44 | 0.16 | 23 | 1.7 | 7.86 | 8.10 | 8.59 |
| AM Cam | 0.217 | 5721 | 106 | 69 | 1.63 | 3.30 | 7.8 | 7.30 | 0.15 | 341 | 7.30 | 0.10 | 32 | 1.3 | 7.89 | 8.04 | 8.53 |
| CK Cam | 0.317 | 5809 | 55 | 74 | 2.15 | 3.12 | 5.9 | 7.59 | 0.12 | 365 | 7.59 | 0.11 | 45 | 1.4 | 8.29 | 8.33 | 8.89 |

Table 3
Stellar Parameters and Abundances

| Object | Phase | T (K) | σ | N | Log(g) (cm/s^2) | $V_t$ (km/s) | $V_M$ (km/s) | Fe I | σ | N | Fe II | Sigma | N | Li | C | N | O |
|---|---|---|---|---|---|---|---|---|---|---|---|---|---|---|---|---|---|
| CK Cam | 0.903 | 6280 | 79 | 55 | 2.42 | 4.07 | 8.9 | 7.54 | 0.09 | 289 | 7.54 | 0.09 | 39 | 1.6 | 8.18 | 8.29 | 8.75 |
| LO Cam | … | 6092 | 66 | 53 | 2.30 | 5.63 | 8.4 | 7.45 | 0.13 | 264 | 7.45 | 0.12 | 39 | 1.7 | 8.09 | 7.96 | 8.77 |
| MN Cam | 0.412 | 5763 | 50 | 63 | 1.93 | 4.55 | 6.8 | 7.51 | 0.12 | 329 | 7.51 | 0.09 | 33 | 1.4 | 8.24 | 8.25 | 8.84 |
| MQ Cam | 0.929 | 5312 | 158 | 52 | 1.46 | 5.32 | 10.3 | 7.38 | 0.20 | 242 | 7.38 | 0.15 | 22 | 0.8 | 7.86 | 8.10 | 8.28 |
| MQ Cam | 0.288 | 6075 | 121 | 52 | 1.78 | 4.44 | 9.6 | 7.40 | 0.17 | 268 | 7.40 | 0.16 | 38 | 1.5 | 7.97 | 8.19 | 8.33 |
| SS Cma | 0.524 | 6462 | 64 | 56 | 2.20 | 4.60 | 8.4 | 7.57 | 0.12 | 264 | 7.58 | 0.11 | 46 | 1.5 | 8.26 | 8.15 | 8.88 |
| TV Cma | 0.746 | 5556 | 52 | 62 | 1.93 | 4.31 | 10.9 | 7.64 | 0.16 | 274 | 7.64 | 0.13 | 11 | 1.0 | 8.22 | 7.71 | 8.81 |
| RS Cas | 0.214 | 5859 | 39 | 63 | 2.00 | 3.63 | 8.2 | 7.68 | 0.17 | 370 | 7.69 | 0.10 | 45 | 1.0 | 8.35 | 8.53 | 8.91 |
| SZ Cas | 0.447 | 6173 | 91 | 45 | 2.08 | 4.74 | 14.3 | 7.60 | 0.14 | 186 | 7.59 | 0.11 | 23 | 1.5 | 8.16 | 8.16 | 8.68 |
| SZ Cas | 0.020 | 5710 | 60 | 57 | 1.77 | 4.68 | 12.2 | 7.52 | 0.11 | 178 | 7.52 | 0.11 | 20 | 1.2 | 8.14 | 8.19 | 8.85 |
| SZ Cas | 0.508 | 6222 | 64 | 48 | 2.22 | 4.89 | 13.9 | 7.60 | 0.17 | 233 | 7.59 | 0.10 | 28 | 1.6 | 8.22 | 8.38 | 8.77 |
| UZ Cas | 0.522 | 5710 | 69 | 67 | 1.89 | 3.20 | 6.7 | 7.45 | 0.17 | 381 | 7.45 | 0.10 | 41 | 1.1 | 7.93 | 8.10 | 8.68 |
| VV Cas | 0.360 | 5764 | 67 | 71 | 1.94 | 3.63 | 6.8 | 7.46 | 0.16 | 406 | 7.46 | 0.11 | 42 | 1.0 | 8.10 | 8.19 | 8.85 |
| VW Cas | 0.377 | 5673 | 44 | 69 | 1.92 | 3.91 | 11.6 | 7.69 | 0.19 | 300 | 7.69 | 0.07 | 24 | 1.0 | 8.19 | 8.37 | 9.02 |
| XY Cas | 0.969 | 6275 | 60 | 53 | 2.33 | 4.20 | 11.8 | 7.61 | 0.13 | 204 | 7.60 | 0.09 | 33 | 1.5 | 8.27 | 8.23 | 8.81 |
| AP Cas | 0.556 | 5301 | 62 | 63 | 1.46 | 4.58 | 10.5 | 7.55 | 0.18 | 295 | 7.55 | 0.16 | 30 | 0.2 | 8.12 | 8.19 | 8.86 |
| AS Cas | 0.947 | 6345 | 125 | 40 | 2.28 | 3.98 | 15.0 | 7.52 | 0.23 | 234 | 7.52 | 0.21 | 35 | 1.6 | 8.19 | 7.82 | 8.62 |
| AW Cas | 0.987 | 6439 | 111 | 54 | 2.72 | 4.40 | 9.6 | 7.56 | 0.16 | 297 | 7.56 | 0.08 | 38 | 1.6 | 8.21 | 8.29 | 8.87 |
| AW Cas | 0.409 | 5787 | 47 | 65 | 2.05 | 3.17 | 5.3 | 7.51 | 0.12 | 374 | 7.51 | 0.10 | 46 | 0.9 | 8.23 | 8.19 | 8.81 |
| AY Cas | 0.711 | 5734 | 84 | 69 | 2.11 | 3.61 | 7.2 | 7.52 | 0.13 | 344 | 7.52 | 0.09 | 38 | 1.3 | 8.09 | 8.16 | 8.62 |
| BF Cas | 0.220 | 6524 | 106 | 32 | 2.27 | 3.74 | 11.8 | 7.45 | 0.18 | 244 | 7.46 | 0.13 | 35 | 2.1 | 8.12 | 8.33 | 8.60 |
| BP Cas | 0.104 | 5826 | 39 | 68 | 2.07 | 3.78 | 7.7 | 7.61 | 0.17 | 384 | 7.61 | 0.15 | 43 | 1.2 | 8.29 | 8.41 | 8.91 |
| BP Cas | 0.997 | 5990 | 38 | 61 | 2.24 | 4.33 | 8.3 | 7.57 | 0.12 | 311 | 7.56 | 0.11 | 37 | 1.0 | 8.25 | 8.34 | 8.86 |
| BV Cas | 0.327 | 5751 | 54 | 68 | 1.99 | 3.29 | 6.5 | 7.52 | 0.10 | 332 | 7.52 | 0.07 | 38 | 0.8 | 8.40 | 8.11 | 8.87 |
| BY Cas | 0.664 | 6110 | 44 | 66 | 2.27 | 3.42 | 9.2 | 7.62 | 0.17 | 325 | 7.62 | 0.10 | 43 | 1.2 | 8.26 | 8.38 | 9.00 |
| CD Cas | 0.086 | 6445 | 51 | 55 | 2.29 | 5.28 | 9.5 | 7.63 | 0.17 | 283 | 7.64 | 0.16 | 55 | 1.3 | 8.30 | 8.42 | 8.94 |
| CE Cas A | 0.699 | 6341 | 73 | 46 | 2.22 | 4.18 | 13.5 | 7.55 | 0.14 | 204 | 7.54 | 0.13 | 37 | 1.3 | 8.23 | 8.32 | 8.71 |
| CE Cas B | 0.199 | 6065 | 80 | 61 | 2.29 | 4.15 | 13.7 | 7.61 | 0.15 | 238 | 7.61 | 0.14 | 34 | 1.2 | 8.28 | 8.26 | 8.81 |
| CF Cas | 0.109 | 5798 | 36 | 60 | 2.01 | 3.84 | 7.0 | 7.52 | 0.10 | 296 | 7.52 | 0.08 | 37 | 1.0 | 8.22 | 8.24 | 8.82 |
| CG Cas | 0.340 | 6080 | 58 | 58 | 2.59 | 3.70 | 8.2 | 7.65 | 0.13 | 300 | 7.65 | 0.10 | 37 | 1.5 | 8.37 | 8.53 | 8.92 |
| CG Cas | 0.138 | 6591 | 79 | 45 | 2.16 | 3.70 | 11.3 | 7.52 | 0.13 | 251 | 7.52 | 0.09 | 42 | 1.3 | 8.27 | 8.43 | 8.75 |
| CT Cas | 0.329 | 5768 | 89 | 65 | 1.83 | 3.42 | 7.3 | 7.42 | 0.15 | 372 | 7.42 | 0.18 | 43 | 1.0 | 8.09 | 8.13 | 8.68 |

Table 3
Stellar Parameters and Abundances

| Object | Phase | T (K) | σ | N | Log(g) (cm/s^2) | $V_t$ (km/s) | $V_M$ (km/s) | Fe I | σ | N | Fe II | Sigma | N | Li | C | N | O |
|---|---|---|---|---|---|---|---|---|---|---|---|---|---|---|---|---|---|
| CT Cas | 0.171 | 6104 | 96 | 60 | 1.91 | 3.40 | 10.6 | 7.48 | 0.18 | 300 | 7.48 | 0.11 | 36 | 1.5 | 8.05 | 8.24 | 8.76 |
| CZ Cas | 0.440 | 5834 | 64 | 62 | 2.04 | 4.02 | 8.8 | 7.57 | 0.13 | 296 | 7.57 | 0.14 | 41 | 1.0 | 8.19 | 8.21 | 8.75 |
| CZ Cas | 0.423 | 5884 | 34 | 57 | 2.09 | 4.09 | 9.5 | 7.56 | 0.12 | 297 | 7.56 | 0.15 | 45 | 1.0 | 8.22 | 8.27 | 8.75 |
| CZ Cas | 0.060 | 6409 | 76 | 49 | 2.30 | 5.10 | 12.5 | 7.57 | 0.15 | 258 | 7.57 | 0.15 | 39 | 1.9 | 8.17 | 8.21 | 8.61 |
| DW Cas | 0.027 | 6025 | 56 | 56 | 2.13 | 4.59 | 8.8 | 7.57 | 0.11 | 268 | 7.57 | 0.10 | 36 | 1.2 | 8.11 | 8.39 | 8.75 |
| DW Cas | 0.901 | 6200 | 59 | 50 | 2.21 | 4.47 | 9.9 | 7.56 | 0.12 | 270 | 7.56 | 0.09 | 35 | 1.2 | 8.14 | 8.42 | 8.79 |
| EX Cas | 0.837 | 5509 | 131 | 70 | 1.53 | 3.73 | 11.6 | 7.43 | 0.12 | 214 | 7.43 | 0.10 | 22 | 1.0 | 7.95 | 8.14 | 8.70 |
| FO Cas | 0.390 | 5820 | 298 | 64 | 1.69 | 3.56 | 8.2 | 6.94 | 0.17 | 253 | 6.94 | 0.11 | 41 | 1.3 | 7.24 | 8.21 | 8.34 |
| FW Cas | 0.319 | 5448 | 161 | 72 | 1.02 | 3.40 | 9.2 | 7.38 | 0.18 | 281 | 7.38 | 0.15 | 29 | 1.0 | 7.80 | … | 8.46 |
| FW Cas | 0.623 | 5722 | 156 | 66 | 1.80 | 5.11 | 12.7 | 7.43 | 0.16 | 228 | 7.43 | 0.16 | 26 | 0.9 | 7.89 | 8.04 | 8.44 |
| GL Cas | 0.206 | 5801 | 106 | 59 | 1.91 | 3.86 | 14.8 | 7.53 | 0.19 | 261 | 7.53 | 0.16 | 28 | 1.0 | 8.08 | 7.84 | 8.61 |
| GM Cas | 0.135 | 5878 | 111 | 59 | 2.06 | 5.22 | 12.0 | 7.40 | 0.15 | 258 | 7.40 | 0.12 | 29 | 1.7 | 8.18 | 8.13 | 8.74 |
| GM Cas | 0.860 | 6157 | 123 | 45 | 2.14 | 5.05 | 14.3 | 7.39 | 0.18 | 226 | 7.39 | 0.18 | 33 | 1.9 | 8.14 | 8.10 | 8.44 |
| GO Cas | 0.200 | 6314 | 80 | 55 | 2.29 | 3.35 | 11.3 | 7.62 | 0.18 | 295 | 7.62 | 0.08 | 36 | 1.3 | 8.33 | 8.49 | 8.91 |
| HK Cas | 0.713 | 6058 | 116 | 59 | 2.23 | 3.53 | 7.4 | 7.95 | 0.20 | 337 | 7.95 | 0.10 | 29 | 1.4 | 9.69 | 9.04 | 9.25 |
| IO Cas | 0.165 | 6088 | 249 | 42 | 1.81 | 3.72 | 12.1 | 7.01 | 0.16 | 194 | 7.01 | 0.13 | 30 | 1.6 | 7.31 | 8.02 | 8.20 |
| KK Cas | 0.342 | 5762 | 48 | 60 | 1.81 | 3.69 | 8.6 | 7.63 | 0.15 | 300 | 7.63 | 0.12 | 31 | 1.6 | 8.30 | 8.65 | 8.89 |
| LT Cas | 0.253 | 5702 | 205 | 66 | 1.63 | 3.46 | 6.6 | 7.15 | 0.14 | 301 | 7.14 | 0.16 | 47 | 1.0 | 7.74 | 8.12 | 8.37 |
| NP Cas | 0.735 | 6299 | 122 | 50 | 2.18 | 4.68 | 13.0 | 7.51 | 0.09 | 183 | 7.51 | 0.15 | 33 | 1.6 | 8.15 | 8.00 | 8.67 |
| NY Cas | 0.863 | 6317 | 257 | 30 | 1.77 | 3.50 | 15.4 | 7.04 | 0.19 | 169 | 7.04 | 0.27 | 30 | 1.9 | 7.51 | 8.45 | 8.23 |
| OP Cas | 0.168 | 5822 | 51 | 62 | 1.93 | 3.97 | 8.1 | 7.64 | 0.14 | 310 | 7.64 | 0.08 | 34 | 1.2 | 8.33 | 8.29 | 8.76 |
| OZ Cas | 0.722 | 5932 | 76 | 56 | 1.94 | 3.72 | 12.7 | 7.56 | 0.20 | 251 | 7.56 | 0.14 | 31 | 1.6 | 8.09 | 8.11 | 8.77 |
| PW Cas | 0.187 | 5887 | 105 | 64 | 2.33 | 3.77 | 6.4 | 7.44 | 0.11 | 313 | 7.44 | 0.12 | 35 | 1.3 | 8.09 | 7.70 | 8.80 |
| V342 Cas | 0.876 | 5757 | 57 | 68 | 1.96 | 3.33 | 7.6 | 7.53 | 0.17 | 368 | 7.53 | 0.11 | 34 | 1.3 | 8.26 | 8.35 | 8.74 |
| V395 Cas | 0.747 | 5596 | 82 | 65 | 1.98 | 4.33 | 13.0 | 7.50 | 0.19 | 261 | 7.50 | 0.13 | 35 | 1.0 | 8.15 | 8.06 | 8.77 |
| V395 Cas | 0.595 | 5616 | 117 | 61 | 1.64 | 4.40 | 15.4 | 7.54 | 0.19 | 226 | 7.54 | 0.15 | 26 | 0.9 | 8.04 | 8.02 | 8.57 |
| V407 Cas | 0.079 | 5794 | 91 | 51 | 1.96 | 3.70 | 8.1 | 7.61 | 0.16 | 298 | 7.61 | 0.22 | 38 | 1.2 | 8.32 | 8.19 | 8.78 |
| V556 Cas | 0.458 | 5643 | 54 | 66 | 1.95 | 4.01 | 6.2 | 7.50 | 0.11 | 321 | 7.50 | 0.05 | 28 | 1.0 | 8.05 | 8.53 | 8.78 |
| V1017 Cas | 0.591 | 5715 | 127 | 68 | 2.02 | 3.88 | 5.9 | 7.32 | 0.12 | 326 | 7.32 | 0.16 | 46 | 1.0 | 8.03 | 8.30 | 8.72 |
| V1019 Cas | 0.223 | 6164 | 84 | 55 | 2.09 | 3.36 | 12.6 | 7.66 | 0.18 | 245 | 7.65 | 0.12 | 29 | 1.7 | 8.17 | 8.29 | 8.86 |
| V1019 Cas | 0.450 | 5960 | 58 | 60 | 2.28 | 3.79 | 10.2 | 7.49 | 0.12 | 276 | 7.49 | 0.13 | 38 | 1.2 | 8.28 | 8.44 | 8.87 |
| V1020 Cas | 0.491 | 6250 | 70 | 50 | 2.45 | 4.22 | 11.7 | 7.65 | 0.15 | 246 | 7.65 | 0.11 | 27 | 1.5 | 8.39 | 8.77 | 8.84 |

Table 3
Stellar Parameters and Abundances

| Object | Phase | T (K) | σ | N | Log(g) (cm/s^2) | $V_t$ (km/s) | $V_M$ (km/s) | Fe I | σ | N | Fe II | Sigma | N | Li | C | N | O |
|---|---|---|---|---|---|---|---|---|---|---|---|---|---|---|---|---|---|
| Delta Cep | 0.764 | 5695 | 52 | 63 | 2.25 | 5.50 | 11.8 | 7.62 | 0.11 | 268 | 7.62 | 0.14 | 27 | 1.0 | 8.28 | 8.35 | 8.92 |
| AK Cep | 0.028 | 6295 | 65 | 58 | 2.12 | 4.23 | 10.3 | 7.55 | 0.15 | 283 | 7.54 | 0.13 | 42 | 1.5 | 8.31 | 8.03 | 8.76 |
| CN Cep | 0.926 | 5466 | 75 | 65 | 1.59 | 3.66 | 7.9 | 7.56 | 0.15 | 321 | 7.56 | 0.19 | 36 | 1.3 | 8.28 | 8.28 | 8.82 |
| DR Cep | 0.696 | 4858 | 190 | 51 | 1.08 | 5.40 | 11.7 | 7.36 | 0.15 | 198 | 7.36 | 0.18 | 20 | 0.5 | 8.28 | … | 8.81 |
| IY Cep | 0.281 | 5349 | 67 | 68 | 1.48 | 5.39 | 11.1 | 7.61 | 0.17 | 258 | 7.61 | 0.16 | 22 | 1.1 | 8.16 | 8.11 | 8.57 |
| MU Cep | 0.973 | 6094 | 84 | 61 | 2.50 | 4.74 | 11.0 | 7.68 | 0.15 | 233 | 7.68 | 0.14 | 35 | 1.3 | 8.25 | 8.19 | 8.69 |
| CD Cyg | 0.146 | 5965 | 73 | 65 | 1.50 | 4.40 | 12.2 | 7.65 | 0.13 | 215 | 7.66 | 0.07 | 17 | 1.7 | 8.34 | 8.44 | 8.83 |
| EP Cyg | 0.300 | 5793 | 113 | 72 | 1.73 | 2.95 | 8.3 | 7.44 | 0.16 | 323 | 7.44 | 0.16 | 37 | 1.2 | 8.07 | 8.16 | 8.65 |
| EU Cyg | 0.243 | 5457 | 74 | 64 | 1.17 | 3.31 | 8.6 | 7.30 | 0.14 | 310 | 7.30 | 0.13 | 40 | 1.2 | 8.01 | 8.00 | 8.58 |
| EX Cyg | 0.746 | 5629 | 51 | 60 | 2.20 | 5.23 | 11.3 | 7.75 | 0.16 | 272 | 7.74 | 0.19 | 31 | 1.5 | 8.25 | 8.44 | 8.98 |
| EZ Cyg | 0.391 | 5150 | 66 | 76 | 1.17 | 3.19 | 7.1 | 7.78 | 0.15 | 323 | 7.78 | 0.13 | 25 | 1.0 | 8.46 | 8.16 | 8.60 |
| GH Cyg | 0.403 | 5633 | 34 | 66 | 1.93 | 3.90 | 7.9 | 7.71 | 0.14 | 346 | 7.71 | 0.10 | 20 | 1.3 | 8.29 | 8.40 | 8.87 |
| GI Cyg | 0.394 | 5610 | 60 | 70 | 1.80 | 3.86 | 10.9 | 7.77 | 0.15 | 267 | 7.77 | 0.18 | 29 | 1.2 | 8.23 | 8.34 | 8.25 |
| GL Cyg | 0.856 | 6091 | 133 | 51 | 2.41 | 4.24 | 11.8 | 7.55 | 0.17 | 276 | 7.55 | 0.11 | 33 | 1.4 | 8.15 | 7.97 | 8.92 |
| IY Cyg | 0.964 | 5477 | 147 | 59 | 1.68 | 5.43 | 13.2 | 7.41 | 0.15 | 189 | 7.41 | 0.15 | 18 | 1.2 | 8.10 | 8.41 | 8.80 |
| KX Cyg | 0.157 | 5814 | 84 | 66 | 1.84 | 6.92 | 10.8 | 7.71 | 0.12 | 203 | 7.71 | 0.13 | 24 | 1.3 | 8.36 | 8.15 | 9.08 |
| V347 Cyg | 0.098 | 5822 | 59 | 69 | 1.93 | 4.16 | 9.6 | 7.75 | 0.16 | 305 | 7.75 | 0.10 | 31 | 1.1 | 8.31 | 8.38 | 8.86 |
| V356 Cyg | 0.697 | 5654 | 55 | 68 | 1.83 | 4.36 | 10.3 | 7.67 | 0.18 | 309 | 7.67 | 0.17 | 34 | 1.0 | 8.16 | 8.14 | 8.68 |
| V396 Cyg | 0.892 | 5266 | 62 | 46 | 0.83 | 6.55 | 16.4 | 7.60 | 0.13 | 131 | 7.60 | 0.20 | 13 | 0.4 | 8.02 | 8.21 | 8.14 |
| V396 Cyg | 0.148 | 5344 | 84 | 76 | 0.88 | 3.88 | 11.3 | 7.62 | 0.16 | 263 | 7.62 | 0.12 | 21 | 0.9 | 8.21 | 8.27 | 8.81 |
| V438 Cyg | 0.961 | 6233 | 45 | 55 | 2.62 | 7.25 | 7.5 | 7.84 | 0.15 | 297 | 7.84 | 0.15 | 28 | 1.8 | 8.44 | 8.63 | 8.99 |
| V438 Cyg | 0.020 | 6321 | 39 | 52 | 2.14 | 5.76 | 8.5 | 7.81 | 0.15 | 274 | 7.81 | 0.10 | 32 | 1.2 | 8.45 | 8.45 | 8.97 |
| V438 Cyg | 0.785 | 5682 | 37 | 60 | 2.06 | 6.46 | 11.6 | 7.84 | 0.14 | 236 | 7.84 | 0.09 | 21 | 1.3 | 8.32 | 7.98 | 8.90 |
| V459 Cyg | 0.634 | 5297 | 35 | 59 | 1.44 | 4.74 | 8.7 | 7.59 | 0.26 | 347 | 7.59 | 0.29 | 25 | 1.0 | 8.12 | 8.29 | 8.93 |
| V492 Cyg | 0.929 | 5861 | 54 | 71 | 2.02 | 4.68 | 10.7 | 7.75 | 0.13 | 256 | 7.76 | 0.13 | 25 | 1.3 | 8.33 | 8.29 | 8.57 |
| V492 Cyg | 0.243 | 5491 | 39 | 70 | 1.77 | 4.60 | 9.4 | 7.74 | 0.15 | 319 | 7.74 | 0.16 | 28 | 1.0 | 8.32 | 8.16 | 8.86 |
| V495 Cyg | 0.010 | 6178 | 67 | 64 | 2.02 | 4.36 | 10.1 | 7.74 | 0.15 | 301 | 7.73 | 0.10 | 56 | 1.0 | 8.28 | 8.09 | 8.69 |
| V514 Cyg | 0.182 | 6102 | 44 | 56 | 2.49 | 4.37 | 8.5 | 7.70 | 0.12 | 290 | 7.70 | 0.08 | 30 | 1.2 | 8.37 | 8.58 | 9.00 |
| V514 Cyg | 0.961 | 6473 | 72 | 49 | 2.36 | 4.27 | 10.3 | 7.63 | 0.13 | 231 | 7.63 | 0.09 | 36 | 1.4 | 8.27 | 8.53 | 8.76 |
| V514 Cyg | 0.921 | 6290 | 71 | 50 | 2.63 | 5.15 | 10.9 | 7.72 | 0.15 | 250 | 7.72 | 0.13 | 35 | 1.4 | 8.30 | 8.29 | 8.91 |
| V520 Cyg | 0.526 | 5629 | 58 | 63 | 1.83 | 4.03 | 12.5 | 7.58 | 0.17 | 240 | 7.58 | 0.13 | 48 | 1.7 | 8.23 | 9.10 | 8.68 |
| V538 Cyg | 0.191 | 5728 | 60 | 60 | 1.77 | 3.92 | 7.1 | 7.55 | 0.16 | 370 | 7.54 | 0.11 | 41 | 1.0 | 8.22 | 8.47 | 8.83 |

Table 3
Stellar Parameters and Abundances

| Object | Phase | T (K) | σ | N | Log(g) (cm/s^2) | $V_t$ (km/s) | $V_M$ (km/s) | Fe I | σ | N | Fe II | Sigma | N | Li | C | N | O |
|---|---|---|---|---|---|---|---|---|---|---|---|---|---|---|---|---|---|
| V547 Cyg | 0.273 | 5866 | 55 | 58 | 2.07 | 3.85 | 5.8 | 7.65 | 0.12 | 350 | 7.65 | 0.08 | 35 | 0.9 | 8.45 | 8.39 | 8.96 |
| V609 Cyg | 0.238 | 5899 | 105 | 66 | 1.76 | 7.25 | 9.0 | 7.72 | 0.17 | 241 | 7.72 | 0.09 | 19 | 1.5 | 8.28 | 8.21 | 8.95 |
| V621 Cyg | 0.832 | 5645 | 83 | 66 | 1.76 | 4.92 | 14.7 | 7.62 | 0.20 | 262 | 7.62 | 0.17 | 28 | 0.9 | 8.11 | 8.59 | 8.78 |
| V1020 Cyg | 0.976 | 6173 | 229 | 55 | 2.39 | 3.69 | 11.5 | 7.79 | 0.24 | 229 | 7.79 | 0.11 | 24 | 1.2 | 8.60 | 8.38 | 8.95 |
| V1025 Cyg | 0.237 | 5649 | 81 | 72 | 1.88 | 4.16 | 7.3 | 7.73 | 0.12 | 304 | 7.73 | 0.13 | 38 | 0.8 | 8.42 | 7.97 | 8.93 |
| V1033 Cyg | 0.982 | 6025 | 34 | 56 | 2.32 | 3.43 | 7.9 | 7.62 | 0.12 | 322 | 7.62 | 0.06 | 29 | 3.2 | 8.64 | 8.44 | 8.90 |
| V1046 Cyg | 0.274 | 5867 | 34 | 56 | 2.25 | 4.26 | 11.6 | 7.73 | 0.16 | 285 | 7.73 | 0.07 | 22 | 1.8 | 8.35 | 8.38 | 8.95 |
| V1364 Cyg | 0.346 | 5056 | 58 | 66 | 1.25 | 3.92 | 6.2 | 7.79 | 0.16 | 292 | 7.79 | 0.16 | 26 | 0.9 | 8.53 | 8.02 | 9.05 |
| V1397 Cyg | 0.030 | 5761 | 87 | 61 | 1.50 | 3.62 | 10.6 | 7.46 | 0.12 | 212 | 7.50 | 0.16 | 31 | 1.8 | 8.07 | 8.28 | 8.63 |
| V1397 Cyg | 0.069 | 5670 | 56 | 63 | 2.05 | 3.94 | 9.9 | 7.55 | 0.12 | 270 | 7.55 | 0.09 | 29 | 1.6 | 8.14 | 8.06 | 8.83 |
| EK Del | 0.274 | 5715 | 487 | 32 | 1.69 | 2.00 | 8.1 | 5.96 | 0.29 | 195 | 5.95 | 0.25 | 42 | 1.0 | 6.57 | 7.36 | 7.80 |
| Zeta Gem | 0.687 | 5602 | 32 | 61 | 2.09 | 6.04 | 9.7 | 7.60 | 0.10 | 245 | 7.61 | 0.10 | 24 | 1.0 | 8.22 | 8.35 | 8.90 |
| RZ Gem | 0.316 | 5831 | 109 | 70 | 1.96 | 3.11 | 5.5 | 7.35 | 0.11 | 362 | 7.35 | 0.10 | 44 | 1.4 | 8.08 | 8.20 | 8.68 |
| RZ Gem | 0.879 | 6387 | 136 | 57 | 2.36 | 3.82 | 12.0 | 7.38 | 0.18 | 248 | 7.38 | 0.12 | 40 | 2.1 | 8.06 | 8.13 | 8.68 |
| RZ Gem | 0.117 | 6333 | 93 | 57 | 2.14 | 3.85 | 8.1 | 7.32 | 0.11 | 273 | 7.32 | 0.10 | 44 | 1.7 | 8.05 | 8.00 | 8.70 |
| RZ Gem | 0.646 | 5563 | 110 | 61 | 1.73 | 3.83 | 9.4 | 7.27 | 0.11 | 251 | 7.27 | 0.11 | 37 | 1.3 | 7.88 | 8.56 | 8.63 |
| W Gem | 0.034 | 6189 | 46 | 56 | 2.21 | 4.43 | 8.4 | 7.57 | 0.10 | 257 | 7.57 | 0.08 | 34 | 1.2 | 8.23 | 8.32 | 8.82 |
| W Gem | 0.399 | 5584 | 46 | 65 | 1.83 | 4.00 | 6.8 | 7.46 | 0.10 | 322 | 7.46 | 0.08 | 32 | 1.0 | 8.17 | 7.62 | 8.85 |
| AA Gem | 0.422 | 5141 | 76 | 65 | 1.23 | 3.92 | 6.6 | 7.34 | 0.11 | 320 | 7.34 | 0.14 | 37 | 0.8 | 8.10 | 7.90 | 8.73 |
| AA Gem | 0.658 | 5190 | 157 | 52 | 1.45 | 5.85 | 13.2 | 7.38 | 0.19 | 234 | 7.38 | 0.15 | 25 | 0.6 | 7.94 | 8.04 | 8.61 |
| AD Gem | 0.412 | 5714 | 103 | 63 | 2.06 | 3.55 | 6.1 | 7.33 | 0.13 | 364 | 7.33 | 0.11 | 46 | 1.0 | 8.01 | 8.12 | 8.72 |
| AD Gem | 0.463 | 5664 | 93 | 62 | 2.01 | 3.49 | 6.2 | 7.38 | 0.11 | 343 | 7.38 | 0.12 | 46 | 1.0 | 7.92 | 8.07 | 8.69 |
| BB Gem | 0.709 | 5910 | 114 | 60 | 2.12 | 3.64 | 9.8 | 7.40 | 0.14 | 279 | 7.40 | 0.11 | 42 | 1.2 | 7.87 | 7.98 | 8.53 |
| BW Gem | 0.219 | 6177 | 138 | 65 | 2.24 | 3.18 | 9.2 | 7.39 | 0.15 | 279 | 7.39 | 0.14 | 40 | 1.6 | 7.95 | 7.99 | 8.77 |
| BW Gem | 0.437 | 5807 | 139 | 66 | 2.25 | 3.40 | 6.3 | 7.25 | 0.11 | 316 | 7.25 | 0.08 | 41 | 1.0 | 7.97 | 8.03 | 8.64 |
| DX Gem | 0.151 | 6316 | 70 | 49 | 2.21 | 3.61 | 13.2 | 7.46 | 0.14 | 210 | 7.46 | 0.12 | 35 | 2.0 | 8.05 | 8.15 | 8.71 |
| BB Her | 0.221 | 5766 | 57 | 65 | 2.10 | 4.39 | 8.7 | 7.78 | 0.11 | 310 | 7.78 | 0.08 | 32 | 1.0 | 8.47 | 8.55 | 8.94 |
| BB Her | 0.104 | 5893 | 66 | 64 | 1.87 | 4.03 | 9.2 | 7.75 | 0.13 | 316 | 7.75 | 0.09 | 31 | 1.4 | 8.40 | 8.46 | 8.94 |
| BB Her | 0.026 | 6068 | 49 | 58 | 2.08 | 4.27 | 10.3 | 7.76 | 0.13 | 327 | 7.76 | 0.08 | 32 | 1.3 | 8.46 | 8.54 | 8.98 |
| V Lac | 0.016 | 6582 | 49 | 45 | 2.30 | 3.84 | 9.2 | 7.56 | 0.11 | 240 | 7.57 | 0.07 | 39 | 1.9 | 8.27 | 8.16 | 8.83 |
| X Lac | 0.133 | 6173 | 39 | 56 | 2.02 | 3.74 | 10.2 | 7.58 | 0.10 | 237 | 7.58 | 0.10 | 38 | 1.3 | 8.23 | 8.35 | 8.78 |
| Y Lac | 0.006 | 6542 | 59 | 42 | 2.30 | 4.14 | 14.0 | 7.53 | 0.14 | 207 | 7.54 | 0.11 | 36 | 1.5 | 8.17 | 8.57 | 8.82 |

Table 3
Stellar Parameters and Abundances

| Object | Phase | T (K) | σ | N | Log(g) (cm/s^2) | $V_t$ (km/s) | $V_M$ (km/s) | Fe I | σ | N | Fe II | Sigma | N | Li | C | N | O |
|---|---|---|---|---|---|---|---|---|---|---|---|---|---|---|---|---|---|
| Z Lac | 0.260 | 5604 | 55 | 65 | 1.53 | 3.36 | 8.5 | 7.60 | 0.11 | 294 | 7.60 | 0.08 | 35 | 1.2 | 8.30 | 8.57 | 8.88 |
| RR Lac | 0.808 | 5706 | 54 | 60 | 1.89 | 4.43 | 13.6 | 7.54 | 0.18 | 277 | 7.54 | 0.10 | 21 | 1.8 | 8.15 | 8.49 | 8.76 |
| BG Lac | 0.210 | 5818 | 31 | 62 | 2.05 | 4.00 | 7.7 | 7.57 | 0.13 | 350 | 7.57 | 0.09 | 37 | 1.2 | 8.27 | 8.30 | 8.90 |
| DF Lac | 0.944 | 6219 | 110 | 46 | 2.23 | 3.99 | 11.0 | 7.56 | 0.20 | 289 | 7.56 | 0.09 | 31 | 1.5 | 8.33 | 8.37 | 8.88 |
| DF Lac | 0.157 | 5804 | 38 | 62 | 2.03 | 3.68 | 7.6 | 7.52 | 0.09 | 277 | 7.52 | 0.12 | 41 | 1.0 | 8.23 | 8.22 | 8.82 |
| FQ Lac | 0.868 | 5776 | 518 | 20 | 1.62 | 3.69 | 6.7 | 7.30 | 0.40 | 147 | 7.30 | 0.30 | 19 | 1.5 | 8.16 | 8.34 | 8.09 |
| V411 Lac | … | 5939 | 96 | 57 | 2.54 | 5.64 | 16.0 | 7.52 | 0.14 | 182 | 7.53 | 0.10 | 21 | 1.6 | 8.17 | 8.66 | 8.75 |
| V473 Lyr | 0.697 | 6098 | 99 | 57 | 2.52 | 4.95 | 6.8 | 7.45 | 0.08 | 251 | 7.45 | 0.10 | 35 | 1.0 | 8.18 | 7.99 | 8.68 |
| V473 Lyr | 0.581 | 6072 | 108 | 55 | 2.29 | 4.35 | 7.7 | 7.40 | 0.09 | 282 | 7.40 | 0.07 | 35 | 1.2 | 8.13 | 8.10 | 8.72 |
| V473 Lyr | 0.204 | 6003 | 63 | 58 | 2.31 | 4.31 | 6.0 | 7.41 | 0.08 | 280 | 7.41 | 0.09 | 39 | 1.1 | 8.17 | 8.15 | 8.81 |
| V473 Lyr | 0.417 | 5969 | 95 | 59 | 2.47 | 4.29 | 7.8 | 7.50 | 0.11 | 283 | 7.50 | 0.07 | 37 | 1.3 | 8.18 | 8.05 | 8.81 |
| V473 Lyr | 0.421 | 5970 | 101 | 62 | 2.50 | 4.50 | 7.5 | 7.47 | 0.11 | 336 | 7.44 | 0.06 | 35 | 1.3 | 8.18 | 8.20 | 8.84 |
| T Mon | 0.102 | 5211 | 58 | 44 | 1.46 | 8.26 | 13.6 | 7.69 | 0.17 | 117 | 7.70 | 0.12 | 8 | 1.1 | 8.18 | 8.53 | 8.97 |
| T Mon | 0.507 | 5162 | 63 | 68 | 1.04 | 3.79 | 8.2 | 7.77 | 0.13 | 260 | 7.77 | 0.12 | 25 | 1.1 | 8.42 | 8.36 | 9.06 |
| TW Mon | 0.710 | 5451 | 138 | 50 | 1.83 | 5.40 | 11.9 | 7.32 | 0.14 | 235 | 7.32 | 0.08 | 23 | 1.2 | 7.95 | 8.15 | 8.76 |
| TX Mon | 0.258 | 5478 | 59 | 71 | 1.28 | 3.51 | 7.8 | 7.46 | 0.12 | 317 | 7.46 | 0.12 | 33 | 1.1 | 8.06 | 8.18 | 8.72 |
| TX Mon | 0.746 | 6035 | 105 | 60 | 1.98 | 4.35 | 10.5 | 7.49 | 0.12 | 242 | 7.49 | 0.07 | 31 | 1.0 | 8.02 | 8.11 | 8.47 |
| TY Mon | 0.809 | 5765 | 121 | 58 | 1.98 | 5.03 | 13.4 | 7.52 | 0.14 | 225 | 7.52 | 0.17 | 26 | 1.0 | 7.95 | 8.75 | 8.72 |
| TZ Mon | 0.149 | 5898 | 37 | 61 | 1.77 | 3.97 | 8.6 | 7.55 | 0.11 | 279 | 7.55 | 0.12 | 37 | 1.4 | 8.21 | 8.21 | 8.76 |
| TZ Mon | 0.470 | 5502 | 55 | 66 | 1.77 | 4.27 | 7.5 | 7.47 | 0.10 | 290 | 7.47 | 0.08 | 32 | 0.7 | 8.18 | 8.34 | 8.82 |
| UY Mon | 0.980 | 6628 | 71 | 51 | 2.65 | 3.58 | 8.1 | 7.46 | 0.11 | 236 | 7.46 | 0.12 | 46 | 0.7 | 8.12 | 8.28 | 8.83 |
| UY Mon | 0.074 | 6543 | 72 | 49 | 2.56 | 3.64 | 7.7 | 7.40 | 0.10 | 263 | 7.39 | 0.09 | 42 | 1.6 | 8.12 | 8.16 | 8.75 |
| UY Mon | 0.379 | 6190 | 68 | 55 | 2.45 | 3.44 | 7.5 | 7.37 | 0.09 | 304 | 7.37 | 0.09 | 47 | 1.5 | 8.17 | 8.19 | 8.76 |
| VZ Mon | 0.518 | 6046 | 178 | 51 | 2.12 | 3.80 | 12.7 | 7.43 | 0.24 | 225 | 7.43 | 0.23 | 29 | 1.6 | 8.12 | 8.06 | 8.71 |
| WW Mon | 0.841 | 5741 | 177 | 57 | 2.04 | 4.76 | 7.7 | 7.36 | 0.13 | 177 | 7.35 | 0.10 | 18 | 1.0 | 7.54 | 8.01 | 8.55 |
| WW Mon | 0.918 | 6318 | 249 | 46 | 2.80 | 4.72 | 12.0 | 7.31 | 0.18 | 214 | 7.31 | 0.09 | 29 | 1.5 | 7.93 | 8.10 | 8.60 |
| WW Mon | 0.561 | 5655 | 161 | 63 | 2.02 | 4.15 | 8.3 | 7.26 | 0.13 | 285 | 7.26 | 0.12 | 34 | 1.1 | 7.91 | 8.16 | 8.66 |
| XX Mon | 0.061 | 6307 | 161 | 60 | 1.96 | 3.97 | 13.4 | 7.51 | 0.13 | 218 | 7.51 | 0.14 | 38 | 1.9 | 8.22 | 8.25 | 8.69 |
| XX Mon | 0.128 | 6190 | 89 | 53 | 2.16 | 4.10 | 9.3 | 7.49 | 0.12 | 247 | 7.49 | 0.12 | 38 | 1.2 | 8.14 | 8.39 | 8.80 |
| YY Mon | 0.375 | 5765 | 179 | 55 | 1.62 | 3.36 | 7.9 | 7.04 | 0.16 | 271 | 7.02 | 0.23 | 39 | 1.0 | 7.67 | 3.81 | 8.45 |
| AA Mon | 0.841 | 5797 | 155 | 53 | 2.26 | 4.90 | 13.7 | 7.41 | 0.18 | 247 | 7.41 | 0.09 | 16 | 1.2 | 8.06 | 8.24 | 8.50 |
| AC Mon | 0.007 | 6222 | 90 | 63 | 1.96 | 4.46 | 10.1 | 7.49 | 0.16 | 307 | 7.49 | 0.10 | 29 | 1.5 | 8.12 | 8.25 | 8.67 |

Table 3
Stellar Parameters and Abundances

| Object | Phase | T (K) | σ | N | Log(g) (cm/s^2) | $V_t$ (km/s) | $V_M$ (km/s) | Fe I | σ | N | Fe II | Sigma | N | Li | C | N | O |
|---|---|---|---|---|---|---|---|---|---|---|---|---|---|---|---|---|---|
| AC Mon | 0.067 | 6080 | 67 | 64 | 2.05 | 5.03 | 10.4 | 7.52 | 0.15 | 312 | 7.52 | 0.12 | 38 | 1.1 | 8.15 | 8.29 | 8.84 |
| AC Mon | 0.037 | 6168 | 101 | 58 | 2.10 | 5.01 | 9.2 | 7.48 | 0.15 | 312 | 7.48 | 0.13 | 38 | 1.5 | 8.06 | 8.21 | 8.74 |
| AC Mon | 0.221 | 5912 | 80 | 67 | 1.88 | 4.01 | 6.8 | 7.41 | 0.15 | 380 | 7.42 | 0.09 | 42 | 1.0 | 8.08 | 8.14 | 8.69 |
| BE Mon | 0.468 | 5756 | 78 | 70 | 2.13 | 4.20 | 9.4 | 7.57 | 0.17 | 342 | 7.57 | 0.16 | 34 | 1.4 | 8.16 | 8.26 | 8.75 |
| BE Mon | 0.812 | 5839 | 128 | 61 | 2.41 | 4.30 | 13.2 | 7.60 | 0.18 | 295 | 7.62 | 0.08 | 23 | 1.2 | 8.12 | 8.30 | 8.85 |
| BE Mon | 0.890 | 6150 | 111 | 60 | 2.23 | 3.99 | 11.9 | 7.56 | 0.14 | 277 | 7.56 | 0.12 | 32 | 1.0 | 8.06 | 8.18 | 8.78 |
| BE Mon | 0.038 | 6427 | 65 | 53 | 2.49 | 3.83 | 10.7 | 7.59 | 0.13 | 272 | 7.59 | 0.12 | 39 | 1.4 | 8.20 | 8.36 | 8.80 |
| BV Mon | 0.159 | 6283 | 100 | 64 | 2.59 | 4.06 | 11.6 | 7.55 | 0.16 | 281 | 7.55 | 0.14 | 42 | 1.7 | 8.26 | 8.33 | 8.82 |
| BV Mon | 0.049 | 6571 | 157 | 38 | 1.93 | 4.02 | 11.0 | 7.26 | 0.16 | 203 | 7.26 | 0.15 | 47 | 1.0 | 7.96 | 8.15 | 8.62 |
| CS Mon | 0.674 | 5392 | 116 | 53 | 1.61 | 4.71 | 16.3 | 7.41 | 0.18 | 247 | 7.41 | 0.10 | 22 | 1.4 | 7.91 | 8.14 | 8.92 |
| CS Mon | 0.945 | 6003 | 119 | 59 | 2.14 | 5.28 | 12.4 | 7.43 | 0.13 | 221 | 7.43 | 0.10 | 27 | 1.5 | 7.99 | 8.05 | 8.67 |
| CU Mon | 0.558 | 5593 | 249 | 49 | 1.83 | 3.95 | 14.3 | 7.25 | 0.23 | 223 | 7.26 | 0.12 | 24 | 1.1 | 7.96 | 8.00 | 8.74 |
| CU Mon | 0.959 | 6382 | 227 | 52 | 2.29 | 3.70 | 13.2 | 7.29 | 0.20 | 240 | 7.29 | 0.10 | 34 | 1.4 | 7.93 | 8.14 | 8.49 |
| CV Mon | 0.589 | 5516 | 85 | 64 | 1.80 | 3.98 | 9.5 | 7.46 | 0.19 | 337 | 7.46 | 0.08 | 38 | 1.4 | 8.14 | 8.20 | 8.80 |
| CV Mon | 0.773 | 5545 | 76 | 64 | 1.50 | 4.11 | 14.0 | 7.50 | 0.17 | 281 | 7.50 | 0.15 | 22 | 1.2 | 8.00 | 8.16 | 8.64 |
| CV Mon | 0.229 | 5922 | 49 | 66 | 1.87 | 3.22 | 8.2 | 7.54 | 0.13 | 304 | 7.53 | 0.09 | 37 | 1.0 | 8.22 | 8.24 | 8.76 |
| CV Mon | 0.490 | 5584 | 74 | 62 | 2.02 | 4.02 | 8.2 | 7.52 | 0.15 | 302 | 7.52 | 0.14 | 36 | 1.4 | 8.23 | 8.18 | 8.78 |
| EE Mon | 0.094 | 6358 | 252 | 45 | 1.70 | 3.79 | 14.8 | 6.98 | 0.18 | 187 | 6.97 | 0.16 | 44 | 1.5 | 7.80 | 8.17 | 8.57 |
| EK Mon | 0.377 | 5644 | 76 | 71 | 1.83 | 3.65 | 8.2 | 7.55 | 0.14 | 329 | 7.56 | 0.10 | 33 | 1.1 | 8.17 | 8.15 | 8.73 |
| EK Mon | 0.324 | 5706 | 53 | 65 | 1.85 | 3.43 | 7.7 | 7.52 | 0.12 | 313 | 7.52 | 0.09 | 35 | 1.6 | 8.13 | 8.15 | 8.76 |
| FG Mon | 0.433 | 5730 | 102 | 66 | 1.76 | 3.04 | 6.1 | 7.36 | 0.15 | 357 | 7.36 | 0.12 | 42 | 1.2 | 7.97 | 8.21 | 8.51 |
| FI Mon | 0.083 | 6244 | 169 | 55 | 2.15 | 3.35 | 6.1 | 7.40 | 0.20 | 233 | 7.40 | 0.08 | 27 | 1.3 | 7.80 | 7.99 | 8.32 |
| FT Mon | 0.967 | 5985 | 136 | 60 | 2.22 | 3.39 | 6.7 | 7.29 | 0.10 | 277 | 7.29 | 0.16 | 51 | 1.3 | 7.94 | 7.96 | 8.58 |
| V446 Mon | 0.441 | 5799 | 255 | 57 | 2.29 | 2.99 | 6.7 | 7.20 | 0.15 | 238 | 7.19 | 0.20 | 46 | 1.8 | 7.91 | 8.06 | 8.63 |
| V447 Mon | 0.328 | 6028 | 247 | 57 | 2.34 | 4.22 | 10.3 | 7.17 | 0.18 | 215 | 7.17 | 0.22 | 38 | 2.0 | 8.05 | 7.91 | 8.74 |
| V447 Mon | 0.205 | 5846 | 185 | 62 | 1.85 | 3.33 | 6.8 | 7.09 | 0.14 | 269 | 7.09 | 0.15 | 39 | 1.6 | 7.80 | 8.32 | 8.51 |
| V465 Mon | 0.092 | 6429 | 94 | 56 | 2.43 | 3.54 | 11.2 | 7.55 | 0.11 | 229 | 7.55 | 0.12 | 41 | 1.5 | 8.22 | 8.12 | 8.76 |
| V465 Mon | 0.820 | 5996 | 78 | 57 | 2.53 | 3.79 | 9.4 | 7.56 | 0.12 | 243 | 7.56 | 0.09 | 33 | 1.2 | 8.19 | 8.22 | 8.89 |
| V465 Mon | 0.244 | 6367 | 67 | 51 | 2.26 | 3.56 | 12.0 | 7.46 | 0.13 | 250 | 7.46 | 0.10 | 39 | 1.3 | 8.17 | 8.16 | 8.76 |
| V484 Mon | 0.497 | 5793 | 206 | 57 | 2.21 | 3.63 | 11.7 | 7.44 | 0.26 | 286 | 7.44 | 0.15 | 27 | 1.0 | 7.85 | 8.42 | 8.45 |
| V495 Mon | 0.021 | 6174 | 150 | 58 | 2.24 | 4.25 | 10.4 | 7.39 | 0.13 | 258 | 7.39 | 0.10 | 33 | 1.7 | 8.00 | 8.21 | 8.59 |
| V504 Mon | 0.864 | 6158 | 119 | 54 | 2.89 | 4.30 | 10.8 | 7.50 | 0.14 | 281 | 7.51 | 0.12 | 36 | 1.5 | 8.16 | 8.21 | 8.82 |

Table 3
Stellar Parameters and Abundances

| Object | Phase | T (K) | σ | N | Log(g) (cm/s^2) | $V_t$ (km/s) | $V_M$ (km/s) | Fe I | σ | N | Fe II | Sigma | N | Li | C | N | O |
|---|---|---|---|---|---|---|---|---|---|---|---|---|---|---|---|---|---|
| V504 Mon | 0.963 | 6408 | 157 | 52 | 2.95 | 4.29 | 13.0 | 7.52 | 0.14 | 211 | 7.52 | 0.10 | 32 | 1.5 | 8.16 | 8.10 | 8.65 |
| V508 Mon | 0.374 | 5617 | 107 | 63 | 1.62 | 3.19 | 8.5 | 7.30 | 0.14 | 338 | 7.30 | 0.10 | 36 | 1.1 | 7.80 | 8.06 | 8.55 |
| V508 Mon | 0.301 | 5706 | 137 | 65 | 1.95 | 3.64 | 7.9 | 7.44 | 0.27 | 420 | 7.44 | 0.15 | 45 | 1.2 | 7.77 | 8.12 | 8.73 |
| V510 Mon | 0.295 | 5614 | 75 | 63 | 1.84 | 4.31 | 7.5 | 7.44 | 0.18 | 331 | 7.45 | 0.13 | 38 | 1.0 | 8.12 | 8.41 | 8.70 |
| V526 Mon | 0.971 | 6551 | 83 | 47 | 2.44 | 3.57 | 6.4 | 7.34 | 0.14 | 330 | 7.34 | 0.11 | 49 | 1.7 | 8.08 | 8.18 | 8.57 |
| V526 Mon | 0.467 | 6624 | 112 | 50 | 2.62 | 3.65 | 8.3 | 7.43 | 0.09 | 188 | 7.43 | 0.14 | 48 | 2.0 | 8.12 | 8.14 | 8.66 |
| V526 Mon | 0.091 | 6257 | 96 | 57 | 2.38 | 3.67 | 6.2 | 7.28 | 0.09 | 261 | 7.28 | 0.11 | 51 | 1.5 | 8.03 | 8.11 | 8.67 |
| V526 Mon | 0.146 | 6646 | 64 | 42 | 2.44 | 3.78 | 6.7 | 7.29 | 0.10 | 253 | 7.29 | 0.10 | 46 | 1.4 | 8.02 | 8.36 | 8.58 |
| CR Ori | 0.458 | 5673 | 138 | 60 | 1.68 | 3.53 | 10.7 | 7.27 | 0.19 | 301 | 7.28 | 0.12 | 27 | 1.2 | 7.49 | 8.25 | 8.35 |
| CR Ori | 0.990 | 6452 | 144 | 46 | 2.59 | 5.54 | 13.3 | 7.38 | 0.16 | 216 | 7.38 | 0.17 | 36 | 1.5 | 7.71 | 8.45 | 8.49 |
| CR Ori | 0.397 | 5690 | 134 | 53 | 1.98 | 3.12 | 10.6 | 7.29 | 0.17 | 255 | 7.29 | 0.12 | 28 | 1.0 | 7.55 | 8.26 | 8.41 |
| DF Ori | 0.256 | 5984 | 170 | 61 | 2.06 | 3.54 | 8.2 | 7.22 | 0.12 | 267 | 7.22 | 0.11 | 38 | 1.7 | 7.92 | 8.02 | 8.54 |
| AU Peg | 0.353 | 5525 | 127 | 57 | 1.81 | 3.75 | 14.7 | 7.96 | 0.20 | 212 | 7.96 | 0.17 | 24 | 1.2 | 8.58 | 8.14 | 9.04 |
| QQ Per | 0.268 | 5164 | 152 | 48 | 0.98 | 3.67 | 12.8 | 6.83 | 0.17 | 242 | 6.83 | 0.15 | 34 | 1.2 | 7.86 | 8.21 | 8.56 |
| SV Per | 0.889 | 5747 | 65 | 66 | 2.17 | 4.68 | 10.0 | 7.56 | 0.11 | 273 | 7.56 | 0.06 | 28 | 1.4 | 8.24 | 8.39 | 8.89 |
| SX Per | 0.904 | 6183 | 129 | 57 | 2.42 | 4.22 | 13.6 | 7.51 | 0.20 | 256 | 7.51 | 0.15 | 38 | 1.4 | 7.98 | 8.21 | 8.33 |
| SX Per | 0.410 | 5784 | 70 | 63 | 2.04 | 3.45 | 8.8 | 7.43 | 0.10 | 274 | 7.42 | 0.09 | 37 | 1.0 | 8.00 | 8.42 | 8.61 |
| UX Per | 0.359 | 6263 | 61 | 56 | 1.97 | 3.31 | 9.1 | 7.45 | 0.12 | 294 | 7.45 | 0.07 | 40 | 1.6 | 8.08 | 8.27 | 8.63 |
| UY Per | 0.666 | 5524 | 55 | 71 | 1.83 | 4.61 | 11.3 | 7.67 | 0.19 | 294 | 7.67 | 0.15 | 28 | 1.1 | 8.34 | 8.34 | 8.97 |
| UY Per | 0.340 | 5866 | 49 | 65 | 2.00 | 3.76 | 9.0 | 7.70 | 0.17 | 356 | 7.70 | 0.15 | 39 | 1.4 | 8.32 | 8.26 | 8.87 |
| VX Per | 0.538 | 6062 | 58 | 60 | 2.20 | 4.89 | 9.1 | 7.58 | 0.11 | 278 | 7.58 | 0.07 | 33 | 1.3 | 8.22 | 8.24 | 8.84 |
| VX Per | 0.471 | 5884 | 77 | 63 | 1.82 | 4.57 | 11.2 | 7.53 | 0.13 | 264 | 7.53 | 0.11 | 31 | 1.1 | 8.13 | 8.24 | 8.65 |
| VY Per | 0.659 | 5496 | 56 | 67 | 1.60 | 3.59 | 11.1 | 7.54 | 0.15 | 272 | 7.53 | 0.12 | 21 | 1.1 | 8.17 | 7.93 | 8.72 |
| AS Per | 0.979 | 6645 | 50 | 42 | 2.42 | 4.02 | 10.4 | 7.64 | 0.15 | 281 | 7.65 | 0.09 | 42 | 1.7 | 8.23 | 8.69 | 8.86 |
| AW Per | 0.812 | 5747 | 43 | 56 | 2.37 | 5.40 | 14.1 | 7.54 | 0.16 | 263 | 7.54 | 0.15 | 26 | 1.1 | 8.18 | 8.42 | 8.91 |
| BM Per | 0.225 | 5792 | 68 | 63 | 1.68 | 5.80 | 10.6 | 7.73 | 0.18 | 251 | 7.73 | 0.11 | 22 | 1.3 | 8.27 | 8.36 | 8.87 |
| CI Per | 0.328 | 6065 | 162 | 15 | 2.01 | 3.18 | 17.5 | 7.18 | 0.27 | 178 | 7.18 | 0.24 | 36 | 1.4 | 7.33 | 8.32 | 8.27 |
| DW Per | 0.086 | 6483 | 118 | 58 | 2.03 | 3.10 | 12.4 | 7.45 | 0.17 | 265 | 7.45 | 0.11 | 37 | 1.7 | 8.14 | 8.17 | 8.63 |
| GP Per | 0.086 | 6142 | 525 | 22 | 1.54 | 3.26 | 13.6 | 6.70 | 0.22 | 151 | 6.70 | 0.12 | 32 | 1.5 | 7.62 | 7.70 | 8.34 |
| HQ Per | 0.456 | 5480 | 157 | 72 | 1.50 | 4.22 | 8.1 | 7.17 | 0.12 | 300 | 7.17 | 0.11 | 33 | 1.0 | 7.84 | 7.75 | 8.55 |
| HQ Per | 0.430 | 5467 | 156 | 62 | 1.18 | 3.99 | 8.2 | 7.08 | 0.12 | 277 | 7.08 | 0.12 | 38 | 1.0 | 7.79 | 7.94 | 8.43 |
| HQ Per | 0.808 | 5645 | 196 | 62 | 1.75 | 5.56 | 12.0 | 7.19 | 0.16 | 260 | 7.19 | 0.09 | 22 | 0.9 | 7.71 | 7.84 | 8.42 |

Table 3
Stellar Parameters and Abundances

| Object | Phase | T (K) | σ | N | Log(g) (cm/s^2) | $V_t$ (km/s) | $V_M$ (km/s) | Fe I | σ | N | Fe II | Sigma | N | Li | C | N | O |
|---|---|---|---|---|---|---|---|---|---|---|---|---|---|---|---|---|---|
| HZ Per | 0.867 | 5616 | 136 | 68 | 1.84 | 4.21 | 12.6 | 7.25 | 0.14 | 220 | 7.25 | 0.10 | 20 | 1.2 | 7.35 | 8.19 | 8.63 |
| MM Per | 0.129 | 5693 | 70 | 64 | 2.00 | 3.57 | 6.4 | 7.43 | 0.10 | 332 | 7.43 | 0.08 | 39 | 1.1 | 8.14 | 8.34 | 8.81 |
| OT Per | 0.032 | 6218 | 216 | 29 | 2.01 | 6.40 | 18.8 | 7.51 | 0.25 | 161 | 7.50 | 0.23 | 22 | 2.0 | 7.82 | 8.03 | 8.47 |
| OT Per | 0.920 | 5551 | 182 | 35 | 1.29 | 6.05 | 18.3 | 7.36 | 0.21 | 169 | 7.37 | 0.20 | 16 | 1.5 | 7.35 | 8.25 | 8.32 |
| X Pup | 0.127 | 6272 | 106 | 51 | 1.77 | 7.05 | 13.0 | 7.58 | 0.17 | 191 | 7.58 | 0.12 | 38 | 1.3 | 8.19 | 8.43 | 8.80 |
| RS Pup | 0.151 | 5018 | 70 | 58 | 1.17 | 5.42 | 8.5 | 7.72 | 0.15 | 201 | 7.72 | 0.08 | 25 | 1.0 | 7.91 | 8.57 | 9.25 |
| VX Pup | 0.233 | 6124 | 74 | 61 | 2.53 | 3.76 | 9.5 | 7.50 | 0.12 | 265 | 7.50 | 0.07 | 42 | 1.5 | 8.08 | 8.15 | 8.79 |
| VZ Pup | 0.767 | 5280 | 75 | 55 | 1.34 | 5.29 | 12.0 | 7.39 | 0.16 | 220 | 7.39 | 0.08 | 19 | 1.0 | 7.86 | 8.40 | 8.77 |
| WX Pup | 0.253 | 5774 | 48 | 61 | 1.95 | 3.98 | 6.0 | 7.46 | 0.10 | 330 | 7.46 | 0.08 | 34 | 0.7 | 8.17 | 8.36 | 8.85 |
| AD Pup | 0.550 | 5120 | 69 | 54 | 1.07 | 4.08 | 8.0 | 7.47 | 0.09 | 242 | 7.47 | 0.18 | 37 | 0.8 | 7.99 | 8.02 | 8.87 |
| AQ Pup | 0.533 | 6458 | 88 | 32 | 1.66 | 7.24 | 18.0 | 7.54 | 0.13 | 140 | 7.54 | 0.12 | 13 | 1.7 | 8.01 | 8.24 | 8.55 |
| BN Pup | 0.113 | 6128 | 42 | 46 | 1.88 | 4.30 | 15.5 | 7.61 | 0.15 | 214 | 7.61 | 0.08 | 24 | 0.9 | 8.14 | 8.52 | 8.93 |
| V335 Pup | 0.064 | 6350 | 49 | 49 | 2.60 | 4.09 | 8.5 | 7.59 | 0.11 | 255 | 7.59 | 0.11 | 39 | 1.7 | 8.34 | 8.39 | 8.88 |
| Y Sct | 0.920 | 5805 | 55 | 64 | 2.14 | 5.24 | 7.8 | 7.74 | 0.10 | 278 | 7.74 | 0.09 | 27 | 1.5 | 8.42 | 8.48 | 9.05 |
| Y Sct | 0.030 | 6063 | 36 | 59 | 2.19 | 5.89 | 7.8 | 7.72 | 0.10 | 267 | 7.72 | 0.11 | 30 | 1.9 | 8.34 | 7.71 | 8.89 |
| Z Sct | 0.347 | 5395 | 113 | 73 | 1.77 | 4.42 | 9.0 | 7.83 | 0.12 | 204 | 7.84 | 0.08 | 18 | 1.0 | 8.57 | … | 9.19 |
| RU Sct | 0.407 | 5441 | 72 | 68 | 1.32 | 4.53 | 6.7 | 7.61 | 0.11 | 286 | 7.61 | 0.08 | 25 | 1.0 | 8.37 | … | 9.00 |
| SS Sct | 0.167 | 5996 | 42 | 59 | 2.03 | 3.50 | 12.0 | 7.62 | 0.14 | 255 | 7.63 | 0.13 | 30 | 1.5 | 8.32 | 8.18 | 8.79 |
| SS Sct | 0.431 | 5657 | 41 | 61 | 2.18 | 4.91 | 11.3 | 7.61 | 0.12 | 230 | 7.61 | 0.16 | 30 | 1.2 | 8.23 | 8.15 | 8.92 |
| SS Sct | 0.416 | 5691 | 52 | 64 | 2.29 | 5.06 | 12.5 | 7.69 | 0.14 | 246 | 7.69 | 0.11 | 29 | 1.4 | 8.23 | 8.42 | 8.88 |
| TY Sct | 0.963 | 6138 | 69 | 57 | 2.44 | 6.89 | 8.0 | 7.87 | 0.13 | 289 | 7.87 | 0.10 | 29 | 1.4 | 8.60 | 8.25 | 9.07 |
| BX Sct | 0.306 | 5878 | 51 | 60 | 2.15 | 3.99 | 9.8 | 7.78 | 0.13 | 277 | 7.79 | 0.14 | 37 | 1.3 | 8.38 | 8.39 | 9.03 |
| CK Sct | 0.952 | 5904 | 58 | 60 | 2.30 | 5.83 | 9.3 | 7.71 | 0.12 | 272 | 7.71 | 0.10 | 29 | 1.0 | 8.36 | 8.35 | 8.92 |
| CM Sct | 0.048 | 6138 | 50 | 56 | 2.38 | 4.47 | 9.7 | 7.65 | 0.12 | 289 | 7.65 | 0.09 | 35 | 1.0 | 8.35 | 8.42 | 8.85 |
| CN Sct | 0.105 | 5935 | 64 | 61 | 2.23 | 5.52 | 7.9 | 7.88 | 0.15 | 270 | 7.88 | 0.28 | 34 | 1.3 | 8.47 | 8.55 | 9.04 |
| CN Sct | 0.092 | 5972 | 90 | 64 | 2.03 | 5.58 | 9.1 | 7.85 | 0.14 | 285 | 7.85 | 0.12 | 33 | 1.2 | 8.48 | 8.55 | 8.95 |
| CN Sct | 0.591 | 5508 | 71 | 65 | 1.51 | 3.78 | 12.6 | 7.76 | 0.14 | 235 | 7.76 | 0.12 | 15 | 1.2 | 8.38 | 8.42 | 8.93 |
| EV Sct | 0.923 | 6456 | 78 | 43 | 2.63 | 4.09 | 14.1 | 7.64 | 0.17 | 208 | 7.64 | 0.09 | 30 | 1.5 | 8.31 | 8.37 | 8.97 |
| EV Sct | 0.809 | 6391 | 84 | 41 | 2.34 | 4.52 | 14.5 | 7.65 | 0.19 | 220 | 7.65 | 0.18 | 35 | 1.5 | 8.13 | 8.24 | 8.87 |
| AA Ser | 0.342 | 5045 | 119 | 68 | 1.26 | 3.52 | 6.4 | 7.91 | 0.16 | 278 | 7.91 | 0.16 | 20 | 1.2 | 8.45 | 8.64 | 9.24 |
| DV Ser | 0.814 | 6364 | 90 | 51 | 1.77 | 5.86 | 11.1 | 7.97 | 0.21 | 244 | 7.97 | 0.10 | 23 | 1.7 | 8.58 | 9.02 | 9.06 |
| DG Sge | 0.069 | 6394 | 60 | 51 | 2.26 | 4.09 | 9.5 | 7.63 | 0.12 | 252 | 7.63 | 0.12 | 41 | 1.7 | 8.37 | 8.52 | 8.90 |

Table 3
Stellar Parameters and Abundances

| Object | Phase | T (K) | σ | N | Log(g) (cm/s^2) | $V_t$ (km/s) | $V_M$ (km/s) | Fe I | σ | N | Fe II | Sigma | N | Li | C | N | O |
|---|---|---|---|---|---|---|---|---|---|---|---|---|---|---|---|---|---|
| GX Sge | 0.196 | 5622 | 114 | 72 | 1.91 | 4.94 | 8.9 | 7.79 | 0.14 | 269 | 7.79 | 0.13 | 25 | 1.4 | 8.43 | 8.49 | 9.01 |
| GY Sge | 0.258 | 5736 | 79 | 60 | 1.78 | 6.98 | 14.9 | 7.87 | 0.18 | 199 | 7.87 | 0.11 | 9 | 1.5 | 8.61 | 8.19 | 8.85 |
| GY Sge | 0.374 | 6012 | 76 | 61 | 1.34 | 4.86 | 13.2 | 7.76 | 0.21 | 229 | 7.76 | 0.10 | 10 | 1.5 | 8.60 | 7.46 | 8.95 |
| GY Sge | 0.767 | 5290 | 58 | 62 | 1.33 | 8.47 | 5.7 | 7.75 | 0.13 | 250 | 7.75 | 0.06 | 16 | 1.2 | 8.64 | 8.28 | 9.19 |
| ST Tau | 0.994 | 6666 | 43 | 46 | 2.43 | 4.04 | 11.1 | 7.50 | 0.10 | 206 | 7.50 | 0.10 | 41 | 1.5 | 8.25 | 8.24 | 8.80 |
| ST Tau | 0.478 | 5727 | 46 | 63 | 2.24 | 3.62 | 7.3 | 7.51 | 0.10 | 316 | 7.52 | 0.07 | 35 | 1.5 | 8.21 | 8.37 | 8.85 |
| AE Tau | 0.248 | 5891 | 109 | 62 | 1.97 | 3.72 | 7.6 | 7.32 | 0.14 | 333 | 7.32 | 0.11 | 42 | 1.5 | 7.99 | 8.08 | 8.69 |
| AV Tau | 0.658 | 5675 | 106 | 64 | 1.95 | 3.85 | 12.5 | 7.43 | 0.19 | 294 | 7.43 | 0.15 | 34 | 1.2 | 8.09 | 8.07 | 8.70 |
| S Vul | 0.144 | 5535 | 95 | 69 | 1.00 | 6.70 | 9.2 | 7.63 | 0.11 | 234 | 7.64 | 0.12 | 22 | 1.1 | 8.29 | 8.43 | 8.82 |
| S Vul | 0.756 | 5394 | 85 | 60 | 1.29 | 8.64 | 12.7 | 7.60 | 0.13 | 180 | 7.60 | 0.04 | 8 | 1.1 | 8.36 | 8.37 | 8.97 |
| U Vul | 0.233 | 6045 | 44 | 62 | 2.25 | 4.36 | 6.5 | 7.69 | 0.10 | 305 | 7.69 | 0.05 | 31 | 1.3 | 8.43 | 8.21 | 9.00 |
| AS Vul | 0.281 | 5279 | 83 | 80 | 1.22 | 3.06 | 7.4 | 7.72 | 0.15 | 319 | 7.72 | 0.19 | 37 | 1.3 | 8.42 | 8.38 | 8.90 |
| DG Vul | 0.969 | 6561 | 77 | 49 | 2.10 | 6.60 | 8.8 | 7.69 | 0.16 | 281 | 7.69 | 0.14 | 32 | 2.0 | 8.32 | 8.53 | 8.87 |
| GSC 0754-1993 | 0.125 | 5649 | 97 | 62 | 2.35 | 4.58 | 6.5 | 7.49 | 0.15 | 323 | 7.49 | 0.14 | 38 | 1.2 | 7.88 | 8.27 | 8.99 |
| GSC 2418-1443 | 0.971 | 5966 | 97 | 58 | 1.80 | 3.75 | 7.9 | 7.36 | 0.12 | 266 | 7.35 | 0.13 | 41 | 1.6 | 8.16 | 7.85 | 8.55 |
| GSC 3706-0233 | 0.425 | 5842 | 106 | 57 | 2.41 | 3.64 | 10.1 | 7.59 | 0.16 | 285 | 7.59 | 0.16 | 37 | 1.3 | 8.24 | 8.18 | 8.92 |
| GSC 3725-0174 | 0.817 | 6452 | 99 | 43 | 2.38 | 4.47 | 10.9 | 7.33 | 0.14 | 231 | 7.33 | 0.16 | 47 | 1.2 | 7.97 | 8.17 | 8.66 |
| GSC 3726-0565 | 0.334 | 5518 | 152 | 50 | 1.64 | 4.80 | 14.4 | 7.28 | 0.19 | 243 | 7.28 | 0.19 | 28 | 1.2 | 7.74 | 7.96 | 8.35 |
| GSC 3729-1127 | 0.065 | 5534 | 85 | 59 | 2.21 | 5.62 | 11.1 | 7.49 | 0.15 | 266 | 7.49 | 0.08 | 21 | 1.6 | 8.20 | 8.22 | 8.99 |
| GSC 3732-0183 | 0.900 | 5514 | 108 | 52 | 1.35 | 4.06 | 12.4 | 7.34 | 0.18 | 264 | 7.34 | 0.13 | 26 | 1.6 | 7.86 | 7.94 | 8.34 |
| GSC 4009-0024 | 0.551 | 5578 | 84 | 62 | 1.85 | 4.19 | 7.2 | 7.34 | 0.12 | 314 | 7.34 | 0.13 | 37 | 1.1 | 8.04 | 8.01 | 8.74 |
| GSC 4038-1585 | 0.160 | 5651 | 47 | 63 | 1.97 | 3.77 | 7.5 | 7.52 | 0.13 | 328 | 7.52 | 0.10 | 33 | 1.4 | 8.19 | 8.23 | 8.80 |
| GSC 4040-1803 | 0.202 | 6482 | 85 | 50 | 2.46 | 3.64 | 11.4 | 7.44 | 0.14 | 225 | 7.44 | 0.11 | 34 | 1.2 | 8.11 | 8.29 | 8.61 |
| GSC 4265-0193 | 0.655 | 6343 | 90 | 51 | 2.31 | 4.42 | 10.0 | 7.49 | 0.15 | 230 | 7.48 | 0.14 | 34 | 1.7 | 8.04 | 8.31 | 8.71 |
| GSC 4265-0569 | 0.168 | 5709 | 96 | 68 | 1.80 | 3.66 | 7.1 | 7.54 | 0.20 | 332 | 7.54 | 0.14 | 37 | 1.1 | 8.28 | 8.14 | 8.87 |
| T Ant | 0.173 | 6289 | 108 | 56 | 2.11 | 4.04 | 11.7 | 7.31 | 0.08 | 219 | 7.31 | 0.09 | 39 | 1.7 | 7.97 | 7.99 | 8.61 |
| L Car | 0.094 | 5265 | 51 | 67 | 1.12 | 5.60 | 9.3 | 7.63 | 0.27 | 243 | 7.63 | 0.30 | 21 | 1.0 | 8.34 | 8.04 | 8.91 |
| SX Car | 0.968 | 6506 | 54 | 47 | 2.62 | 4.98 | 10.5 | 7.55 | 0.11 | 279 | 7.55 | 0.25 | 48 | 1.5 | 8.30 | 8.40 | 8.82 |
| U Car | 0.889 | 4944 | 96 | 55 | 0.89 | 5.22 | 14.1 | 7.54 | 0.15 | 153 | 7.54 | 0.05 | 6 | 1.0 | 8.62 | 8.41 | 8.76 |
| V Car | 0.943 | 5920 | 99 | 65 | 2.12 | 4.95 | 10.7 | 7.54 | 0.11 | 307 | 7.54 | 0.09 | 39 | 1.6 | 8.07 | 8.26 | 8.72 |
| UW Car | 0.035 | 6599 | 53 | 42 | 2.28 | 4.20 | 11.5 | 7.59 | 0.14 | 285 | 7.59 | 0.12 | 48 | 1.5 | 8.28 | 8.45 | 8.87 |

Table 3
Stellar Parameters and Abundances

| Object | Phase | T (K) | σ | N | Log(g) (cm/s^2) | $V_t$ (km/s) | $V_M$ (km/s) | Fe I | σ | N | Fe II | Sigma | N | Li | C | N | O |
|---|---|---|---|---|---|---|---|---|---|---|---|---|---|---|---|---|---|
| UX Car | 0.902 | 6426 | 76 | 56 | 2.61 | 4.64 | 9.3 | 7.55 | 0.09 | 304 | 7.55 | 0.09 | 51 | 1.2 | 8.32 | 8.42 | 8.79 |
| UY Car | 0.946 | 6377 | 41 | 48 | 2.37 | 4.64 | 13.2 | 7.63 | 0.12 | 262 | 7.62 | 0.09 | 33 | 2.0 | 8.31 | 8.37 | 8.85 |
| UZ Car | 0.044 | 6076 | 52 | 63 | 1.95 | 4.03 | 12.7 | 7.63 | 0.12 | 283 | 7.63 | 0.12 | 36 | 1.6 | 8.27 | 8.25 | 8.86 |
| VY Car | 0.169 | 4914 | 73 | 58 | 0.75 | 3.31 | 14.3 | 7.52 | 0.15 | 123 | 7.52 | 0.15 | 11 | 0.9 | 8.24 | 8.58 | 8.86 |
| WW Car | 0.852 | 5877 | 88 | 60 | 2.10 | 5.03 | 13.1 | 7.50 | 0.11 | 262 | 7.50 | 0.10 | 34 | 1.6 | 8.16 | 8.16 | 8.69 |
| WZ Car | 0.009 | 5766 | 88 | 53 | 1.26 | 5.48 | 15.0 | 7.55 | 0.14 | 222 | 7.55 | 0.13 | 19 | 2.0 | 8.05 | 8.10 | 8.60 |
| XX Car | 0.692 | 5936 | 78 | 70 | 1.74 | 4.57 | 9.7 | 7.70 | 0.10 | 268 | 7.69 | 0.10 | 31 | 1.1 | 8.15 | 8.58 | 8.93 |
| XY Car | 0.242 | 5747 | 76 | 71 | 1.71 | 4.39 | 11.7 | 7.57 | 0.10 | 252 | 7.57 | 0.07 | 24 | 1.5 | 8.22 | 8.30 | 8.87 |
| XZ Car | 0.072 | 6160 | 29 | 51 | 1.75 | 5.06 | 13.4 | 7.69 | 0.09 | 200 | 7.69 | 0.05 | 22 | 1.7 | 8.27 | 8.35 | 8.72 |
| YZ Car | 0.860 | 5644 | 43 | 68 | 1.66 | 5.21 | 10.0 | 7.50 | 0.09 | 277 | 7.50 | 0.07 | 25 | 0.9 | 8.22 | 8.27 | 8.85 |
| AQ Car | 0.998 | 5818 | 29 | 65 | 1.99 | 4.68 | 8.0 | 7.53 | 0.11 | 404 | 7.54 | 0.07 | 33 | 1.0 | 8.27 | 8.23 | 8.87 |
| CN Car | 0.066 | 6328 | 38 | 55 | 2.51 | 4.42 | 10.5 | 7.71 | 0.12 | 296 | 7.71 | 0.09 | 35 | 1.3 | 8.42 | 8.44 | 8.96 |
| CY Car | 0.095 | 6057 | 34 | 63 | 2.27 | 4.06 | 8.7 | 7.61 | 0.10 | 342 | 7.61 | 0.10 | 45 | 1.2 | 8.33 | 8.38 | 8.88 |
| DY Car | 0.956 | 6521 | 49 | 51 | 2.45 | 4.73 | 10.6 | 7.57 | 0.11 | 265 | 7.58 | 0.12 | 51 | 1.3 | 8.32 | 8.44 | 8.82 |
| ER Car | 0.872 | 5860 | 50 | 65 | 2.06 | 4.92 | 11.8 | 7.65 | 0.11 | 252 | 7.65 | 0.11 | 29 | 1.1 | 8.22 | 8.16 | 8.84 |
| FI Car | 0.215 | 5209 | 64 | 72 | 1.33 | 4.06 | 12.0 | 7.81 | 0.15 | 256 | 7.81 | 0.14 | 19 | 1.0 | 8.34 | 7.96 | 8.94 |
| FR Car | 0.011 | 5906 | 45 | 63 | 2.20 | 5.80 | 9.7 | 7.61 | 0.12 | 353 | 7.61 | 0.09 | 28 | 1.4 | 7.94 | 8.43 | 8.89 |
| GH Car | 0.965 | 6353 | 40 | 53 | 2.28 | 4.70 | 11.2 | 7.72 | 0.12 | 295 | 7.72 | 0.11 | 35 | 1.7 | 8.32 | 8.41 | 8.82 |
| GX Car | 0.996 | 6312 | 51 | 55 | 2.39 | 4.72 | 11.1 | 7.64 | 0.14 | 325 | 7.64 | 0.11 | 42 | 1.3 | 8.23 | 8.39 | 8.89 |
| HW Car | 0.947 | 5679 | 28 | 63 | 1.96 | 4.85 | 8.3 | 7.59 | 0.10 | 361 | 7.59 | 0.23 | 30 | 1.5 | 8.27 | 8.29 | 8.88 |
| IO Car | 0.072 | 6119 | 61 | 62 | 2.46 | 6.44 | 4.0 | 7.63 | 0.13 | 411 | 7.63 | 0.12 | 50 | 1.5 | 8.40 | 8.51 | 8.98 |
| IT Car | 0.950 | 5791 | 31 | 62 | 1.86 | 4.15 | 10.4 | 7.64 | 0.10 | 293 | 7.64 | 0.07 | 34 | 1.4 | 8.24 | 8.35 | 8.82 |
| V397 Car | 0.903 | 6026 | 66 | 62 | 2.54 | 4.94 | 10.8 | 7.65 | 0.12 | 329 | 7.65 | 0.10 | 38 | 1.3 | 8.30 | 8.32 | 8.92 |
| V Cen #1 | 0.781 | 5592 | 80 | 62 | 2.01 | 5.36 | 11.8 | 7.47 | 0.16 | 304 | 7.47 | 0.08 | 23 | 1.3 | 8.17 | 8.18 | 8.79 |
| V Cen #2 | 0.952 | 6442 | 49 | 50 | 2.57 | 4.49 | 10.6 | 7.58 | 0.11 | 300 | 7.58 | 0.07 | 46 | 1.8 | 8.29 | 8.40 | 8.80 |
| QY Cen | 0.004 | 6164 | 82 | 60 | 2.04 | 7.33 | 13.1 | 7.74 | 0.15 | 249 | 7.74 | 0.13 | 29 | 1.7 | 8.35 | 8.46 | 8.86 |
| XX Cen | 0.116 | 5941 | 36 | 64 | 1.96 | 5.69 | 10.4 | 7.68 | 0.11 | 314 | 7.68 | 0.12 | 34 | 1.3 | 8.28 | 8.38 | 8.88 |
| AY Cen | 0.999 | 5920 | 27 | 61 | 2.13 | 4.32 | 8.8 | 7.58 | 0.12 | 373 | 7.58 | 0.09 | 36 | 1.5 | 8.31 | 8.36 | 8.85 |
| AZ Cen | 0.947 | 6400 | 56 | 54 | 2.61 | 4.18 | 7.2 | 7.59 | 0.09 | 346 | 7.59 | 0.14 | 62 | 1.6 | 8.40 | 8.48 | 8.97 |
| BB Cen | 0.183 | 6178 | 65 | 58 | 2.44 | 4.86 | 6.6 | 7.72 | 0.10 | 375 | 7.72 | 0.10 | 47 | 1.6 | 8.47 | 8.58 | 9.02 |
| KK Cen | 0.012 | 5899 | 45 | 63 | 1.88 | 4.78 | 10.0 | 7.74 | 0.11 | 303 | 7.74 | 0.08 | 27 | 1.2 | 8.40 | 8.35 | 8.87 |
| KN Cen | 0.488 | 4880 | 158 | 69 | 0.99 | 5.58 | 9.0 | 7.91 | 0.15 | 206 | 7.91 | 0.16 | 12 | 1.0 | 8.41 | 8.41 | 9.02 |

Table 3
Stellar Parameters and Abundances

| Object | Phase | T (K) | σ | N | Log(g) (cm/s^2) | $V_t$ (km/s) | $V_M$ (km/s) | Fe I | σ | N | Fe II | Sigma | N | Li | C | N | O |
|---|---|---|---|---|---|---|---|---|---|---|---|---|---|---|---|---|---|
| MZ Cen | 0.207 | 5818 | 81 | 72 | 2.13 | 5.80 | 11.1 | 7.85 | 0.11 | 258 | 7.85 | 0.09 | 27 | 1.3 | 8.48 | 8.51 | 8.96 |
| V339 Cen | 0.142 | 5928 | 31 | 62 | 2.06 | 4.58 | 8.3 | 7.64 | 0.10 | 343 | 7.64 | 0.08 | 36 | 1.0 | 8.40 | 8.39 | 8.94 |
| V378 Cen | 0.131 | 6155 | 40 | 62 | 2.28 | 4.86 | 6.2 | 7.58 | 0.12 | 397 | 7.58 | 0.11 | 49 | 1.5 | 8.27 | 8.45 | 8.92 |
| V381 Cen | 0.896 | 6217 | 82 | 60 | 2.33 | 5.42 | 9.5 | 7.52 | 0.10 | 306 | 7.52 | 0.08 | 43 | 1.5 | 8.22 | 8.24 | 8.80 |
| V419 Cen | 0.905 | 6295 | 60 | 46 | 2.18 | 5.77 | 15.0 | 7.64 | 0.09 | 176 | 7.64 | 0.11 | 26 | 1.8 | 8.23 | 8.24 | 8.81 |
| V496 Cen | 0.025 | 6189 | 33 | 59 | 2.32 | 4.10 | 10.7 | 7.59 | 0.10 | 297 | 7.60 | 0.08 | 35 | 1.0 | 8.30 | 8.37 | 8.90 |
| V659 Cen | 0.856 | 6072 | 34 | 62 | 2.29 | 5.25 | 10.1 | 7.59 | 0.11 | 299 | 7.59 | 0.08 | 36 | 1.3 | 8.25 | 8.26 | 8.88 |
| V737 Cen | 0.967 | 5863 | 25 | 62 | 2.17 | 5.26 | 8.7 | 7.64 | 0.09 | 318 | 7.64 | 0.08 | 32 | 1.5 | 8.33 | 8.34 | 8.92 |
| AV Cir | 0.207 | 6156 | 50 | 61 | 2.53 | 3.61 | 6.6 | 7.67 | 0.10 | 389 | 7.67 | 0.12 | 54 | 1.4 | 8.44 | 8.46 | 9.02 |
| AX Cir #1 | 0.745 | 5669 | 81 | 61 | 2.07 | 4.74 | 13.2 | 7.46 | 0.12 | 306 | 7.47 | 0.11 | 33 | 1.5 | 8.07 | 8.12 | 8.76 |
| AX Cir #2 | 0.119 | 5774 | 41 | 63 | 2.11 | 3.90 | 9.7 | 7.52 | 0.12 | 379 | 7.52 | 0.11 | 42 | 1.5 | 8.17 | 8.26 | 8.83 |
| BP Cir | 0.033 | 6543 | 67 | 53 | 2.62 | 3.36 | 8.2 | 7.52 | 0.10 | 351 | 7.52 | 0.11 | 54 | 1.4 | 8.28 | 8.34 | 8.85 |
| R Cru | 0.100 | 6204 | 28 | 58 | 2.24 | 4.20 | 8.6 | 7.63 | 0.10 | 342 | 7.63 | 0.09 | 42 | 1.8 | 8.36 | 8.34 | 8.91 |
| S Cru | 0.983 | 6471 | 39 | 51 | 2.42 | 4.08 | 10.9 | 7.61 | 0.10 | 264 | 7.62 | 0.09 | 44 | 1.4 | 8.33 | 8.42 | 8.88 |
| T Cru | 0.958 | 5916 | 37 | 62 | 2.19 | 5.26 | 10.2 | 7.64 | 0.11 | 333 | 7.64 | 0.09 | 32 | 1.3 | 8.31 | 8.29 | 8.87 |
| X Cru | 0.125 | 5935 | 38 | 61 | 2.08 | 4.25 | 10.5 | 7.65 | 0.11 | 313 | 7.65 | 0.10 | 36 | 1.3 | 8.28 | 8.38 | 8.94 |
| VW Cru | 0.175 | 5864 | 38 | 65 | 2.13 | 4.57 | 10.3 | 7.69 | 0.11 | 301 | 7.69 | 0.13 | 40 | 1.5 | 8.29 | 8.25 | 8.86 |
| AD Cru | 0.030 | 5988 | 36 | 64 | 2.07 | 3.89 | 8.9 | 7.61 | 0.12 | 354 | 7.62 | 0.11 | 45 | 1.3 | 8.28 | 8.33 | 8.91 |
| AG Cru | 0.039 | 6627 | 48 | 47 | 2.40 | 3.91 | 10.3 | 7.58 | 0.12 | 304 | 7.58 | 0.11 | 52 | 1.8 | 8.34 | 8.51 | 8.91 |
| BG Cru | 0.150 | 6305 | 46 | 32 | 2.68 | 3.25 | 17.0 | 7.42 | 0.15 | 148 | 7.42 | 0.15 | 28 | 1.9 | 8.25 | 8.46 | 9.02 |
| GH Lup | 0.142 | 5455 | 30 | 64 | 1.72 | 4.62 | 9.2 | 7.63 | 0.11 | 343 | 7.63 | 0.08 | 32 | 1.2 | 8.29 | 8.17 | 8.87 |
| R Mus | 0.184 | 5978 | 47 | 65 | 2.10 | 4.53 | 10.5 | 7.65 | 0.10 | 286 | 7.65 | 0.12 | 39 | 1.7 | 8.30 | 8.33 | 8.92 |
| S Mus | 0.106 | 5751 | 25 | 63 | 1.97 | 4.75 | 10.0 | 7.57 | 0.11 | 354 | 7.57 | 0.13 | 42 | 1.3 | 8.23 | 8.25 | 8.82 |
| RT Mus | 0.204 | 6224 | 48 | 61 | 2.53 | 3.48 | 7.7 | 7.62 | 0.10 | 366 | 7.61 | 0.09 | 49 | 1.6 | 8.32 | 8.42 | 8.94 |
| TZ Mus | 0.024 | 6344 | 62 | 48 | 2.28 | 4.93 | 14.0 | 7.60 | 0.11 | 218 | 7.60 | 0.14 | 39 | 1.4 | 8.26 | 8.36 | 8.84 |
| UU Mus | 0.041 | 6151 | 46 | 59 | 2.15 | 6.86 | 12.9 | 7.69 | 0.11 | 243 | 7.69 | 0.12 | 28 | 1.4 | 8.23 | 8.36 | 8.81 |
| S Nor | 0.124 | 5862 | 35 | 66 | 2.00 | 4.63 | 10.0 | 7.63 | 0.11 | 332 | 7.64 | 0.20 | 35 | 1.5 | 8.33 | 8.37 | 8.90 |
| U Nor | 0.476 | 5397 | 86 | 73 | 1.55 | 3.82 | 6.4 | 7.69 | 0.11 | 466 | 7.69 | 0.13 | 48 | 0.8 | 8.54 | 8.55 | 9.09 |
| SY Nor | 0.164 | 5597 | 111 | 72 | 1.86 | 4.85 | 10.2 | 7.84 | 0.08 | 231 | 7.84 | 0.10 | 27 | 1.4 | 8.58 | 8.57 | 9.13 |
| TW Nor | 0.094 | 5946 | 83 | 67 | 2.02 | 5.37 | 12.3 | 7.83 | 0.12 | 254 | 7.84 | 0.11 | 23 | 1.7 | 8.50 | 8.47 | 8.98 |
| GU Nor | 0.145 | 6016 | 33 | 60 | 2.41 | 3.80 | 7.9 | 7.77 | 0.12 | 438 | 7.77 | 0.13 | 48 | 1.6 | 8.51 | 8.50 | 9.03 |
| V340 Nor | 0.430 | 5704 | 74 | 63 | 1.97 | 5.37 | 13.9 | 7.66 | 0.15 | 258 | 7.66 | 0.11 | 26 | 1.2 | 8.37 | 8.33 | 8.96 |

Table 3
Stellar Parameters and Abundances

| Object | Phase | T (K) | σ | N | Log(g) (cm/s^2) | $V_t$ (km/s) | $V_M$ (km/s) | Fe I | σ | N | Fe II | Sigma | N | Li | C | N | O |
|---|---|---|---|---|---|---|---|---|---|---|---|---|---|---|---|---|---|
| BF Oph | 0.033 | 6236 | 45 | 57 | 2.37 | 4.12 | 9.6 | 7.64 | 0.11 | 333 | 7.64 | 0.11 | 45 | 1.3 | 8.32 | 8.43 | 8.90 |
| AP Pup | 0.052 | 6238 | 39 | 55 | 2.35 | 4.84 | 10.5 | 7.58 | 0.12 | 331 | 7.58 | 0.09 | 38 | 1.2 | 8.25 | 8.29 | 8.81 |
| AT Pup | 0.658 | 6447 | 62 | 55 | 2.26 | 4.59 | 10.0 | 7.55 | 0.12 | 305 | 7.55 | 0.13 | 51 | 1.2 | 8.23 | 8.39 | 8.79 |
| CE Pup | 0.804 | 5783 | 111 | 67 | 1.37 | 6.70 | 12.0 | 7.53 | 0.19 | 273 | 7.52 | 0.13 | 26 | 1.2 | 7.58 | 8.43 | 8.75 |
| MY Pup | 0.114 | 6301 | 48 | 55 | 2.65 | 5.34 | 6.5 | 7.54 | 0.10 | 385 | 7.54 | 0.12 | 52 | 1.5 | 8.34 | 8.42 | 8.97 |
| NT Pup | 0.219 | 5547 | 84 | 64 | 1.39 | 4.45 | 9.6 | 7.48 | 0.15 | 339 | 7.48 | 0.10 | 29 | 0.9 | 7.15 | 8.54 | 8.61 |
| RV Sco | 0.914 | 6168 | 53 | 57 | 2.26 | 4.59 | 10.8 | 7.61 | 0.11 | 322 | 7.61 | 0.08 | 39 | 1.3 | 8.20 | 8.40 | 8.78 |
| V482 Sco | 0.113 | 6108 | 38 | 60 | 2.26 | 4.13 | 9.5 | 7.70 | 0.11 | 343 | 7.70 | 0.10 | 40 | 1.5 | 8.37 | 8.45 | 8.92 |
| V636 Sco | 0.969 | 5340 | 34 | 67 | 1.71 | 4.46 | 10.9 | 7.60 | 0.11 | 298 | 7.60 | 0.09 | 23 | 1.2 | 8.35 | 8.32 | 8.92 |
| V950 Sco | 0.126 | 6306 | 34 | 53 | 2.63 | 4.43 | 7.2 | 7.71 | 0.10 | 352 | 7.71 | 0.10 | 50 | 1.7 | 8.43 | 8.52 | 9.00 |
| R TrA | 0.167 | 6111 | 58 | 63 | 2.47 | 4.01 | 12.6 | 7.69 | 0.11 | 251 | 7.69 | 0.14 | 38 | 1.2 | 8.19 | 8.24 | 8.92 |
| S TrA | 0.176 | 5961 | 54 | 64 | 2.23 | 5.40 | 10.7 | 7.71 | 0.11 | 312 | 7.71 | 0.12 | 31 | 1.5 | 8.31 | 8.35 | 8.91 |
| LR TrA | 0.086 | 5929 | 52 | 48 | 2.31 | 5.13 | 14.9 | 7.81 | 0.12 | 204 | 7.81 | 0.09 | 19 | 1.5 | 8.31 | 8.18 | 8.90 |
| T Vel | 0.827 | 5717 | 84 | 65 | 2.04 | 4.57 | 11.7 | 7.54 | 0.13 | 314 | 7.54 | 0.10 | 28 | 1.5 | 8.10 | 8.16 | 8.76 |
| V Vel | 0.043 | 6374 | 67 | 53 | 2.15 | 3.93 | 10.6 | 7.42 | 0.11 | 298 | 7.42 | 0.10 | 42 | 1.8 | 8.07 | 8.10 | 8.68 |
| RY Vel | 0.412 | 5446 | 84 | 72 | 1.37 | 4.95 | 5.2 | 7.59 | 0.10 | 380 | 7.59 | 0.07 | 28 | 1.3 | 8.40 | 8.39 | 8.98 |
| RZ Vel | 0.808 | 5249 | 46 | 51 | 1.02 | 4.72 | 11.5 | 7.54 | 0.13 | 207 | 7.54 | 0.11 | 12 | 1.3 | 8.12 | 8.17 | 8.78 |
| ST Vel | 0.978 | 6244 | 51 | 60 | 2.17 | 4.63 | 11.0 | 7.55 | 0.12 | 323 | 7.55 | 0.10 | 37 | 1.4 | 8.28 | 8.31 | 8.78 |
| SV Vel | 0.939 | 6061 | 37 | 61 | 1.84 | 4.56 | 9.6 | 7.62 | 0.11 | 327 | 7.62 | 0.08 | 30 | 1.0 | 8.27 | 8.40 | 8.77 |
| SW Vel | 0.607 | 6637 | 69 | 28 | 1.86 | 5.57 | 15.0 | 7.50 | 0.13 | 184 | 7.50 | 0.12 | 29 | 1.7 | 7.99 | 8.48 | 8.73 |
| SX Vel | 0.852 | 6277 | 41 | 56 | 2.26 | 5.05 | 8.9 | 7.56 | 0.13 | 353 | 7.56 | 0.10 | 44 | 1.5 | 8.30 | 8.36 | 8.90 |
| XX Vel | 0.011 | 6517 | 72 | 52 | 2.36 | 4.45 | 10.4 | 7.61 | 0.13 | 287 | 7.61 | 0.10 | 42 | 1.6 | 8.37 | 8.45 | 8.90 |
| AE Vel | 0.732 | 5447 | 46 | 62 | 1.98 | 5.90 | 9.9 | 7.64 | 0.13 | 322 | 7.64 | 0.11 | 25 | 1.2 | 8.23 | 8.38 | 8.93 |
| AH Vel | 0.383 | 6015 | 44 | 62 | 2.34 | 4.26 | 9.3 | 7.69 | 0.12 | 359 | 7.69 | 0.10 | 42 | 1.0 | 8.20 | 8.51 | 8.89 |
| BG Vel | 0.849 | 5749 | 81 | 68 | 1.98 | 6.22 | 10.3 | 7.53 | 0.12 | 330 | 7.53 | 0.11 | 33 | 1.1 | 8.08 | 8.12 | 8.69 |
| CS Vel | 0.678 | 5436 | 48 | 62 | 2.12 | 4.97 | 11.3 | 7.62 | 0.12 | 246 | 7.62 | 0.10 | 21 | 1.5 | 8.37 | 8.20 | 8.98 |
| CX Vel | 0.003 | 6219 | 66 | 56 | 2.21 | 4.50 | 11.6 | 7.66 | 0.14 | 305 | 7.66 | 0.10 | 35 | 1.8 | 8.31 | 8.45 | 8.84 |
| DK Vel | 0.009 | 6444 | 51 | 52 | 2.69 | 3.64 | 8.0 | 7.68 | 0.12 | 363 | 7.68 | 0.11 | 53 | 1.8 | 8.46 | 8.47 | 8.98 |
| DR Vel | 0.165 | 5467 | 53 | 67 | 1.47 | 3.50 | 8.8 | 7.68 | 0.12 | 362 | 7.68 | 0.08 | 33 | 1.1 | 8.33 | 8.19 | 8.91 |
| EX Vel | 0.954 | 5886 | 46 | 63 | 1.93 | 4.48 | 10.4 | 7.57 | 0.13 | 328 | 7.57 | 0.09 | 33 | 1.8 | 8.20 | 8.38 | 8.81 |
| FG Vel | 0.756 | 5418 | 73 | 62 | 1.83 | 6.27 | 11.6 | 7.52 | 0.13 | 273 | 7.52 | 0.07 | 17 | 1.4 | 8.15 | 8.10 | 8.85 |
| FN Vel | 0.221 | 5822 | 37 | 65 | 1.96 | 4.11 | 11.3 | 7.65 | 0.12 | 299 | 7.65 | 0.15 | 39 | 1.2 | 8.24 | 8.18 | 8.78 |

Notes:

All abundances are relative to $\log \varepsilon_H = 12.00$.

Phase: Phase of the observation. Phase 0.0 is maximum light.
T: Effective temperature determined using the line ratio method of Kovtyukh (2007). Sigma ($\sigma$) the standard deviation of determination. N is the number of ratios utilized.
log g: The base 10 logarithm of the surface gravity (cm s$^{-2}$)
$V_t$: Microturbulent velocity in km s$^{-1}$.
$V_M$: Macroturbulent velocity in km s$^{-1}$.
Fe I: The iron abundance determined from Fe I lines. The two columns that follow are the standard deviation of the abundance and N is the number of individual lines utilized.
Fe II: The iron abundance determined from Fe II lines. The two columns that follow are the standard deviation of the abundance and N is the number of individual lines utilized.
Li: Upper limit for the lithium abundance except for V1033 Cyg which is a determined value.
C: The carbon abundance from a combination of C I lines.
N: The nitrogen abundance from N I lines at 744 and 768 nm.
O: The oxygen abundance from a combination of O I lines at 616 nm and the [O I] line at 630 nm.

**Table 4**
Average [x/H] Ratios for Cepheids

| Cepheid | C | N | O | Na | Mg | Al | Si | Ca | Ti | Cr | Mn | Fe | Co | Ni | Y | La | Ce | Pr | Nd | Sm | Eu |
|---|---|---|---|---|---|---|---|---|---|---|---|---|---|---|---|---|---|---|---|---|---|
| U Aql | -0.11 | 0.31 | 0.25 | 0.32 | 0.09 | 0.27 | 0.16 | -0.01 | 0.09 | 0.05 | -0.13 | 0.17 | -0.02 | -0.02 | 0.24 | 0.36 | 0.21 | -0.04 | 0.26 | 0.15 | 0.16 |
| TT Aql | 0.03 | 0.37 | 0.35 | 0.37 | 0.27 | 0.24 | 0.24 | 0.02 | 0.10 | 0.14 | -0.05 | 0.22 | 0.01 | 0.02 | 0.28 | 0.33 | 0.32 | 0.03 | 0.25 | 0.30 | 0.17 |
| BC Aql | -0.32 | -0.02 | 0.16 | -0.26 | -0.06 | -0.48 | -0.10 | 0.01 | 0.22 | -0.21 | -0.48 | -0.28 | -0.01 | -0.20 | -0.11 | -0.19 | -0.35 | -0.75 | -0.60 | -0.29 | 0.10 |
| EV Aql | -0.11 | 0.36 | 0.31 | 0.16 | 0.03 | 0.10 | 0.05 | -0.15 | -0.02 | 0.00 | -0.29 | 0.06 | -0.12 | -0.18 | 0.19 | 0.34 | 0.18 | -0.01 | 0.19 | 0.22 | 0.31 |
| FM Aql | -0.13 | 0.28 | 0.29 | 0.28 | 0.28 | 0.27 | 0.19 | 0.04 | 0.21 | 0.13 | -0.06 | 0.24 | 0.01 | 0.01 | 0.31 | 0.65 | 0.29 | 0.08 | 0.39 | 0.25 | 0.22 |
| FN Aql | -1.37 | 0.43 | 0.08 | 0.19 | -0.10 | 0.01 | -0.03 | -0.13 | -0.01 | -0.12 | -0.30 | -0.06 | -0.08 | -0.19 | 0.05 | 0.13 | 0.02 | -0.20 | 0.07 | 0.04 | 0.05 |
| KL Aql | 0.12 | 0.66 | 0.42 | 0.46 | 0.28 | 0.32 | 0.28 | 0.13 | 0.24 | 0.18 | 0.01 | 0.33 | 0.13 | 0.11 | 0.38 | 0.39 | 0.35 | 0.07 | 0.36 | 0.28 | 0.34 |
| V336 Aql | -0.11 | 0.39 | 0.23 | 0.26 | 0.11 | 0.25 | 0.18 | 0.04 | 0.19 | 0.07 | -0.11 | 0.18 | 0.06 | 0.01 | 0.27 | 0.30 | 0.15 | 0.00 | 0.24 | 0.14 | 0.21 |
| V493 Aql | -0.32 | 0.16 | 0.01 | 0.32 | 0.14 | 0.01 | 0.07 | 0.01 | 0.06 | -0.02 | -0.15 | 0.03 | 0.04 | -0.06 | 0.02 | 0.25 | -0.02 | -0.72 | 0.05 | 0.26 | -0.01 |
| V526 Aql | 0.90 | 1.11 | 0.49 | 0.63 | 0.22 | 0.58 | 0.38 | 0.16 | 0.42 | 0.27 | 0.23 | 0.50 | 0.40 | 0.22 | 0.43 | 0.56 | 0.36 | 0.21 | 0.46 | 0.56 | 0.24 |
| V916 Aql | 0.02 | 0.60 | 0.31 | 0.55 | 0.41 | 0.63 | 0.39 | 0.22 | 0.24 | 0.31 | 0.19 | 0.39 | 0.20 | 0.22 | 0.34 | 0.32 | 0.36 | 0.07 | 0.27 | 0.40 | 0.19 |
| V1344 Aql | -0.18 | 0.39 | 0.26 | 0.21 | 0.03 | 0.21 | 0.12 | -0.04 | 0.08 | 0.03 | -0.14 | 0.15 | 0.00 | -0.08 | 0.16 | 0.28 | 0.13 | -0.09 | 0.17 | 0.09 | 0.03 |
| V1359 Aql | 0.32 | 0.55 | 0.28 | -0.19 | -0.03 | 0.01 | 0.32 | -0.58 | -0.42 | -0.36 | -0.35 | 0.28 | -0.03 | -0.14 | 0.38 | 0.24 | 0.25 | 0.17 | 0.28 | 0.32 | 0.12 |
| Y Aur | -0.29 | 0.28 | -0.03 | 0.23 | 0.11 | 0.14 | 0.08 | 0.00 | 0.08 | -0.04 | -0.23 | 0.03 | -0.11 | -0.09 | 0.14 | 0.39 | 0.23 | -0.12 | 0.25 | 0.09 | 0.09 |
| RT Aur | -0.04 | 0.56 | 0.30 | 0.37 | 0.17 | 0.14 | 0.19 | 0.09 | 0.21 | 0.06 | -0.12 | 0.13 | 0.17 | 0.02 | 0.26 | 0.29 | 0.16 | -0.08 | 0.17 | 0.12 | 0.19 |
| RX Aur | -0.20 | 0.37 | 0.14 | 0.27 | 0.22 | 0.10 | 0.15 | 0.06 | 0.23 | 0.07 | -0.03 | 0.10 | 0.14 | -0.02 | 0.21 | 0.49 | 0.17 | -0.26 | 0.33 | 0.03 | 0.24 |
| SY Aur | -0.29 | 0.43 | 0.18 | 0.19 | -0.04 | 0.05 | 0.03 | -0.10 | 0.07 | -0.06 | -0.28 | 0.04 | -0.07 | -0.15 | 0.18 | 0.29 | 0.19 | 0.00 | 0.27 | 0.16 | 0.11 |
| YZ Aur | -0.33 | ... | 0.05 | -0.01 | -0.02 | -0.14 | -0.17 | -0.32 | -0.30 | -0.26 | -0.40 | -0.30 | -0.47 | -0.38 | -0.35 | -0.03 | -0.26 | -0.54 | -0.17 | -0.26 | -0.32 |
| AN Aur | -0.39 | 0.06 | 0.03 | 0.07 | -0.11 | -0.08 | -0.09 | -0.19 | -0.02 | -0.16 | -0.37 | -0.10 | -0.11 | -0.22 | 0.09 | 0.27 | 0.18 | 0.01 | 0.20 | 0.15 | -0.04 |
| AO Aur | -0.51 | -0.03 | -0.15 | -0.07 | -0.27 | -0.18 | -0.20 | -0.30 | -0.19 | -0.31 | -0.50 | -0.27 | -0.32 | -0.37 | -0.18 | -0.03 | -0.12 | -0.38 | -0.06 | -0.10 | -0.01 |
| AS Aur | -0.54 | 0.16 | 0.04 | -0.06 | -0.17 | -0.04 | -0.11 | -0.17 | -0.09 | -0.19 | -0.41 | -0.20 | -0.25 | -0.25 | -0.16 | 0.02 | -0.05 | -0.35 | 0.00 | -0.04 | -0.06 |
| AX Aur | -0.45 | 0.29 | -0.01 | 0.14 | -0.02 | 0.00 | 0.01 | -0.09 | -0.02 | -0.14 | -0.34 | -0.04 | -0.03 | -0.17 | 0.05 | 0.21 | 0.11 | -0.22 | 0.06 | 0.07 | 0.03 |
| BK Aur | -0.38 | 0.41 | 0.28 | 0.51 | -0.10 | 0.15 | -0.07 | -0.19 | 0.06 | -0.16 | -0.13 | -0.04 | -0.20 | -0.20 | 0.06 | 0.82 | 0.27 | -0.01 | 0.30 | 0.58 | 0.47 |
| CO Aur | -0.22 | 0.41 | 0.12 | 0.19 | -0.12 | 0.00 | 0.07 | -0.01 | 0.11 | -0.01 | -0.21 | -0.01 | -0.01 | -0.09 | 0.08 | 0.10 | 0.13 | -0.12 | 0.14 | 0.08 | 0.13 |
| CY Aur | -0.49 | 0.05 | 0.00 | 0.06 | -0.16 | -0.04 | -0.10 | -0.20 | -0.03 | -0.18 | -0.38 | -0.12 | -0.19 | -0.21 | 0.02 | 0.36 | 0.13 | -0.09 | 0.15 | 0.18 | 0.19 |
| ER Aur | -0.49 | 0.07 | -0.10 | -0.01 | -0.31 | -0.17 | -0.18 | -0.29 | -0.05 | -0.34 | -0.48 | -0.27 | -0.08 | -0.33 | -0.10 | 0.08 | -0.03 | -0.16 | 0.01 | 0.06 | -0.18 |
| EW Aur | -0.69 | -0.06 | -0.32 | -0.27 | -0.40 | -0.20 | -0.31 | -0.44 | -0.24 | -0.54 | -0.59 | -0.54 | -0.02 | -0.39 | -0.55 | -0.13 | -0.54 | -0.58 | -0.38 | -0.78 | -0.24 |
| FF Aur | -0.77 | ... | -0.37 | -0.52 | -0.37 | -0.64 | -0.33 | -0.60 | -0.06 | -0.28 | -0.66 | -0.51 | 0.01 | -0.41 | -0.47 | 0.00 | -0.11 | ... | -0.33 | -1.07 | ... |
| GT Aur | -0.31 | 0.23 | -0.14 | 0.01 | -0.04 | -0.02 | 0.05 | 0.00 | 0.15 | 0.01 | -0.20 | -0.02 | 0.15 | -0.06 | 0.03 | 0.14 | 0.12 | -0.24 | 0.18 | 0.07 | -0.02 |
| GV Aur | -0.53 | 0.15 | -0.22 | 0.04 | -0.09 | -0.03 | -0.05 | -0.09 | -0.10 | -0.20 | -0.31 | -0.21 | -0.14 | -0.21 | -0.21 | 0.08 | -0.04 | -0.44 | -0.05 | -0.10 | -0.07 |
| IN Aur | -0.65 | -0.06 | -0.25 | -0.02 | -0.18 | -0.13 | -0.18 | -0.11 | -0.14 | -0.12 | -0.51 | -0.28 | -0.18 | -0.29 | -0.26 | -0.07 | -0.05 | -0.26 | -0.09 | -0.12 | 0.06 |
| V335 Aur | -0.59 | 0.07 | -0.07 | 0.05 | -0.16 | -0.12 | -0.08 | -0.22 | -0.05 | -0.26 | -0.28 | -0.30 | 0.10 | -0.22 | -0.21 | ... | -0.20 | -0.51 | -0.04 | -0.11 | ... |
| RW Cam | -0.36 | ... | 0.07 | 0.07 | 0.13 | 0.33 | 0.13 | -0.28 | -0.20 | -0.17 | -0.49 | 0.11 | -0.20 | -0.19 | -0.06 | 0.38 | -0.14 | -0.16 | 0.05 | -0.08 | 0.15 |
| RX Cam | -0.31 | 0.26 | 0.05 | 0.22 | 0.20 | 0.16 | 0.11 | 0.03 | 0.14 | -0.07 | -0.14 | 0.11 | -0.07 | -0.08 | 0.16 | 0.34 | 0.17 | -0.24 | 0.22 | 0.09 | 0.12 |
| TV Cam | -0.35 | 0.28 | 0.08 | 0.19 | -0.07 | -0.03 | 0.02 | -0.05 | 0.08 | -0.04 | -0.23 | 0.04 | -0.01 | -0.14 | 0.14 | 0.29 | 0.24 | 0.00 | 0.29 | 0.24 | -0.08 |
| AB Cam | -0.39 | 0.08 | -0.11 | 0.09 | -0.11 | -0.05 | -0.05 | -0.15 | -0.03 | -0.12 | -0.31 | -0.08 | -0.17 | -0.25 | 0.05 | 0.22 | 0.14 | -0.09 | 0.21 | 0.17 | -0.03 |

**Table 4**
Average [x/H] Ratios for Cepheids

| Cepheid | C | N | O | Na | Mg | Al | Si | Ca | Ti | Cr | Mn | Fe | Co | Ni | Y | La | Ce | Pr | Nd | Sm | Eu |
|---|---|---|---|---|---|---|---|---|---|---|---|---|---|---|---|---|---|---|---|---|---|
| AC Cam | -0.51 | 0.26 | -0.19 | 0.18 | -0.15 | -0.01 | -0.02 | -0.03 | 0.06 | -0.09 | -0.32 | -0.13 | -0.01 | -0.19 | 0.04 | -0.04 | -0.08 | -0.35 | 0.03 | -0.27 | 0.71 |
| AD Cam | -0.99 | -0.02 | -0.11 | 0.07 | -0.06 | -0.07 | -0.15 | -0.20 | -0.16 | -0.24 | -0.44 | -0.25 | -0.32 | -0.34 | -0.25 | 0.08 | 0.05 | -0.40 | -0.07 | -0.25 | -0.16 |
| AM Cam | -0.58 | 0.08 | -0.13 | 0.08 | -0.01 | -0.10 | -0.02 | -0.01 | 0.03 | -0.10 | -0.31 | -0.13 | -0.15 | -0.16 | -0.01 | 0.21 | 0.15 | -0.43 | 0.01 | 0.29 | 0.08 |
| CK Cam | -0.21 | 0.32 | 0.13 | 0.19 | 0.15 | 0.10 | 0.12 | 0.07 | 0.13 | 0.02 | -0.17 | 0.07 | -0.03 | -0.05 | 0.12 | 0.19 | 0.11 | -0.14 | 0.13 | 0.07 | 0.12 |
| LO Cam | -0.36 | 0.16 | 0.09 | 0.12 | -0.06 | -0.14 | -0.07 | -0.19 | 0.05 | -0.15 | -0.29 | -0.05 | -0.09 | -0.23 | 0.16 | 0.30 | 0.20 | -0.18 | 0.22 | -0.10 | 0.30 |
| MN Cam | -0.21 | 0.35 | 0.15 | 0.18 | 0.01 | 0.04 | 0.02 | -0.08 | 0.00 | -0.10 | -0.25 | 0.01 | -0.10 | -0.17 | 0.10 | 0.24 | 0.13 | -0.04 | 0.23 | 0.10 | 0.06 |
| MQ Cam | -0.53 | 0.18 | -0.39 | 0.18 | 0.00 | -0.01 | -0.09 | -0.18 | -0.03 | -0.14 | -0.30 | -0.11 | -0.13 | -0.26 | -0.11 | 0.19 | 0.01 | -0.16 | 0.10 | -0.02 | -0.12 |
| SS CMa | -0.19 | 0.57 | 0.19 | 0.26 | -0.04 | 0.07 | 0.14 | -0.02 | 0.28 | 0.10 | -0.14 | 0.07 | 0.12 | 0.03 | 0.22 | 0.22 | 0.29 | 0.21 | 0.26 | 0.15 | 0.14 |
| TV CMa | -0.23 | 0.52 | 0.12 | 0.25 | 0.06 | 0.26 | 0.18 | 0.09 | 0.14 | 0.09 | -0.10 | 0.14 | 0.06 | 0.02 | 0.29 | 0.29 | 0.26 | -0.10 | 0.27 | 0.14 | 0.07 |
| RS Cas | -0.10 | 0.50 | 0.22 | 0.36 | 0.11 | 0.10 | 0.19 | 0.13 | 0.22 | 0.16 | -0.12 | 0.18 | 0.04 | 0.06 | 0.26 | 0.24 | 0.36 | 0.07 | 0.34 | 0.25 | 0.26 |
| SZ Cas | -0.28 | 0.18 | 0.08 | 0.34 | 0.20 | 0.08 | 0.14 | 0.07 | 0.22 | 0.00 | -0.06 | 0.07 | 0.04 | -0.03 | 0.15 | 0.46 | 0.30 | -0.14 | 0.31 | 0.20 | 0.28 |
| UZ Cas | -0.52 | 0.27 | -0.01 | 0.22 | -0.23 | 0.01 | -0.02 | -0.03 | 0.01 | -0.04 | -0.32 | -0.05 | -0.14 | -0.19 | 0.06 | 0.15 | 0.15 | -0.11 | 0.19 | 0.14 | 0.03 |
| VV Cas | -0.35 | 0.17 | 0.16 | 0.15 | -0.15 | 0.00 | 0.04 | -0.09 | 0.03 | -0.06 | -0.30 | -0.04 | -0.13 | -0.16 | 0.06 | 0.14 | 0.24 | -0.05 | 0.21 | 0.11 | 0.21 |
| VW Cas | -0.25 | 0.20 | 0.33 | 0.20 | 0.24 | 0.21 | 0.14 | 0.10 | 0.22 | 0.14 | -0.12 | 0.19 | 0.00 | 0.04 | 0.30 | 0.34 | 0.41 | 0.15 | 0.38 | 0.09 | 0.28 |
| XY Cas | -0.18 | 0.40 | 0.12 | 0.37 | 0.16 | 0.12 | 0.14 | 0.10 | 0.19 | 0.07 | -0.09 | 0.10 | 0.27 | -0.02 | 0.18 | 0.35 | 0.07 | -0.08 | 0.13 | -0.07 | 0.29 |
| AP Cas | -0.33 | 0.20 | 0.17 | 0.24 | -0.02 | 0.11 | 0.05 | -0.06 | -0.04 | -0.04 | -0.32 | 0.05 | -0.18 | -0.14 | 0.03 | 0.18 | 0.10 | -0.25 | 0.12 | 0.10 | 0.04 |
| AS Cas | -0.26 | -0.17 | -0.07 | 0.12 | 0.38 | -0.10 | 0.07 | 0.11 | 0.17 | -0.09 | -0.07 | 0.02 | 0.05 | -0.09 | -0.12 | 0.35 | 0.07 | -0.41 | 0.07 | ... | ... |
| AW Cas | -0.23 | 0.25 | 0.15 | 0.06 | 0.05 | 0.02 | 0.03 | -0.04 | 0.10 | -0.03 | -0.18 | 0.03 | -0.04 | -0.11 | 0.11 | 0.23 | 0.13 | -0.15 | 0.15 | 0.03 | 0.08 |
| AY Cas | -0.36 | 0.17 | -0.07 | 0.13 | 0.09 | 0.07 | 0.08 | 0.03 | 0.05 | -0.04 | -0.22 | 0.02 | -0.18 | -0.11 | 0.11 | 0.23 | 0.13 | -0.15 | 0.20 | 0.06 | 0.10 |
| BF Cas | -0.33 | 0.34 | -0.08 | 0.11 | -0.13 | -0.04 | 0.10 | -0.02 | 0.20 | -0.06 | -0.25 | -0.05 | 0.09 | -0.12 | 0.10 | 0.14 | 0.01 | 0.12 | 0.05 | 0.01 | ... |
| BP Cas | -0.18 | 0.38 | 0.20 | 0.24 | 0.03 | 0.06 | 0.09 | 0.02 | 0.12 | 0.00 | -0.18 | 0.09 | -0.04 | -0.07 | 0.18 | 0.26 | 0.27 | -0.02 | 0.27 | 0.19 | 0.19 |
| BV Cas | -0.05 | 0.12 | 0.18 | 0.09 | 0.05 | 0.12 | 0.06 | -0.02 | 0.04 | -0.01 | -0.21 | 0.02 | -0.09 | -0.11 | 0.14 | 0.22 | 0.09 | -0.10 | 0.16 | 0.17 | 0.09 |
| BY Cas | -0.19 | 0.39 | 0.31 | 0.26 | 0.18 | 0.11 | 0.21 | 0.09 | 0.24 | 0.13 | -0.08 | 0.12 | 0.05 | 0.03 | 0.25 | 0.22 | 0.29 | 0.01 | 0.23 | -0.02 | 0.21 |
| CD Cas | -0.15 | 0.43 | 0.25 | 0.29 | ... | 0.11 | 0.21 | 0.06 | 0.26 | 0.13 | -0.04 | 0.13 | 0.14 | 0.03 | 0.21 | 0.17 | 0.14 | 0.08 | 0.24 | 0.38 | 0.45 |
| CE Cas A | -0.22 | 0.33 | 0.02 | 0.14 | 0.41 | 0.18 | 0.14 | 0.06 | 0.20 | 0.02 | -0.12 | 0.05 | 0.15 | 0.00 | 0.10 | 0.49 | 0.17 | -0.40 | -0.04 | 0.00 | ... |
| CE Cas B | -0.17 | 0.27 | 0.12 | 0.00 | 0.30 | 0.20 | 0.11 | 0.06 | 0.21 | 0.02 | -0.15 | 0.11 | 0.13 | -0.03 | 0.20 | 0.59 | 0.23 | -0.28 | 0.23 | 0.13 | 0.25 |
| CF Cas | -0.23 | 0.25 | 0.13 | 0.20 | 0.10 | 0.16 | 0.06 | 0.01 | 0.06 | 0.01 | -0.23 | 0.02 | -0.07 | -0.10 | 0.12 | 0.20 | 0.18 | -0.10 | 0.16 | 0.08 | 0.10 |
| CG Cas | -0.13 | 0.49 | 0.15 | 0.28 | 0.21 | 0.03 | 0.13 | 0.04 | 0.16 | 0.01 | -0.18 | 0.09 | 0.04 | -0.04 | 0.18 | 0.33 | 0.14 | -0.20 | 0.16 | 0.01 | 0.03 |
| CT Cas | -0.38 | 0.20 | 0.03 | 0.26 | -0.08 | 0.04 | 0.02 | -0.01 | 0.06 | -0.05 | -0.27 | -0.05 | -0.08 | -0.14 | -0.02 | 0.31 | 0.02 | -0.37 | 0.07 | -0.18 | 0.03 |
| CZ Cas | -0.26 | 0.24 | 0.02 | 0.16 | 0.09 | 0.09 | 0.09 | 0.00 | 0.09 | 0.00 | -0.18 | 0.07 | 0.00 | -0.08 | 0.14 | 0.24 | 0.08 | -0.14 | 0.15 | 0.08 | 0.18 |
| DW Cas | -0.32 | 0.41 | 0.08 | 0.25 | 0.23 | 0.15 | 0.12 | 0.04 | 0.14 | 0.03 | -0.25 | 0.06 | 0.02 | -0.05 | 0.15 | 0.31 | 0.14 | -0.08 | 0.15 | 0.02 | 0.09 |
| EX Cas | -0.50 | 0.15 | 0.01 | 0.28 | 0.01 | 0.13 | 0.07 | 0.15 | 0.13 | 0.01 | -0.16 | -0.07 | -0.09 | -0.12 | -0.03 | 0.48 | 0.19 | -0.38 | 0.12 | -0.02 | -0.29 |
| FO Cas | -1.21 | 0.22 | -0.35 | 0.07 | -0.34 | -0.12 | -0.27 | -0.46 | -0.30 | -0.52 | -0.60 | -0.56 | -0.36 | -0.50 | -0.52 | -0.30 | -0.36 | -0.51 | -0.35 | -0.69 | ... |
| FW Cas | -0.61 | 0.05 | -0.24 | -0.03 | -0.16 | 0.02 | -0.07 | -0.05 | 0.04 | -0.12 | -0.28 | -0.09 | -0.12 | -0.18 | -0.10 | 0.22 | 0.11 | -0.36 | 0.06 | 0.16 | -0.10 |
| GL Cas | -0.37 | -0.15 | -0.08 | 0.32 | 0.37 | 0.19 | 0.07 | 0.15 | 0.08 | 0.01 | -0.14 | 0.03 | 0.06 | -0.07 | 0.00 | 0.29 | 0.14 | -0.23 | 0.02 | -0.05 | 0.11 |
| GM Cas | -0.29 | 0.12 | -0.10 | 0.15 | -0.18 | -0.18 | -0.10 | -0.10 | 0.01 | -0.13 | -0.13 | -0.10 | -0.10 | -0.21 | -0.03 | 0.26 | 0.04 | -0.33 | 0.05 | 0.24 | 0.26 |
| GO Cas | -0.12 | 0.50 | 0.22 | 0.33 | 0.01 | 0.09 | 0.19 | 0.13 | 0.29 | 0.13 | -0.02 | 0.12 | 0.19 | 0.04 | 0.17 | 0.37 | 0.12 | -0.06 | 0.16 | 0.07 | 0.01 |

**Table 4**
Average [x/H] Ratios for Cepheids

| Cepheid | C | N | O | Na | Mg | Al | Si | Ca | Ti | Cr | Mn | Fe | Co | Ni | Y | La | Ce | Pr | Nd | Sm | Eu |
|---|---|---|---|---|---|---|---|---|---|---|---|---|---|---|---|---|---|---|---|---|---|
| HK Cas | 1.24 | 1.05 | 0.56 | 0.81 | 0.59 | 0.72 | 0.52 | 0.33 | 0.52 | 0.37 | 0.29 | 0.45 | 0.56 | 0.34 | 0.37 | 0.35 | 0.31 | 0.23 | 0.39 | 0.32 | ... |
| IO Cas | -1.14 | 0.03 | -0.49 | 0.09 | -0.12 | -0.05 | -0.17 | -0.24 | -0.16 | -0.40 | -0.57 | -0.49 | -0.31 | -0.37 | -0.48 | 0.11 | -0.53 | -0.59 | -0.40 | -0.41 | -0.30 |
| KK Cas | -0.15 | 0.42 | 0.20 | 0.34 | 0.04 | 0.17 | 0.15 | 0.14 | 0.19 | 0.12 | -0.06 | 0.13 | 0.02 | 0.01 | 0.21 | 0.33 | 0.24 | -0.04 | 0.33 | 0.15 | -0.10 |
| LT Cas | -0.71 | -0.03 | -0.32 | -0.09 | -0.29 | -0.23 | -0.28 | -0.35 | -0.25 | -0.39 | -0.63 | -0.36 | -0.36 | -0.48 | -0.42 | -0.25 | -0.25 | -0.54 | -0.19 | -0.20 | -0.17 |
| NP Cas | -0.30 | 0.20 | -0.02 | 0.07 | 0.01 | 0.01 | 0.07 | 0.04 | 0.12 | -0.08 | -0.14 | 0.01 | 0.16 | -0.03 | 0.05 | 0.39 | 0.05 | -0.24 | 0.17 | 0.07 | 0.28 |
| NY Cas | -0.94 | 0.01 | -0.46 | -0.12 | -0.40 | 0.09 | -0.21 | -0.40 | -0.22 | -0.44 | -0.42 | -0.46 | 0.07 | -0.35 | -0.59 | 0.01 | -0.55 | -0.54 | -0.37 | 0.05 | -0.49 |
| OP Cas | -0.12 | 0.46 | 0.07 | 0.49 | 0.14 | 0.27 | 0.19 | 0.15 | 0.15 | 0.09 | -0.13 | 0.14 | 0.02 | 0.03 | 0.15 | 0.14 | 0.10 | -0.09 | 0.19 | 0.11 | 0.03 |
| OZ Cas | -0.36 | 0.26 | 0.08 | 0.07 | -0.07 | 0.03 | 0.13 | 0.07 | 0.14 | 0.00 | 0.27 | 0.06 | -0.03 | -0.03 | -0.03 | 0.50 | 0.21 | -0.19 | 0.09 | 0.13 | -0.08 |
| PW Cas | -0.36 | 0.12 | 0.11 | 0.12 | -0.04 | -0.06 | -0.03 | -0.14 | 0.01 | -0.16 | -0.28 | -0.06 | -0.20 | -0.24 | 0.11 | 0.28 | 0.22 | -0.02 | 0.20 | 0.19 | 0.01 |
| V342 Cas | -0.19 | 0.22 | 0.05 | 0.17 | -0.08 | 0.09 | 0.08 | 0.04 | 0.10 | 0.00 | -0.18 | 0.03 | -0.07 | -0.10 | 0.11 | 0.16 | 0.13 | -0.06 | 0.19 | -0.02 | 0.03 |
| V395 Cas | -0.35 | 0.11 | -0.02 | 0.28 | 0.07 | 0.16 | 0.11 | 0.01 | 0.10 | 0.03 | -0.11 | 0.02 | -0.09 | -0.03 | 0.07 | 0.30 | 0.09 | -0.35 | 0.11 | -0.26 | 0.26 |
| V407 Cas | -0.13 | 0.28 | 0.09 | 0.36 | -0.01 | 0.20 | 0.11 | 0.06 | 0.11 | 0.08 | -0.11 | 0.11 | -0.01 | -0.03 | 0.08 | 0.19 | 0.05 | -0.09 | 0.17 | 0.06 | 0.25 |
| V556 Cas | -0.40 | 0.40 | 0.09 | 0.24 | 0.05 | 0.06 | 0.02 | -0.05 | 0.04 | -0.09 | -0.24 | -0.01 | -0.13 | -0.16 | 0.10 | 0.25 | 0.19 | -0.11 | 0.19 | 0.06 | -0.12 |
| V1017 Cas | -0.42 | 0.11 | 0.03 | 0.05 | -0.21 | -0.14 | -0.16 | -0.25 | -0.17 | -0.25 | -0.49 | -0.18 | -0.31 | -0.32 | -0.15 | 0.06 | -0.02 | -0.31 | 0.00 | 0.06 | 0.02 |
| V1019 Cas | -0.23 | 0.30 | 0.17 | 0.22 | 0.12 | 0.08 | 0.11 | 0.08 | 0.18 | -0.01 | -0.17 | 0.07 | 0.01 | -0.03 | 0.18 | 0.51 | 0.18 | -0.10 | 0.22 | 0.03 | 0.36 |
| V1020 Cas | -0.06 | 0.45 | 0.15 | 0.29 | 0.20 | 0.25 | 0.17 | 0.09 | 0.32 | 0.13 | 0.01 | 0.15 | 0.17 | 0.06 | 0.30 | 0.53 | 0.37 | -0.09 | 0.23 | 0.16 | 0.19 |
| Delta Cep | -0.17 | 0.36 | 0.23 | 0.28 | 0.10 | 0.19 | 0.14 | 0.01 | 0.11 | 0.02 | -0.13 | 0.12 | -0.04 | -0.04 | 0.22 | 0.43 | 0.16 | -0.18 | 0.24 | 0.08 | 0.14 |
| AK Cep | -0.14 | 0.04 | 0.07 | 0.08 | 0.21 | 0.08 | 0.12 | 0.03 | 0.20 | 0.02 | -0.17 | 0.05 | 0.07 | -0.06 | 0.16 | 0.31 | 0.13 | -0.01 | 0.17 | 0.23 | -0.19 |
| CN Cep | -0.17 | 0.29 | 0.13 | 0.29 | 0.12 | -0.15 | 0.06 | -0.03 | 0.03 | -0.01 | -0.25 | 0.06 | -0.11 | -0.09 | 0.12 | 0.19 | 0.05 | -0.02 | 0.23 | 0.32 | 0.24 |
| DR Cep | -0.17 | ... | 0.12 | 0.02 | -0.36 | -0.06 | -0.10 | -0.47 | -0.30 | -0.25 | -0.57 | -0.14 | -0.38 | -0.41 | -0.11 | 0.43 | -0.31 | -0.24 | 0.07 | 0.19 | 0.07 |
| IY Cep | -0.33 | 0.07 | -0.15 | 0.19 | 0.01 | 0.14 | 0.07 | -0.02 | -0.02 | -0.01 | -0.20 | 0.11 | -0.15 | -0.16 | -0.05 | 0.29 | -0.08 | -0.22 | 0.02 | 0.00 | 0.35 |
| MU Cep | -0.20 | 0.20 | 0.00 | 0.20 | 0.46 | -0.03 | 0.11 | -0.01 | 0.27 | 0.02 | -0.17 | 0.18 | 0.00 | -0.05 | 0.23 | 0.70 | 0.27 | 0.06 | 0.27 | 0.48 | 0.43 |
| CD Cyg | -0.11 | 0.45 | 0.14 | 0.34 | 0.18 | 0.34 | 0.27 | 0.22 | 0.31 | 0.20 | 0.09 | 0.15 | 0.23 | 0.15 | 0.23 | 0.40 | 0.17 | -0.07 | 0.23 | 0.12 | 0.39 |
| EP Cyg | -0.38 | 0.17 | -0.04 | 0.12 | -0.07 | 0.01 | -0.01 | -0.01 | 0.01 | -0.08 | -0.25 | -0.06 | -0.13 | -0.13 | -0.09 | 0.08 | -0.08 | -0.43 | 0.07 | -0.16 | -0.09 |
| EU Cyg | -0.44 | 0.01 | -0.11 | 0.01 | -0.10 | 0.07 | -0.04 | -0.17 | -0.09 | -0.12 | -0.36 | -0.20 | -0.28 | -0.28 | -0.22 | 0.12 | 0.02 | -0.30 | 0.03 | -0.01 | -0.03 |
| EX Cyg | -0.20 | 0.45 | 0.29 | 0.44 | -0.13 | 0.21 | 0.20 | 0.01 | 0.17 | 0.02 | -0.10 | 0.25 | 0.08 | 0.00 | 0.28 | 0.35 | 0.23 | 0.13 | 0.38 | 0.21 | 0.48 |
| EZ Cyg | 0.01 | 0.17 | -0.09 | 0.60 | 0.45 | 0.44 | 0.36 | 0.25 | 0.25 | 0.32 | 0.18 | 0.28 | 0.14 | 0.21 | 0.35 | 0.38 | 0.55 | 0.04 | 0.30 | 0.45 | 0.25 |
| GH Cyg | -0.16 | 0.41 | 0.18 | 0.24 | -0.02 | 0.14 | 0.17 | 0.01 | 0.15 | 0.14 | -0.10 | 0.21 | -0.02 | 0.01 | 0.32 | 0.32 | 0.20 | -0.02 | 0.27 | 0.02 | 0.36 |
| GI Cyg | -0.22 | 0.35 | -0.44 | 0.34 | 0.49 | 0.32 | 0.22 | 0.21 | 0.27 | 0.17 | 0.12 | 0.27 | 0.12 | 0.14 | 0.23 | 0.46 | 0.38 | 0.01 | 0.34 | 0.23 | 0.26 |
| GL Cyg | -0.30 | -0.02 | 0.23 | 0.16 | -0.14 | 0.07 | 0.04 | 0.00 | 0.17 | -0.04 | -0.18 | 0.05 | 0.06 | -0.07 | 0.10 | 0.22 | 0.16 | -0.16 | 0.16 | -0.13 | 0.20 |
| IY Cyg | -0.35 | -0.17 | 0.11 | 0.08 | -0.11 | 0.03 | -0.10 | -0.23 | 0.05 | -0.13 | -0.48 | -0.09 | -0.47 | -0.29 | 0.05 | 0.33 | 0.07 | -0.24 | 0.18 | 0.25 | -0.28 |
| KX Cyg | -0.09 | 0.16 | 0.39 | 0.05 | 0.09 | 0.18 | 0.15 | -0.05 | 0.17 | 0.02 | -0.08 | 0.21 | 0.02 | -0.03 | 0.29 | 0.51 | 0.20 | 0.05 | 0.33 | 0.27 | 0.23 |
| V347 Cyg | -0.14 | 0.36 | 0.17 | 0.36 | 0.30 | 0.14 | 0.22 | 0.12 | 0.28 | 0.14 | 0.00 | 0.25 | 0.06 | 0.05 | 0.33 | 0.36 | 0.29 | -0.03 | 0.36 | 0.38 | 0.16 |
| V356 Cyg | -0.29 | 0.39 | -0.01 | 0.33 | 0.02 | 0.22 | 0.19 | 0.09 | 0.12 | 0.05 | -0.01 | 0.17 | -0.03 | -0.01 | 0.24 | 0.34 | 0.28 | -0.10 | 0.20 | 0.24 | 0.29 |
| V396 Cyg | -0.33 | 0.13 | -0.22 | 0.30 | 0.19 | 0.32 | 0.17 | 0.29 | 0.10 | 0.10 | -0.07 | 0.11 | 0.02 | 0.01 | 0.00 | 0.19 | 0.01 | -0.19 | 0.05 | 0.19 | 0.00 |
| V438 Cyg | -0.05 | 0.59 | 0.26 | 0.38 | 0.26 | 0.28 | 0.24 | 0.10 | 0.31 | 0.16 | -0.01 | 0.33 | 0.14 | 0.09 | 0.40 | 0.44 | 0.28 | 0.12 | 0.35 | 0.17 | 0.24 |
| V459 Cyg | -0.32 | 0.33 | 0.24 | 0.11 | -0.27 | 0.17 | 0.14 | -0.03 | 0.01 | 0.07 | -0.20 | 0.09 | -0.03 | -0.07 | 0.05 | 0.19 | -0.01 | -0.23 | 0.12 | 0.14 | -0.03 |

**Table 4**
Average [x/H] Ratios for Cepheids

| Cepheid | C | N | O | Na | Mg | Al | Si | Ca | Ti | Cr | Mn | Fe | Co | Ni | Y | La | Ce | Pr | Nd | Sm | Eu |
|---|---|---|---|---|---|---|---|---|---|---|---|---|---|---|---|---|---|---|---|---|---|
| V492 Cyg | -0.12 | 0.37 | 0.03 | 0.32 | 0.18 | 0.23 | 0.17 | 0.05 | 0.18 | 0.10 | -0.08 | 0.24 | 0.02 | 0.03 | 0.24 | 0.38 | 0.25 | -0.01 | 0.26 | 0.11 | 0.19 |
| V495 Cyg | -0.17 | 0.65 | 0.00 | 0.44 | ... | 0.20 | 0.26 | 0.11 | 0.27 | 0.17 | 0.06 | 0.24 | 0.12 | 0.14 | 0.23 | 0.14 | 0.22 | 0.09 | 0.26 | 0.31 | 0.24 |
| V514 Cyg | -0.14 | 0.52 | 0.20 | 0.32 | 0.18 | 0.13 | 0.17 | 0.07 | 0.25 | 0.08 | -0.08 | 0.18 | 0.19 | 0.03 | 0.25 | 0.35 | 0.22 | -0.03 | 0.29 | 0.05 | 0.35 |
| V520 Cyg | -0.22 | 0.30 | -0.01 | 0.26 | -0.05 | 0.24 | 0.17 | 0.10 | 0.19 | 0.11 | -0.11 | 0.08 | 0.04 | 0.04 | 0.11 | 0.12 | 0.23 | -0.20 | 0.18 | 0.07 | 0.13 |
| V538 Cyg | -0.23 | 0.37 | 0.14 | 0.25 | -0.06 | 0.14 | 0.13 | 0.03 | 0.12 | 0.08 | -0.17 | 0.05 | -0.06 | -0.06 | 0.08 | 0.12 | 0.20 | -0.08 | 0.18 | 0.18 | 0.10 |
| V547 Cyg | 0.00 | 0.48 | 0.27 | 0.26 | 0.24 | 0.24 | 0.16 | 0.05 | 0.17 | 0.07 | -0.10 | 0.15 | 0.02 | -0.01 | 0.24 | 0.26 | 0.24 | -0.09 | 0.24 | 0.12 | 0.18 |
| V609 Cyg | -0.17 | 0.54 | 0.26 | 0.32 | 0.04 | 0.21 | 0.20 | 0.02 | 0.24 | 0.10 | -0.07 | 0.22 | 0.07 | 0.03 | 0.36 | 0.50 | 0.37 | 0.21 | 0.38 | 0.32 | 0.53 |
| V621 Cyg | -0.34 | 0.22 | 0.09 | 0.31 | 0.22 | 0.11 | 0.15 | 0.02 | 0.11 | -0.01 | -0.14 | 0.12 | -0.03 | -0.05 | 0.07 | 0.47 | 0.17 | -0.16 | 0.12 | 0.18 | 0.41 |
| V1020 Cyg | 0.15 | 0.78 | 0.26 | 0.44 | -0.08 | 0.20 | 0.18 | 0.12 | 0.45 | 0.20 | 0.22 | 0.29 | 0.49 | 0.15 | 0.46 | 0.55 | 0.58 | 0.14 | 0.40 | ... | ... |
| V1025 Cyg | -0.03 | 0.39 | 0.24 | 0.24 | 0.16 | 0.25 | 0.20 | 0.06 | 0.15 | 0.18 | -0.02 | 0.23 | 0.12 | 0.05 | 0.24 | 0.24 | 0.15 | -0.15 | 0.22 | 0.33 | 0.33 |
| V1033 Cyg | 0.19 | -0.02 | 0.21 | -0.01 | 0.12 | 0.14 | 0.14 | 0.06 | 0.17 | 0.08 | -0.09 | 0.12 | 0.08 | 0.00 | 0.23 | 0.34 | 0.20 | -0.01 | 0.25 | 0.09 | 0.06 |
| V1046 Cyg | -0.10 | 0.45 | 0.26 | 0.37 | 0.28 | 0.24 | 0.23 | 0.08 | 0.27 | 0.12 | -0.05 | 0.23 | 0.11 | 0.05 | 0.34 | 0.43 | 0.28 | 0.09 | 0.38 | 0.28 | 0.33 |
| V1364 Cyg | 0.08 | 0.33 | 0.36 | 0.42 | 0.28 | 0.38 | 0.31 | 0.10 | 0.10 | 0.17 | -0.05 | 0.29 | 0.01 | 0.03 | 0.26 | 0.32 | 0.32 | 0.02 | 0.25 | 0.27 | 0.15 |
| V1397 Cyg | -0.34 | 0.16 | 0.04 | 0.17 | 0.30 | 0.15 | 0.11 | 0.01 | 0.10 | -0.02 | -0.13 | 0.01 | -0.02 | -0.02 | 0.07 | 0.23 | 0.14 | -0.23 | 0.16 | 0.09 | 0.02 |
| EK Del | -1.88 | -0.64 | -0.89 | -1.62 | -1.53 | -0.75 | -1.17 | -1.24 | -1.02 | -1.64 | -1.72 | -1.54 | -0.85 | -1.28 | -1.40 | -1.28 | -1.35 | -1.28 | -1.39 | ... | -0.68 |
| Zeta Gem | -0.23 | 0.41 | 0.21 | 0.31 | 0.06 | 0.17 | 0.07 | -0.10 | 0.05 | -0.01 | -0.22 | 0.10 | -0.08 | -0.12 | 0.20 | 0.33 | 0.24 | -0.10 | 0.22 | 0.20 | 0.04 |
| RZ Gem | -0.43 | 0.14 | -0.02 | 0.03 | -0.16 | -0.08 | -0.07 | -0.14 | -0.04 | -0.18 | -0.36 | -0.17 | -0.15 | -0.24 | -0.05 | 0.12 | 0.02 | -0.24 | 0.01 | -0.13 | -0.04 |
| W Gem | -0.25 | 0.37 | 0.15 | 0.23 | -0.01 | 0.08 | 0.06 | -0.01 | 0.07 | -0.05 | -0.26 | 0.02 | -0.06 | -0.13 | 0.16 | 0.27 | 0.22 | -0.07 | 0.22 | 0.14 | 0.13 |
| AA Gem | -0.43 | -0.02 | -0.02 | 0.07 | -0.11 | 0.06 | -0.08 | -0.29 | -0.22 | -0.20 | -0.48 | -0.14 | -0.34 | -0.34 | -0.09 | 0.27 | 0.00 | -0.28 | 0.10 | 0.02 | 0.01 |
| AD Gem | -0.49 | 0.11 | 0.02 | 0.01 | -0.10 | -0.04 | -0.08 | -0.16 | -0.10 | -0.22 | -0.43 | -0.15 | -0.22 | -0.28 | -0.07 | 0.10 | 0.05 | -0.24 | 0.07 | 0.00 | -0.02 |
| BB Gem | -0.58 | -0.01 | -0.16 | 0.09 | 0.02 | 0.05 | 0.01 | -0.04 | -0.02 | -0.16 | -0.38 | -0.10 | -0.07 | -0.21 | -0.08 | 0.23 | 0.05 | -0.55 | 0.03 | 0.01 | 0.01 |
| BW Gem | -0.49 | 0.02 | 0.01 | -0.05 | -0.08 | -0.05 | -0.09 | -0.16 | -0.08 | -0.23 | -0.40 | -0.18 | -0.12 | -0.27 | -0.08 | 0.09 | -0.06 | -0.27 | 0.03 | 0.04 | 0.05 |
| DX Gem | -0.40 | 0.16 | 0.02 | 0.07 | 0.08 | 0.03 | 0.09 | 0.04 | 0.14 | -0.06 | -0.12 | -0.04 | 0.10 | -0.10 | 0.03 | 0.29 | 0.08 | -0.42 | 0.11 | -0.06 | -0.07 |
| BB Her | 0.00 | 0.53 | 0.27 | 0.48 | 0.23 | 0.29 | 0.25 | 0.14 | 0.23 | 0.14 | 0.01 | 0.26 | 0.09 | 0.12 | 0.29 | 0.26 | 0.14 | -0.01 | 0.20 | 0.13 | 0.27 |
| V Lac | -0.18 | 0.41 | 0.14 | 0.32 | 0.16 | 0.11 | 0.19 | 0.09 | 0.24 | 0.03 | -0.16 | 0.06 | 0.24 | -0.02 | 0.22 | 0.33 | 0.15 | -0.17 | 0.25 | 0.07 | 0.33 |
| X Lac | -0.22 | 0.37 | 0.09 | 0.36 | 0.12 | 0.11 | 0.16 | 0.08 | 0.18 | 0.02 | -0.12 | 0.08 | 0.13 | -0.02 | 0.12 | 0.25 | 0.17 | -0.11 | 0.20 | 0.08 | -0.10 |
| Y Lac | -0.28 | 0.31 | 0.13 | 0.15 | 0.16 | 0.09 | 0.14 | 0.13 | 0.32 | 0.01 | -0.07 | 0.03 | 0.20 | -0.03 | 0.16 | 0.46 | 0.11 | -0.23 | 0.14 | 0.18 | 0.03 |
| Z Lac | -0.15 | 0.36 | 0.19 | 0.50 | 0.19 | 0.22 | 0.19 | 0.14 | 0.17 | 0.14 | -0.07 | 0.10 | 0.00 | 0.04 | 0.18 | 0.43 | 0.28 | -0.04 | 0.34 | 0.16 | 0.03 |
| RR Lac | -0.30 | 0.22 | 0.07 | 0.31 | 0.22 | 0.19 | 0.12 | -0.04 | 0.08 | 0.01 | -0.22 | 0.04 | -0.08 | -0.09 | 0.07 | 0.25 | 0.09 | -0.13 | 0.16 | 0.04 | -0.03 |
| BG Lac | -0.18 | 0.32 | 0.21 | 0.23 | 0.06 | 0.12 | 0.10 | 0.01 | 0.09 | -0.01 | -0.20 | 0.07 | -0.05 | -0.06 | 0.15 | 0.24 | 0.14 | -0.16 | 0.14 | 0.11 | 0.12 |
| DF Lac | -0.17 | 0.31 | 0.16 | 0.21 | -0.12 | 0.18 | 0.07 | 0.02 | 0.13 | 0.01 | -0.16 | 0.04 | -0.04 | -0.07 | 0.13 | 0.26 | 0.08 | -0.16 | 0.15 | -0.02 | -0.01 |
| FQ Lac | -0.29 | 0.35 | -0.60 | 0.11 | 0.42 | -0.50 | -0.05 | 0.04 | 0.15 | -0.27 | 0.19 | -0.20 | -0.03 | -0.18 | -0.74 | ... | 0.08 | 0.28 | 0.08 | 0.23 | ... |
| V411 Lac | -0.28 | 0.20 | 0.06 | 0.34 | 0.08 | 0.08 | 0.04 | -0.07 | 0.12 | -0.08 | -0.12 | 0.02 | -0.03 | -0.12 | 0.18 | 0.48 | 0.16 | -0.24 | 0.24 | ... | 0.20 |
| V473 Lyr | -0.28 | 0.10 | 0.08 | -0.01 | -0.07 | -0.02 | 0.00 | -0.10 | -0.01 | -0.13 | -0.28 | -0.06 | -0.08 | -0.17 | 0.05 | 0.14 | 0.04 | -0.24 | 0.05 | -0.04 | -0.02 |
| T Mon | -0.15 | 0.47 | 0.33 | 0.43 | 0.39 | 0.45 | 0.28 | 0.23 | 0.15 | 0.20 | -0.05 | 0.23 | 0.07 | 0.10 | 0.30 | 0.54 | 0.16 | 0.07 | 0.35 | 0.33 | 0.36 |
| TW Mon | -0.50 | -0.28 | 0.07 | 0.10 | -0.21 | 0.07 | -0.12 | -0.22 | -0.18 | -0.23 | -0.43 | -0.18 | -0.31 | -0.35 | -0.07 | 0.23 | 0.00 | -0.31 | 0.02 | -0.02 | -0.23 |
| TX Mon | -0.41 | 0.17 | -0.09 | 0.15 | 0.01 | 0.08 | 0.06 | 0.00 | 0.03 | -0.05 | -0.28 | -0.03 | -0.09 | -0.12 | 0.03 | 0.25 | 0.11 | -0.16 | 0.14 | 0.01 | -0.02 |

**Table 4**
Average [x/H] Ratios for Cepheids

| Cepheid | C | N | O | Na | Mg | Al | Si | Ca | Ti | Cr | Mn | Fe | Co | Ni | Y | La | Ce | Pr | Nd | Sm | Eu |
|---|---|---|---|---|---|---|---|---|---|---|---|---|---|---|---|---|---|---|---|---|---|
| TY Mon | -0.50 | 0.12 | 0.03 | 0.28 | -0.04 | 0.16 | 0.04 | -0.02 | 0.03 | -0.12 | -0.23 | 0.02 | -0.09 | -0.15 | -0.06 | 0.46 | 0.14 | -0.53 | 0.26 | 0.10 | 0.02 |
| TZ Mon | -0.25 | 0.24 | 0.10 | 0.22 | 0.04 | 0.08 | 0.06 | -0.04 | 0.06 | -0.04 | -0.23 | 0.01 | -0.10 | -0.12 | 0.10 | 0.33 | 0.17 | -0.13 | 0.17 | 0.06 | 0.06 |
| UY Mon | -0.31 | 0.24 | 0.09 | 0.11 | -0.03 | -0.12 | -0.02 | -0.12 | 0.05 | -0.15 | -0.30 | -0.09 | 0.08 | -0.20 | 0.09 | 0.24 | 0.09 | -0.20 | 0.14 | 0.03 | 0.09 |
| VZ Mon | -0.33 | -0.06 | 0.02 | 0.11 | 0.01 | -0.02 | -0.12 | -0.21 | 0.03 | 0.06 | -0.03 | -0.07 | 0.00 | -0.23 | 0.05 | 0.13 | 0.20 | -0.07 | 0.04 | ... | ... |
| WW Mon | -0.66 | -0.08 | -0.09 | -0.07 | -0.13 | -0.11 | -0.14 | -0.20 | -0.09 | -0.27 | -0.40 | -0.19 | -0.17 | -0.30 | -0.14 | 0.10 | 0.11 | -0.16 | 0.05 | 0.00 | 0.00 |
| XX Mon | -0.27 | 0.35 | 0.05 | 0.16 | 0.00 | 0.00 | 0.06 | -0.03 | 0.12 | -0.05 | -0.25 | 0.00 | 0.04 | -0.10 | 0.06 | 0.24 | 0.09 | -0.02 | 0.14 | 0.09 | 0.22 |
| YY Mon | -0.78 | -0.28 | -0.24 | -0.19 | -0.10 | -0.23 | -0.38 | -0.36 | -0.28 | -0.38 | -0.61 | -0.46 | -0.17 | -0.44 | -0.51 | -0.24 | -0.65 | -0.55 | -0.27 | -0.34 | -0.76 |
| AA Mon | -0.39 | 0.25 | -0.19 | 0.21 | -0.19 | -0.13 | -0.04 | -0.17 | 0.01 | -0.27 | -0.44 | -0.09 | -0.42 | -0.21 | 0.09 | 0.40 | 0.03 | -0.16 | 0.09 | ... | 0.16 |
| AC Mon | -0.35 | 0.23 | 0.05 | 0.19 | -0.05 | 0.02 | 0.03 | -0.08 | 0.08 | -0.06 | -0.29 | -0.03 | -0.02 | -0.14 | 0.07 | 0.21 | 0.17 | -0.14 | 0.17 | 0.09 | 0.14 |
| BE Mon | -0.32 | 0.29 | 0.11 | 0.24 | 0.15 | 0.11 | 0.14 | 0.08 | 0.14 | 0.02 | -0.15 | 0.08 | -0.06 | -0.03 | 0.11 | 0.19 | 0.16 | -0.19 | 0.17 | -0.02 | 0.16 |
| BV Mon | -0.34 | 0.25 | 0.03 | 0.08 | -0.01 | -0.10 | 0.01 | -0.12 | 0.05 | -0.16 | -0.35 | -0.10 | 0.02 | -0.19 | -0.07 | 0.15 | -0.02 | -0.39 | -0.07 | 0.04 | 0.18 |
| CS Mon | -0.50 | 0.11 | 0.11 | 0.06 | -0.07 | -0.01 | -0.07 | -0.21 | -0.05 | -0.15 | -0.34 | -0.08 | -0.17 | -0.23 | 0.02 | 0.34 | 0.12 | -0.31 | 0.13 | -0.01 | 0.25 |
| CU Mon | -0.50 | 0.08 | -0.07 | 0.03 | -0.23 | -0.20 | -0.16 | -0.19 | -0.06 | -0.08 | -0.47 | -0.23 | 0.16 | -0.26 | -0.14 | 0.22 | 0.11 | -0.34 | 0.08 | -0.01 | 0.00 |
| CV Mon | -0.30 | 0.20 | 0.05 | 0.17 | 0.05 | 0.09 | 0.05 | 0.01 | 0.06 | -0.02 | -0.22 | 0.01 | -0.15 | -0.11 | 0.06 | 0.27 | 0.15 | -0.25 | 0.15 | 0.13 | 0.03 |
| EE Mon | -0.65 | 0.18 | -0.12 | -0.27 | -0.53 | -0.65 | -0.38 | -0.48 | -0.37 | -0.54 | -0.83 | -0.52 | -0.13 | -0.52 | -0.58 | -0.02 | -0.58 | -0.61 | -0.60 | -0.51 | ... |
| EK Mon | -0.30 | 0.16 | 0.06 | 0.22 | 0.01 | 0.14 | 0.11 | 0.10 | 0.09 | 0.01 | -0.19 | 0.04 | -0.05 | -0.07 | 0.09 | 0.21 | 0.13 | -0.19 | 0.21 | 0.05 | 0.13 |
| FG Mon | -0.48 | 0.22 | -0.18 | 0.04 | 0.05 | 0.01 | -0.06 | -0.10 | -0.03 | -0.11 | -0.31 | -0.14 | -0.16 | -0.21 | -0.08 | 0.12 | -0.12 | -0.21 | 0.07 | -0.06 | -0.21 |
| FI Mon | -0.65 | 0.00 | -0.37 | -0.02 | -0.24 | -0.10 | 0.06 | -0.08 | 0.15 | -0.10 | -0.03 | -0.11 | 0.12 | -0.15 | 0.05 | 0.58 | -0.03 | -0.19 | 0.04 | 0.56 | ... |
| FT Mon | -0.51 | -0.03 | -0.11 | -0.04 | -0.17 | -0.14 | -0.15 | -0.29 | -0.15 | -0.33 | -0.49 | -0.21 | -0.14 | -0.34 | -0.15 | 0.03 | -0.07 | -0.38 | -0.02 | -0.10 | 0.04 |
| V446 Mon | -0.54 | -0.01 | -0.06 | 0.03 | -0.35 | -0.18 | -0.24 | -0.33 | -0.12 | -0.34 | -0.68 | -0.30 | -0.23 | -0.40 | -0.34 | -0.06 | -0.04 | -0.35 | -0.10 | -0.15 | ... |
| V447 Mon | -0.53 | 0.00 | -0.06 | -0.07 | -0.40 | -0.16 | -0.26 | -0.35 | -0.16 | -0.37 | -0.40 | -0.37 | -0.17 | -0.41 | -0.33 | -0.58 | -0.14 | -0.49 | -0.11 | -0.45 | 0.03 |
| V465 Mon | -0.26 | 0.22 | 0.11 | 0.15 | 0.02 | 0.03 | 0.08 | -0.01 | 0.15 | -0.02 | -0.23 | 0.02 | 0.01 | -0.07 | 0.12 | 0.45 | 0.22 | -0.04 | 0.25 | 0.17 | 0.29 |
| V484 Mon | -0.60 | 0.21 | -0.24 | -0.10 | -0.13 | -0.01 | -0.10 | -0.13 | 0.09 | -0.12 | -0.26 | -0.06 | -0.07 | -0.11 | 0.15 | 0.48 | 0.38 | -0.36 | 0.21 | -0.17 | 0.21 |
| V495 Mon | -0.45 | 0.10 | -0.10 | 0.03 | -0.18 | -0.13 | -0.10 | -0.10 | 0.03 | -0.14 | -0.35 | -0.11 | -0.18 | -0.22 | -0.03 | 0.24 | -0.03 | -0.11 | 0.11 | -0.08 | -0.27 |
| V504 Mon | -0.29 | 0.22 | 0.04 | 0.07 | -0.02 | 0.04 | -0.02 | -0.09 | 0.13 | -0.12 | -0.31 | 0.01 | 0.13 | -0.14 | 0.22 | 0.41 | 0.25 | 0.09 | 0.35 | 0.22 | ... |
| V508 Mon | -0.66 | 0.09 | -0.05 | 0.19 | -0.09 | -0.03 | -0.05 | -0.10 | -0.07 | -0.18 | -0.34 | -0.13 | -0.20 | -0.23 | -0.09 | 0.04 | 0.04 | -0.22 | 0.04 | -0.19 | -0.05 |
| V510 Mon | -0.33 | 0.13 | 0.01 | -0.03 | -0.11 | 0.05 | -0.08 | -0.17 | -0.03 | -0.10 | -0.39 | -0.06 | -0.24 | -0.23 | 0.01 | 0.19 | 0.11 | -0.03 | 0.22 | 0.24 | 0.03 |
| V526 Mon | -0.39 | 0.17 | -0.07 | 0.05 | -0.14 | -0.21 | -0.06 | -0.16 | 0.00 | -0.18 | -0.36 | -0.17 | -0.04 | -0.23 | -0.04 | 0.10 | 0.03 | -0.16 | 0.05 | 0.00 | 0.06 |
| CR Ori | -0.87 | 0.33 | -0.28 | 0.23 | -0.32 | -0.10 | -0.14 | -0.18 | -0.02 | -0.25 | -0.31 | -0.19 | -0.16 | -0.27 | -0.13 | 0.17 | 0.06 | -0.25 | 0.01 | -0.16 | 0.02 |
| DF Ori | -0.53 | 0.03 | -0.15 | 0.00 | -0.25 | -0.23 | -0.16 | -0.21 | -0.08 | -0.30 | -0.43 | -0.28 | -0.26 | -0.31 | -0.17 | -0.02 | -0.16 | -0.31 | -0.05 | -0.07 | -0.39 |
| AU Peg | 0.13 | 0.15 | 0.35 | 0.56 | 0.33 | 0.82 | 0.50 | 0.20 | 0.41 | 0.31 | 0.36 | 0.46 | 0.44 | 0.40 | 0.08 | 0.12 | -0.12 | -0.40 | 0.24 | 0.07 | ... |
| QQ Per | -0.59 | -0.29 | -0.13 | -0.64 | -0.21 | -0.27 | -0.33 | -0.64 | -0.51 | -0.79 | -0.91 | -0.67 | -0.54 | -0.62 | -0.92 | -0.40 | -0.96 | -1.07 | -0.74 | -0.67 | -0.36 |
| SV Per | -0.22 | 0.38 | 0.20 | 0.17 | -0.03 | 0.06 | 0.05 | -0.09 | 0.06 | -0.07 | -0.25 | 0.06 | -0.07 | -0.14 | 0.19 | 0.45 | 0.21 | -0.10 | 0.24 | 0.19 | 0.16 |
| SX Per | -0.46 | 0.30 | -0.22 | 0.10 | -0.01 | 0.08 | -0.01 | -0.03 | 0.03 | -0.07 | -0.32 | -0.03 | -0.10 | -0.17 | 0.05 | 0.28 | 0.13 | -0.10 | 0.19 | -0.03 | 0.15 |
| UX Per | -0.37 | 0.22 | -0.06 | 0.22 | 0.03 | -0.07 | 0.05 | 0.04 | 0.11 | -0.07 | -0.24 | -0.05 | 0.10 | -0.09 | 0.02 | 0.20 | 0.08 | -0.26 | 0.07 | 0.06 | -0.11 |
| UY Per | -0.12 | 0.27 | 0.23 | 0.26 | 0.07 | 0.21 | 0.15 | 0.05 | 0.19 | 0.10 | -0.07 | 0.18 | 0.04 | 0.03 | 0.23 | 0.38 | 0.29 | -0.07 | 0.29 | 0.19 | 0.28 |
| VX Per | -0.27 | 0.31 | 0.06 | 0.27 | 0.05 | 0.09 | 0.07 | 0.03 | 0.12 | -0.01 | -0.19 | 0.06 | 0.02 | -0.09 | 0.13 | 0.29 | 0.19 | -0.13 | 0.18 | 0.09 | 0.05 |

**Table 4**
Average [x/H] Ratios for Cepheids

| Cepheid | C | N | O | Na | Mg | Al | Si | Ca | Ti | Cr | Mn | Fe | Co | Ni | Y | La | Ce | Pr | Nd | Sm | Eu |
|---|---|---|---|---|---|---|---|---|---|---|---|---|---|---|---|---|---|---|---|---|---|
| VY Per | -0.28 | 0.16 | 0.03 | 0.35 | 0.06 | 0.25 | 0.13 | 0.06 | 0.15 | 0.10 | -0.13 | 0.04 | -0.13 | 0.03 | 0.09 | 0.45 | 0.15 | -0.18 | 0.19 | 0.08 | -0.01 |
| AS Per | -0.23 | 0.70 | 0.18 | 0.32 | 0.46 | 0.17 | 0.17 | 0.11 | 0.23 | 0.10 | -0.01 | 0.14 | 0.19 | 0.04 | 0.18 | 0.32 | 0.07 | -0.25 | 0.10 | -0.06 | ... |
| AW Per | -0.27 | 0.43 | 0.22 | 0.39 | 0.12 | 0.20 | 0.04 | -0.17 | 0.03 | -0.03 | -0.23 | 0.04 | -0.08 | -0.12 | 0.21 | 0.46 | 0.20 | -0.22 | 0.23 | 0.11 | 0.00 |
| BM Per | -0.18 | 0.37 | 0.18 | 0.19 | 0.13 | 0.19 | 0.15 | 0.01 | 0.23 | 0.08 | -0.01 | 0.23 | 0.09 | -0.02 | 0.31 | 0.54 | 0.27 | 0.21 | 0.39 | 0.31 | 0.36 |
| CI Per | -1.12 | 0.33 | -0.42 | ... | ... | -0.34 | -0.30 | -0.26 | -0.15 | -0.38 | -0.47 | -0.32 | -0.10 | -0.39 | -0.32 | -0.32 | -0.45 | -0.55 | -0.26 | -0.32 | -0.55 |
| DW Per | -0.31 | 0.17 | -0.06 | 0.12 | 0.05 | -0.17 | 0.10 | 0.03 | 0.13 | -0.06 | -0.22 | -0.05 | 0.25 | -0.08 | -0.05 | 0.20 | 0.02 | -0.29 | 0.10 | -0.22 | 0.28 |
| GP Per | -0.83 | -0.29 | -0.35 | -0.43 | -0.27 | -0.34 | -0.39 | -0.70 | -0.37 | -0.84 | -0.69 | -0.80 | -0.05 | -0.75 | -1.03 | -0.72 | -0.79 | -0.79 | -0.79 | ... | -0.45 |
| HQ Per | -0.67 | -0.15 | -0.22 | -0.08 | -0.21 | -0.22 | -0.23 | -0.30 | -0.24 | -0.36 | -0.54 | -0.35 | -0.42 | -0.42 | -0.35 | -0.11 | -0.20 | -0.46 | -0.14 | -0.19 | -0.18 |
| HZ Per | -1.10 | 0.20 | -0.06 | 0.03 | -0.13 | -0.10 | -0.17 | -0.16 | -0.04 | -0.22 | -0.29 | -0.25 | -0.27 | -0.26 | -0.20 | 0.26 | -0.02 | -0.33 | 0.04 | 0.48 | 0.20 |
| MM Per | -0.31 | 0.13 | 0.12 | 0.07 | -0.06 | -0.02 | -0.03 | -0.06 | -0.02 | -0.11 | -0.30 | -0.07 | -0.16 | -0.19 | 0.05 | 0.18 | 0.11 | -0.15 | 0.17 | 0.03 | -0.01 |
| OT Per | -0.86 | 0.17 | -0.30 | 0.25 | -0.29 | -0.08 | -0.05 | -0.07 | 0.20 | -0.17 | -0.30 | -0.07 | -0.02 | -0.21 | -0.05 | 0.55 | 0.12 | -0.10 | 0.16 | 0.60 | 0.46 |
| X Pup | -0.26 | 0.36 | 0.11 | 0.13 | ... | 0.14 | 0.12 | -0.09 | 0.31 | 0.03 | -0.07 | 0.08 | 0.17 | -0.01 | 0.15 | 0.34 | 0.19 | -0.04 | 0.28 | 0.22 | 0.28 |
| RS Pup | -0.54 | 0.54 | 0.56 | 0.61 | ... | 0.28 | 0.37 | 0.19 | 0.11 | 0.26 | -0.13 | 0.22 | 0.08 | 0.13 | 0.29 | 0.51 | 0.30 | 0.10 | 0.41 | 0.39 | 0.38 |
| VX Pup | -0.37 | 0.25 | 0.10 | 0.16 | -0.09 | -0.04 | 0.09 | 0.01 | 0.14 | 0.01 | -0.24 | 0.00 | -0.15 | -0.09 | 0.15 | 0.14 | 0.27 | -0.07 | 0.18 | -0.12 | 0.05 |
| VZ Pup | -0.59 | 0.07 | 0.08 | 0.11 | -0.23 | 0.08 | -0.06 | -0.30 | -0.08 | -0.06 | -0.31 | -0.11 | -0.26 | -0.25 | 0.05 | 0.32 | 0.12 | -0.06 | 0.24 | 0.17 | 0.09 |
| WX Pup | -0.28 | 0.17 | 0.16 | 0.19 | -0.15 | 0.04 | 0.01 | -0.12 | 0.01 | -0.08 | -0.30 | -0.04 | -0.16 | -0.17 | 0.12 | 0.04 | 0.20 | -0.04 | 0.15 | 0.08 | 0.11 |
| AD Pup | -0.46 | 0.03 | 0.18 | 0.03 | -0.07 | -0.06 | 0.01 | -0.18 | -0.12 | -0.06 | -0.39 | -0.03 | -0.26 | -0.25 | -0.06 | 0.11 | 0.15 | -0.14 | 0.12 | -0.12 | 0.07 |
| AQ Pup | -0.44 | 0.25 | -0.14 | 0.21 | ... | 0.25 | 0.14 | -0.01 | 0.27 | -0.04 | -0.05 | 0.04 | 0.21 | 0.02 | 0.08 | 0.18 | 0.13 | -0.32 | 0.27 | ... | 0.04 |
| BN Pup | -0.31 | 0.53 | 0.24 | 0.28 | 0.84 | 0.18 | 0.21 | 0.11 | 0.27 | 0.20 | 0.08 | 0.11 | 0.08 | 0.08 | 0.23 | 0.33 | 0.41 | -0.13 | 0.31 | 0.21 | 0.11 |
| V335 Pup | -0.11 | 0.47 | 0.19 | 0.17 | -0.04 | 0.03 | 0.10 | -0.04 | 0.20 | 0.04 | -0.20 | 0.09 | 0.05 | -0.03 | 0.26 | 0.25 | 0.33 | 0.02 | 0.29 | 0.14 | 0.24 |
| Y Sct | -0.07 | 0.53 | 0.28 | 0.38 | 0.08 | 0.28 | 0.16 | 0.03 | 0.19 | 0.08 | -0.10 | 0.23 | 0.12 | 0.02 | 0.30 | 0.32 | 0.20 | 0.03 | 0.30 | 0.18 | 0.14 |
| Z Sct | 0.12 | 0.58 | 0.50 | 0.59 | 0.36 | 0.43 | 0.32 | 0.09 | 0.19 | 0.20 | 0.10 | 0.33 | 0.17 | 0.14 | 0.29 | 0.35 | 0.24 | 0.04 | 0.35 | 0.31 | 0.36 |
| RU Sct | -0.08 | 0.35 | 0.31 | 0.33 | 0.07 | 0.17 | 0.15 | 0.00 | 0.07 | 0.08 | -0.14 | 0.11 | -0.04 | -0.05 | 0.14 | 0.17 | 0.08 | -0.16 | 0.09 | 0.07 | 0.10 |
| SS Sct | -0.19 | 0.17 | 0.17 | 0.15 | 0.24 | 0.18 | 0.14 | 0.06 | 0.14 | 0.03 | -0.08 | 0.14 | -0.08 | -0.04 | 0.16 | 0.48 | 0.15 | -0.24 | 0.20 | 0.05 | -0.06 |
| TY Sct | 0.15 | 0.76 | 0.38 | 0.45 | 0.29 | 0.40 | 0.26 | 0.07 | 0.28 | 0.14 | 0.05 | 0.37 | 0.28 | 0.14 | 0.38 | 0.31 | 0.16 | 0.08 | 0.29 | 0.29 | 0.14 |
| BX Sct | -0.07 | 0.40 | 0.34 | 0.46 | 0.27 | 0.27 | 0.21 | 0.15 | 0.27 | 0.16 | -0.01 | 0.28 | 0.10 | 0.09 | 0.32 | 0.36 | 0.24 | 0.09 | 0.39 | 0.30 | 0.53 |
| CK Sct | -0.09 | 0.36 | 0.23 | 0.32 | 0.10 | 0.25 | 0.13 | 0.00 | 0.21 | 0.04 | -0.05 | 0.21 | 0.17 | 0.01 | 0.27 | 0.37 | 0.23 | -0.09 | 0.26 | 0.20 | 0.23 |
| CM Sct | -0.10 | 0.43 | 0.16 | 0.33 | 0.03 | 0.19 | 0.15 | 0.04 | 0.20 | 0.04 | -0.10 | 0.15 | 0.10 | 0.01 | 0.21 | 0.17 | 0.06 | -0.06 | 0.13 | 0.05 | ... |
| CN Sct | -0.01 | 0.52 | 0.28 | 0.56 | 0.32 | 0.45 | 0.32 | 0.17 | 0.28 | 0.21 | 0.14 | 0.33 | 0.21 | 0.20 | 0.30 | 0.19 | 0.13 | -0.01 | 0.22 | 0.07 | 0.05 |
| EV Sct | -0.23 | 0.32 | 0.23 | 0.20 | 0.21 | 0.31 | 0.20 | 0.19 | 0.28 | 0.05 | 0.13 | 0.15 | 0.11 | 0.04 | 0.22 | 0.44 | 0.23 | -0.11 | 0.23 | 0.07 | 0.41 |
| AA Ser | 0.00 | 0.65 | 0.55 | 0.72 | ... | 0.61 | 0.52 | 0.44 | 0.40 | 0.46 | 0.38 | 0.41 | 0.32 | 0.34 | 0.43 | 0.42 | 0.13 | 0.03 | 0.32 | 0.40 | 0.34 |
| DV Ser | 0.13 | 1.03 | 0.37 | 0.72 | 0.67 | 0.59 | 0.46 | 0.37 | 0.55 | 0.38 | 0.29 | 0.47 | 0.43 | 0.37 | 0.39 | 0.43 | 0.23 | 0.10 | 0.27 | 0.46 | 0.34 |
| DG Sge | -0.08 | 0.53 | 0.21 | 0.38 | 0.14 | 0.10 | 0.18 | 0.09 | 0.22 | 0.08 | -0.01 | 0.13 | 0.12 | 0.05 | 0.21 | 0.23 | 0.02 | -0.26 | 0.14 | 0.20 | 0.12 |
| GX Sge | -0.02 | 0.50 | 0.32 | 0.27 | 0.16 | 0.31 | 0.21 | 0.01 | 0.17 | 0.13 | -0.04 | 0.29 | 0.02 | 0.06 | 0.37 | 0.43 | 0.28 | 0.16 | 0.34 | 0.33 | 0.33 |
| GY Sge | 0.17 | -0.02 | 0.31 | 0.29 | 0.15 | 0.24 | 0.26 | 0.17 | 0.27 | 0.22 | 0.02 | 0.29 | 0.11 | 0.15 | 0.48 | 0.56 | 0.15 | 0.07 | 0.40 | 0.35 | 0.37 |
| ST Tau | -0.22 | 0.34 | 0.14 | 0.29 | 0.01 | 0.06 | 0.06 | -0.01 | 0.11 | -0.05 | -0.22 | 0.00 | -0.06 | -0.11 | 0.13 | 0.29 | 0.15 | -0.09 | 0.16 | 0.07 | 0.11 |
| AE Tau | -0.46 | 0.09 | 0.00 | 0.05 | -0.15 | -0.06 | -0.11 | -0.14 | -0.07 | -0.20 | -0.46 | -0.18 | -0.27 | -0.25 | -0.10 | 0.04 | 0.00 | -0.42 | -0.02 | -0.02 | -0.12 |

**Table 4**
Average [x/H] Ratios for Cepheids

| Cepheid | C | N | O | Na | Mg | Al | Si | Ca | Ti | Cr | Mn | Fe | Co | Ni | Y | La | Ce | Pr | Nd | Sm | Eu |
|---|---|---|---|---|---|---|---|---|---|---|---|---|---|---|---|---|---|---|---|---|---|
| AV Tau | -0.36 | 0.08 | 0.01 | -0.03 | -0.06 | -0.15 | -0.05 | -0.06 | -0.02 | -0.11 | -0.19 | -0.07 | -0.18 | -0.19 | 0.01 | 0.17 | 0.11 | -0.16 | 0.15 | -0.12 | -0.25 |
| S Vul | -0.13 | 0.42 | 0.21 | 0.22 | 0.06 | 0.27 | 0.11 | 0.03 | 0.11 | 0.05 | -0.14 | 0.12 | -0.05 | -0.02 | 0.23 | 0.35 | 0.08 | -0.05 | 0.21 | 0.23 | 0.23 |
| U Vul | -0.02 | 0.51 | 0.31 | 0.28 | 0.09 | 0.17 | 0.16 | 0.01 | 0.22 | 0.08 | -0.09 | 0.19 | 0.07 | 0.00 | 0.35 | 0.39 | 0.29 | 0.10 | 0.32 | 0.23 | 0.21 |
| AS Vul | -0.03 | 0.39 | 0.21 | 0.62 | 0.29 | 0.42 | 0.30 | 0.27 | 0.16 | 0.22 | 0.09 | 0.22 | 0.13 | 0.12 | 0.27 | 0.34 | 0.20 | -0.01 | 0.39 | 0.24 | 0.26 |
| DG Vul | -0.13 | 0.54 | 0.18 | 0.42 | 0.26 | 0.21 | 0.21 | 0.05 | 0.29 | 0.14 | 0.00 | 0.19 | 0.16 | 0.09 | 0.26 | 0.21 | 0.13 | -0.05 | 0.24 | 0.08 | 0.35 |
| GSC 0754-1993 | -0.57 | 0.28 | 0.30 | 0.02 | -0.16 | -0.06 | -0.07 | -0.26 | -0.06 | -0.19 | -0.38 | -0.01 | -0.34 | -0.23 | 0.10 | 0.31 | 0.25 | 0.02 | 0.29 | 0.20 | 0.04 |
| GSC 2418-1443 | -0.29 | -0.14 | -0.14 | -0.06 | -0.19 | -0.13 | -0.09 | -0.15 | -0.06 | -0.22 | -0.36 | -0.15 | -0.27 | -0.26 | -0.08 | 0.13 | 0.03 | -0.31 | 0.00 | -0.25 | -0.02 |
| GSC 3706-0233 | -0.21 | 0.19 | 0.23 | -0.07 | 0.29 | 0.07 | 0.07 | -0.07 | 0.14 | -0.04 | -0.27 | 0.09 | -0.06 | -0.12 | 0.15 | 0.43 | 0.15 | 0.01 | 0.22 | 0.23 | 0.54 |
| GSC 3725-0174 | -0.48 | 0.18 | -0.03 | 0.11 | 0.04 | -0.28 | -0.06 | -0.16 | 0.09 | -0.24 | -0.31 | -0.17 | -0.06 | -0.22 | -0.06 | 0.30 | 0.05 | -0.11 | 0.08 | -0.17 | 0.10 |
| GSC 3726-0565 | -0.71 | -0.03 | -0.34 | 0.10 | 0.10 | -0.07 | -0.09 | -0.27 | -0.21 | -0.27 | -0.34 | -0.22 | -0.27 | -0.32 | -0.25 | 0.28 | -0.08 | -0.63 | -0.08 | -0.16 | -0.33 |
| GSC 3729-1127 | -0.25 | 0.23 | 0.30 | 0.10 | 0.08 | 0.03 | -0.01 | -0.15 | -0.03 | -0.12 | -0.30 | -0.01 | -0.25 | -0.20 | 0.18 | 0.38 | 0.19 | -0.02 | 0.27 | 0.16 | 0.12 |
| GSC 3732-0183 | -0.59 | -0.05 | -0.35 | 0.08 | -0.02 | -0.12 | -0.07 | -0.07 | -0.03 | -0.27 | -0.37 | -0.16 | -0.28 | -0.26 | -0.23 | 0.23 | -0.05 | -0.48 | -0.06 | -0.60 | -0.11 |
| GSC 4009-0024 | -0.41 | 0.02 | 0.04 | -0.04 | -0.18 | -0.19 | -0.13 | -0.24 | -0.16 | -0.25 | -0.44 | -0.16 | -0.24 | -0.32 | -0.09 | 0.09 | 0.07 | -0.25 | 0.11 | 0.08 | -0.03 |
| GSC 4038-1585 | -0.26 | 0.24 | 0.11 | 0.23 | 0.02 | 0.06 | 0.05 | 0.00 | 0.03 | -0.03 | -0.25 | 0.02 | -0.10 | -0.16 | 0.09 | 0.30 | 0.19 | -0.10 | 0.22 | 0.13 | 0.27 |
| GSC 4040-1803 | -0.34 | 0.30 | -0.08 | 0.12 | 0.16 | -0.03 | 0.07 | 0.04 | 0.16 | -0.02 | -0.16 | -0.06 | 0.13 | -0.07 | 0.10 | 0.32 | 0.08 | -0.55 | 0.21 | -0.24 | ... |
| GSC 4265-0193 | -0.41 | 0.32 | 0.02 | 0.35 | 0.24 | -0.11 | 0.09 | 0.02 | 0.20 | 0.01 | -0.13 | -0.02 | 0.00 | -0.07 | 0.10 | 0.34 | 0.34 | -0.21 | 0.20 | 0.27 | 0.45 |
| GSC 4265-0569 | -0.17 | 0.15 | 0.18 | 0.20 | 0.16 | 0.04 | 0.06 | -0.03 | 0.18 | 0.07 | -0.09 | 0.04 | 0.03 | -0.03 | 0.13 | 0.35 | 0.37 | 0.04 | 0.39 | 0.20 | 0.28 |
| T Ant | -0.48 | 0.00 | -0.08 | -0.04 | -0.20 | -0.09 | -0.09 | -0.18 | -0.06 | -0.19 | -0.43 | -0.20 | -0.17 | -0.25 | -0.06 | 0.19 | 0.02 | -0.21 | 0.04 | -0.04 | 0.08 |
| L Car | -0.11 | 0.04 | 0.22 | -0.01 | 0.01 | 0.11 | 0.11 | -0.13 | -0.06 | 0.01 | -0.26 | 0.13 | -0.10 | -0.14 | 0.01 | 0.08 | -0.06 | -0.21 | 0.01 | -0.02 | 0.05 |
| SX Car | -0.15 | 0.41 | 0.13 | 0.29 | 0.24 | 0.02 | 0.14 | 0.01 | 0.20 | 0.00 | -0.15 | 0.05 | 0.17 | -0.06 | 0.22 | 0.39 | 0.06 | -0.30 | 0.19 | 0.10 | 0.03 |
| U Car | 0.17 | 0.42 | 0.07 | 0.43 | ... | 0.41 | 0.23 | 0.10 | 0.05 | 0.07 | -0.16 | 0.04 | -0.06 | 0.10 | 0.19 | 0.51 | -0.09 | -0.12 | 0.24 | 0.52 | 0.16 |
| V Car | -0.38 | 0.27 | 0.03 | 0.20 | -0.08 | 0.12 | 0.11 | 0.00 | 0.08 | -0.02 | -0.17 | 0.04 | -0.11 | -0.11 | 0.09 | 0.22 | 0.07 | -0.19 | 0.14 | 0.07 | -0.11 |
| UW Car | -0.17 | 0.46 | 0.18 | 0.18 | 0.26 | 0.05 | 0.14 | 0.09 | 0.19 | 0.05 | -0.16 | 0.09 | -0.04 | -0.04 | 0.17 | 0.27 | 0.10 | -0.15 | 0.07 | 0.01 | 0.22 |
| UX Car | -0.14 | 0.43 | 0.10 | 0.27 | 0.03 | 0.03 | 0.13 | 0.02 | 0.16 | -0.01 | -0.17 | 0.05 | 0.02 | -0.04 | 0.17 | 0.26 | 0.05 | -0.20 | 0.06 | -0.19 | 0.15 |
| UY Car | -0.14 | 0.38 | 0.16 | 0.23 | 0.46 | 0.12 | 0.19 | 0.17 | 0.28 | 0.09 | -0.07 | 0.13 | 0.04 | 0.02 | 0.23 | 0.33 | 0.10 | -0.26 | 0.13 | -0.01 | ... |
| UZ Car | -0.18 | 0.26 | 0.17 | 0.20 | 0.38 | 0.17 | 0.21 | 0.19 | 0.19 | 0.14 | -0.08 | 0.13 | 0.15 | 0.05 | 0.16 | 0.42 | 0.08 | -0.14 | 0.16 | 0.11 | 0.03 |
| VY Car | -0.21 | 0.59 | 0.17 | 0.85 | ... | 0.71 | 0.28 | 0.22 | 0.20 | 0.40 | -0.10 | 0.02 | 0.06 | 0.25 | 0.12 | 0.09 | -0.14 | -0.08 | 0.35 | 0.30 | 0.05 |
| WW Car | -0.29 | 0.17 | 0.00 | 0.20 | 0.04 | 0.12 | 0.05 | 0.00 | 0.05 | -0.13 | -0.22 | 0.00 | -0.10 | -0.14 | 0.05 | 0.14 | 0.01 | -0.27 | 0.03 | -0.10 | 0.01 |
| WZ Car | -0.40 | 0.11 | -0.09 | 0.43 | 0.45 | 0.19 | 0.10 | 0.06 | 0.07 | 0.00 | -0.07 | 0.05 | -0.03 | -0.03 | 0.01 | 0.37 | 0.03 | -0.20 | 0.05 | 0.04 | 0.26 |
| XX Car | -0.30 | 0.59 | 0.24 | 0.36 | 0.17 | 0.31 | 0.19 | 0.08 | 0.21 | 0.10 | -0.09 | 0.20 | 0.06 | 0.02 | 0.38 | 0.42 | 0.25 | 0.17 | 0.40 | 0.28 | 0.26 |
| XY Car | -0.23 | 0.31 | 0.18 | 0.20 | 0.11 | 0.16 | 0.13 | 0.01 | 0.07 | 0.03 | -0.14 | 0.07 | -0.11 | -0.05 | 0.15 | 0.26 | 0.14 | -0.14 | 0.15 | 0.11 | 0.32 |
| XZ Car | -0.18 | 0.36 | 0.03 | 0.44 | 0.28 | 0.29 | 0.27 | 0.22 | 0.29 | 0.15 | 0.10 | 0.19 | 0.25 | 0.13 | 0.23 | 0.32 | 0.17 | -0.03 | 0.19 | -0.03 | 0.38 |
| YZ Car | -0.23 | 0.28 | 0.16 | 0.16 | 0.15 | 0.12 | 0.07 | -0.03 | 0.07 | 0.01 | -0.17 | 0.00 | -0.08 | -0.11 | 0.20 | 0.30 | 0.16 | -0.08 | 0.17 | -0.07 | 0.05 |
| AQ Car | -0.18 | 0.24 | 0.18 | 0.15 | -0.02 | 0.00 | 0.05 | -0.09 | 0.04 | -0.05 | -0.27 | 0.03 | -0.07 | -0.13 | 0.16 | 0.21 | 0.08 | -0.12 | 0.15 | 0.08 | 0.18 |
| CN Car | -0.03 | 0.45 | 0.27 | 0.42 | 0.18 | 0.10 | 0.20 | 0.14 | 0.27 | 0.10 | -0.02 | 0.21 | 0.12 | 0.04 | 0.34 | 0.45 | 0.24 | 0.06 | 0.28 | 0.04 | 0.43 |
| CY Car | -0.13 | 0.39 | 0.19 | 0.25 | 0.08 | 0.17 | 0.13 | 0.05 | 0.15 | 0.02 | -0.11 | 0.11 | -0.04 | -0.04 | 0.18 | 0.25 | 0.09 | -0.02 | 0.17 | 0.11 | 0.07 |

**Table 4**
Average [x/H] Ratios for Cepheids

| Cepheid | C | N | O | Na | Mg | Al | Si | Ca | Ti | Cr | Mn | Fe | Co | Ni | Y | La | Ce | Pr | Nd | Sm | Eu |
|---|---|---|---|---|---|---|---|---|---|---|---|---|---|---|---|---|---|---|---|---|---|
| DY Car | -0.13 | 0.45 | 0.13 | 0.28 | 0.11 | 0.08 | 0.09 | -0.05 | 0.21 | 0.02 | -0.06 | 0.07 | 0.09 | -0.06 | 0.18 | 0.21 | 0.01 | 0.01 | 0.25 | 0.06 | ... |
| ER Car | -0.23 | 0.16 | 0.15 | 0.22 | 0.21 | 0.17 | 0.13 | 0.05 | 0.15 | 0.04 | -0.15 | 0.15 | 0.01 | -0.04 | 0.17 | 0.30 | 0.18 | -0.11 | 0.20 | -0.13 | 0.14 |
| FI Car | -0.11 | -0.03 | 0.25 | 0.21 | 0.24 | 0.28 | 0.28 | 0.14 | 0.16 | 0.27 | 0.06 | 0.31 | 0.08 | 0.06 | 0.25 | 0.52 | 0.35 | 0.14 | 0.44 | 0.39 | 0.44 |
| FO Car | 0.13 | 0.31 | -0.66 | -0.62 | -0.16 | -0.30 | -0.20 | -1.11 | -1.05 | -0.78 | -1.05 | -0.44 | -0.78 | -0.96 | -0.78 | -1.01 | -1.06 | -0.97 | -0.78 | -0.87 | -0.66 |
| FR Car | -0.51 | 0.44 | 0.20 | 0.14 | 0.00 | 0.11 | 0.08 | -0.08 | 0.07 | -0.07 | -0.21 | 0.11 | -0.05 | -0.12 | 0.20 | 0.30 | 0.17 | 0.01 | 0.21 | 0.14 | 0.34 |
| GH Car | -0.13 | 0.42 | 0.13 | 0.41 | 0.23 | 0.15 | 0.21 | 0.11 | 0.30 | 0.14 | -0.04 | 0.22 | 0.08 | 0.07 | 0.35 | 0.41 | 0.26 | 0.07 | 0.30 | 0.24 | 0.47 |
| GX Car | -0.22 | 0.40 | 0.20 | 0.19 | 0.33 | 0.15 | 0.17 | 0.07 | 0.22 | 0.06 | -0.12 | 0.14 | 0.11 | -0.02 | 0.23 | 0.30 | 0.16 | -0.05 | 0.26 | 0.03 | 0.04 |
| HW Car | -0.18 | 0.30 | 0.19 | 0.24 | -0.04 | 0.09 | 0.05 | -0.08 | 0.02 | -0.04 | -0.24 | 0.09 | -0.08 | -0.13 | 0.15 | 0.21 | 0.11 | -0.06 | 0.17 | 0.09 | 0.04 |
| IO Car | -0.05 | 0.52 | 0.29 | 0.15 | -0.09 | 0.05 | 0.04 | -0.17 | 0.10 | -0.07 | -0.23 | 0.13 | -0.02 | -0.15 | 0.28 | 0.36 | 0.18 | 0.17 | 0.25 | 0.20 | 0.21 |
| IT Car | -0.21 | 0.36 | 0.13 | 0.36 | 0.25 | 0.25 | 0.19 | 0.13 | 0.19 | 0.13 | -0.11 | 0.14 | 0.10 | 0.05 | 0.17 | 0.26 | 0.17 | -0.10 | 0.22 | 0.10 | 0.13 |
| V397 Car | -0.15 | 0.33 | 0.23 | 0.22 | 0.14 | 0.14 | 0.14 | 0.05 | 0.13 | 0.06 | -0.10 | 0.15 | 0.08 | -0.02 | 0.20 | 0.38 | 0.14 | -0.06 | 0.23 | 0.19 | 0.43 |
| V Cen | -0.22 | 0.30 | 0.11 | 0.28 | 0.02 | 0.10 | 0.09 | -0.07 | 0.09 | -0.06 | -0.20 | 0.03 | -0.05 | -0.11 | 0.15 | 0.31 | 0.08 | -0.16 | 0.14 | -0.06 | 0.17 |
| QY Cen | -0.10 | 0.47 | 0.17 | 0.41 | -0.17 | 0.21 | 0.24 | 0.11 | 0.31 | 0.11 | 0.09 | 0.24 | 0.25 | 0.07 | 0.28 | 0.51 | 0.12 | 0.02 | 0.23 | 0.25 | 0.60 |
| XX Cen | -0.17 | 0.39 | 0.19 | 0.24 | 0.10 | 0.21 | 0.17 | 0.01 | 0.18 | 0.03 | -0.11 | 0.18 | 0.03 | -0.01 | 0.28 | 0.28 | 0.17 | 0.06 | 0.25 | 0.17 | 0.02 |
| AY Cen | -0.14 | 0.37 | 0.16 | 0.26 | -0.01 | 0.12 | 0.08 | 0.00 | 0.09 | -0.01 | -0.18 | 0.08 | 0.01 | -0.08 | 0.17 | 0.21 | 0.12 | -0.09 | 0.15 | 0.07 | 0.12 |
| AZ Cen | -0.05 | 0.48 | 0.28 | 0.24 | 0.02 | 0.03 | 0.10 | -0.03 | 0.16 | 0.00 | -0.17 | 0.09 | 0.05 | -0.09 | 0.23 | 0.27 | 0.14 | 0.01 | 0.17 | 0.13 | 0.06 |
| BB Cen | 0.02 | 0.59 | 0.33 | 0.30 | 0.11 | 0.18 | 0.19 | 0.01 | 0.22 | 0.08 | -0.07 | 0.22 | 0.14 | 0.02 | 0.34 | 0.35 | 0.20 | 0.06 | 0.26 | -0.01 | 0.46 |
| KK Cen | -0.05 | 0.36 | 0.18 | 0.52 | 0.18 | 0.27 | 0.24 | 0.16 | 0.26 | 0.17 | -0.01 | 0.24 | 0.12 | 0.10 | 0.31 | 0.35 | 0.25 | 0.01 | 0.27 | 0.23 | 0.29 |
| KN Cen | -0.04 | 0.42 | 0.33 | 0.55 | ... | 0.45 | 0.42 | 0.31 | 0.20 | 0.25 | 0.12 | 0.41 | 0.21 | 0.31 | 0.22 | 0.33 | 0.08 | -0.04 | 0.16 | 0.39 | 0.41 |
| MZ Cen | 0.03 | 0.52 | 0.27 | 0.36 | 0.16 | 0.28 | 0.30 | 0.11 | 0.27 | 0.19 | 0.07 | 0.35 | 0.21 | 0.14 | 0.33 | 0.41 | 0.16 | 0.08 | 0.30 | 0.15 | 0.55 |
| V339 Cen | -0.05 | 0.40 | 0.25 | 0.29 | 0.06 | 0.16 | 0.12 | 0.02 | 0.13 | 0.04 | -0.12 | 0.14 | -0.03 | -0.03 | 0.21 | 0.23 | 0.06 | -0.08 | 0.17 | -0.13 | 0.14 |
| V378 Cen | -0.18 | 0.46 | 0.23 | 0.25 | -0.05 | 0.06 | 0.07 | -0.09 | 0.09 | -0.04 | -0.21 | 0.08 | 0.06 | -0.10 | 0.23 | 0.28 | 0.12 | -0.03 | 0.22 | 0.17 | 0.44 |
| V381 Cen | -0.23 | 0.25 | 0.11 | 0.20 | 0.00 | 0.08 | 0.08 | -0.03 | 0.10 | -0.04 | -0.16 | 0.02 | 0.04 | -0.08 | 0.11 | 0.20 | 0.03 | -0.13 | 0.01 | -0.02 | -0.06 |
| V419 Cen | -0.22 | 0.25 | 0.12 | 0.36 | 0.42 | 0.33 | 0.20 | 0.16 | 0.29 | 0.05 | 0.04 | 0.14 | 0.16 | 0.03 | 0.25 | 0.50 | 0.24 | -0.25 | 0.28 | -0.01 | 0.18 |
| V496 Cen | -0.15 | 0.38 | 0.21 | 0.36 | 0.12 | 0.13 | 0.14 | 0.06 | 0.20 | 0.06 | -0.09 | 0.09 | 0.12 | -0.02 | 0.20 | 0.28 | 0.16 | -0.06 | 0.15 | 0.09 | 0.13 |
| V659 Cen | -0.20 | 0.27 | 0.19 | 0.19 | 0.00 | 0.14 | 0.09 | -0.05 | 0.09 | -0.06 | -0.17 | 0.09 | -0.07 | -0.11 | 0.20 | 0.34 | 0.14 | -0.15 | 0.17 | 0.12 | 0.10 |
| V737 Cen | -0.12 | 0.35 | 0.23 | 0.25 | 0.01 | 0.16 | 0.11 | -0.04 | 0.10 | 0.01 | -0.18 | 0.14 | -0.01 | -0.06 | 0.20 | 0.21 | 0.12 | -0.10 | 0.16 | 0.10 | 0.15 |
| AV Cir | -0.01 | 0.47 | 0.33 | 0.21 | 0.09 | 0.13 | 0.15 | -0.01 | 0.17 | 0.02 | -0.13 | 0.17 | 0.09 | -0.04 | 0.32 | 0.32 | 0.21 | 0.11 | 0.24 | 0.19 | 0.13 |
| AX Cir | -0.33 | 0.20 | 0.11 | 0.17 | -0.08 | 0.03 | 0.01 | -0.12 | -0.06 | -0.12 | -0.31 | -0.01 | -0.13 | -0.20 | 0.05 | 0.26 | 0.06 | -0.20 | 0.09 | 0.00 | 0.06 |
| BP Cir | -0.17 | 0.35 | 0.16 | 0.12 | 0.11 | -0.05 | 0.10 | -0.06 | 0.13 | -0.07 | -0.19 | 0.02 | 0.02 | -0.13 | 0.14 | 0.22 | 0.03 | -0.17 | 0.03 | -0.06 | -0.15 |
| R Cru | -0.09 | 0.35 | 0.22 | 0.23 | 0.09 | 0.14 | 0.14 | 0.05 | 0.17 | 0.03 | -0.14 | 0.13 | 0.07 | -0.03 | 0.24 | 0.24 | 0.13 | -0.04 | 0.18 | -0.12 | 0.26 |
| S Cru | -0.12 | 0.43 | 0.19 | 0.23 | 0.15 | 0.10 | 0.16 | 0.07 | 0.18 | 0.02 | -0.10 | 0.11 | 0.17 | -0.03 | 0.20 | 0.32 | 0.11 | -0.14 | 0.11 | 0.10 | -0.05 |
| T Cru | -0.14 | 0.30 | 0.18 | 0.22 | 0.02 | 0.17 | 0.11 | -0.01 | 0.12 | 0.02 | -0.18 | 0.14 | -0.02 | -0.07 | 0.18 | 0.29 | 0.10 | -0.06 | 0.15 | 0.11 | 0.01 |
| X Cru | -0.17 | 0.39 | 0.25 | 0.36 | 0.18 | 0.22 | 0.16 | 0.08 | 0.13 | 0.05 | -0.11 | 0.15 | -0.03 | -0.01 | 0.16 | 0.29 | 0.14 | -0.07 | 0.16 | 0.11 | 0.00 |
| VW Cru | -0.16 | 0.26 | 0.17 | 0.29 | 0.07 | 0.18 | 0.15 | 0.01 | 0.13 | 0.09 | -0.09 | 0.19 | 0.01 | -0.01 | 0.26 | 0.31 | 0.18 | 0.00 | 0.25 | 0.25 | 0.38 |
| AD Cru | -0.17 | 0.34 | 0.22 | 0.31 | 0.24 | 0.07 | 0.14 | 0.09 | 0.15 | 0.05 | -0.15 | 0.11 | -0.01 | -0.03 | 0.16 | 0.26 | 0.14 | -0.03 | 0.19 | -0.09 | 0.40 |
| AG Cru | -0.11 | 0.52 | 0.22 | 0.31 | 0.14 | -0.03 | 0.14 | 0.04 | 0.15 | 0.02 | -0.07 | 0.08 | 0.08 | -0.04 | 0.16 | 0.26 | 0.01 | -0.07 | 0.08 | -0.01 | -0.03 |

**Table 4**
Average [x/H] Ratios for Cepheids

| Cepheid | C | N | O | Na | Mg | Al | Si | Ca | Ti | Cr | Mn | Fe | Co | Ni | Y | La | Ce | Pr | Nd | Sm | Eu |
|---|---|---|---|---|---|---|---|---|---|---|---|---|---|---|---|---|---|---|---|---|---|
| BG Cru | -0.20 | 0.47 | 0.33 | -0.13 | 0.30 | 0.13 | -0.07 | -0.16 | 0.03 | -0.21 | -0.27 | -0.08 | 0.11 | -0.26 | 0.10 | 0.47 | 0.09 | -0.50 | 0.28 | ... | ... |
| GH Lup | -0.16 | 0.18 | 0.18 | 0.25 | 0.12 | 0.23 | 0.15 | -0.01 | 0.02 | 0.05 | -0.16 | 0.13 | -0.04 | -0.07 | 0.13 | 0.23 | 0.10 | -0.17 | 0.16 | 0.03 | 0.02 |
| R Mus | -0.15 | 0.34 | 0.23 | 0.37 | 0.23 | 0.18 | 0.14 | 0.03 | 0.15 | 0.06 | -0.10 | 0.15 | 0.08 | 0.00 | 0.24 | 0.29 | 0.15 | -0.01 | 0.20 | 0.14 | 0.31 |
| S Mus | -0.22 | 0.26 | 0.13 | 0.29 | 0.03 | 0.15 | 0.08 | -0.04 | 0.04 | -0.04 | -0.21 | 0.07 | -0.04 | -0.12 | 0.15 | 0.23 | 0.11 | -0.13 | 0.18 | 0.08 | 0.14 |
| RT Mus | -0.14 | 0.42 | 0.25 | 0.20 | 0.04 | 0.07 | 0.14 | 0.03 | 0.13 | 0.00 | -0.17 | 0.12 | -0.02 | -0.08 | 0.23 | 0.24 | 0.14 | -0.02 | 0.19 | 0.02 | -0.09 |
| TZ Mus | -0.19 | 0.37 | 0.15 | 0.14 | 0.43 | 0.05 | 0.17 | 0.05 | 0.21 | 0.07 | -0.09 | 0.10 | 0.25 | 0.01 | 0.18 | 0.53 | 0.09 | 0.12 | 0.08 | 0.09 | 0.17 |
| UU Mus | -0.22 | 0.37 | 0.12 | 0.27 | 0.19 | 0.16 | 0.17 | 0.06 | 0.25 | 0.10 | -0.10 | 0.19 | 0.13 | 0.01 | 0.34 | 0.48 | 0.24 | -0.06 | 0.27 | 0.25 | 0.50 |
| S Nor | -0.12 | 0.38 | 0.21 | 0.35 | 0.14 | 0.20 | 0.16 | 0.05 | 0.16 | 0.09 | -0.11 | 0.13 | 0.01 | 0.01 | 0.21 | 0.27 | 0.11 | -0.16 | 0.19 | 0.10 | 0.08 |
| U Nor | 0.09 | 0.56 | 0.40 | 0.31 | 0.08 | 0.15 | 0.14 | -0.06 | 0.04 | 0.02 | -0.15 | 0.19 | -0.03 | -0.06 | 0.23 | 0.24 | 0.15 | 0.02 | 0.25 | 0.22 | 0.15 |
| SY Nor | 0.13 | 0.58 | 0.44 | 0.27 | 0.15 | 0.30 | 0.27 | 0.03 | 0.19 | 0.16 | 0.01 | 0.34 | 0.10 | 0.10 | 0.33 | 0.27 | 0.21 | 0.10 | 0.29 | 0.23 | 0.33 |
| TW Nor | 0.05 | 0.48 | 0.29 | 0.38 | 0.27 | 0.31 | 0.35 | 0.21 | 0.34 | 0.26 | 0.16 | 0.33 | 0.31 | 0.18 | 0.34 | 0.53 | 0.20 | 0.09 | 0.29 | 0.27 | 0.37 |
| GU Nor | 0.06 | 0.51 | 0.34 | 0.35 | 0.34 | 0.20 | 0.24 | 0.10 | 0.23 | 0.13 | -0.03 | 0.27 | 0.15 | 0.07 | 0.30 | 0.28 | 0.18 | -0.02 | 0.24 | 0.17 | 0.33 |
| V340 Nor | -0.08 | 0.34 | 0.27 | 0.32 | 0.12 | 0.26 | 0.16 | 0.06 | 0.13 | 0.08 | -0.01 | 0.16 | 0.02 | -0.01 | 0.15 | 0.36 | 0.06 | -0.33 | 0.12 | 0.00 | 0.26 |
| BF Oph | -0.13 | 0.44 | 0.21 | 0.31 | 0.05 | 0.07 | 0.18 | 0.04 | 0.18 | 0.05 | -0.12 | 0.14 | 0.04 | -0.01 | 0.24 | 0.25 | 0.14 | -0.05 | 0.18 | 0.09 | -0.02 |
| AP Pup | -0.20 | 0.30 | 0.12 | 0.12 | 0.17 | 0.10 | 0.11 | 0.00 | 0.15 | -0.02 | -0.18 | 0.08 | 0.02 | -0.08 | 0.22 | 0.31 | 0.13 | -0.11 | 0.21 | 0.14 | 0.02 |
| AT Pup | -0.22 | 0.40 | 0.10 | 0.25 | 0.04 | -0.04 | 0.08 | -0.07 | 0.17 | -0.02 | -0.18 | 0.05 | 0.04 | -0.08 | 0.19 | 0.29 | 0.18 | 0.11 | 0.19 | -0.11 | 0.56 |
| CE Pup | -0.87 | 0.44 | 0.06 | 0.18 | 0.31 | 0.07 | 0.02 | -0.08 | 0.03 | -0.11 | -0.21 | 0.03 | -0.12 | -0.17 | 0.08 | 0.25 | 0.11 | -0.09 | 0.21 | 0.17 | 0.15 |
| MY Pup | -0.11 | 0.43 | 0.28 | 0.15 | 0.04 | -0.06 | 0.00 | -0.17 | 0.08 | -0.09 | -0.24 | 0.04 | -0.07 | -0.17 | 0.26 | 0.36 | 0.20 | 0.06 | 0.25 | -0.04 | 0.29 |
| NT Pup | -1.30 | 0.55 | -0.08 | 0.34 | -0.15 | 0.01 | -0.05 | -0.20 | -0.09 | -0.15 | -0.38 | -0.02 | -0.21 | -0.23 | 0.03 | 0.21 | 0.09 | -0.09 | 0.19 | 0.12 | 0.14 |
| RV Sco | -0.25 | 0.40 | 0.09 | 0.41 | 0.01 | 0.20 | 0.16 | 0.11 | 0.20 | 0.05 | -0.04 | 0.11 | 0.10 | 0.01 | 0.13 | 0.26 | 0.10 | -0.12 | 0.10 | -0.19 | 0.27 |
| V482 Sco | -0.08 | 0.46 | 0.23 | 0.38 | 0.15 | 0.23 | 0.19 | 0.08 | 0.20 | 0.08 | -0.09 | 0.20 | 0.14 | 0.04 | 0.27 | 0.32 | 0.20 | 0.05 | 0.25 | 0.19 | 0.35 |
| V636 Sco | -0.10 | 0.33 | 0.23 | 0.31 | 0.09 | 0.29 | 0.11 | -0.02 | 0.01 | 0.06 | -0.06 | 0.10 | -0.11 | -0.09 | 0.14 | 0.28 | 0.14 | -0.14 | 0.08 | -0.25 | 0.02 |
| V950 Sco | -0.02 | 0.53 | 0.31 | 0.27 | 0.06 | 0.11 | 0.15 | -0.02 | 0.18 | 0.06 | -0.12 | 0.21 | 0.07 | -0.03 | 0.33 | 0.35 | 0.20 | 0.06 | 0.26 | 0.19 | 0.42 |
| R TrA | -0.26 | 0.24 | 0.23 | 0.49 | 0.21 | 0.12 | 0.23 | 0.10 | 0.22 | 0.11 | 0.02 | 0.19 | 0.05 | -0.01 | 0.29 | 0.51 | 0.29 | -0.10 | 0.32 | 0.31 | 0.47 |
| S TrA | -0.15 | 0.36 | 0.22 | 0.41 | 0.28 | 0.16 | 0.14 | -0.03 | 0.14 | 0.03 | -0.13 | 0.21 | 0.06 | -0.03 | 0.28 | 0.35 | 0.17 | 0.03 | 0.21 | 0.20 | 0.30 |
| LR TrA | -0.15 | 0.19 | 0.21 | 0.15 | 0.47 | 0.36 | 0.28 | 0.22 | 0.28 | 0.13 | 0.11 | 0.31 | 0.24 | 0.14 | 0.35 | 0.56 | 0.16 | -0.04 | 0.27 | 0.25 | 0.30 |
| T Vel | -0.35 | 0.17 | 0.07 | 0.19 | 0.04 | 0.16 | 0.12 | 0.06 | 0.09 | -0.01 | -0.15 | 0.04 | -0.11 | -0.08 | 0.10 | 0.43 | 0.18 | -0.17 | 0.15 | 0.06 | -0.06 |
| V Vel | -0.38 | 0.11 | -0.01 | 0.09 | -0.12 | -0.11 | 0.00 | -0.05 | 0.06 | -0.11 | -0.29 | -0.08 | -0.15 | -0.18 | 0.02 | 0.22 | 0.02 | -0.08 | 0.05 | -0.03 | ... |
| RY Vel | -0.05 | 0.40 | 0.29 | 0.20 | 0.01 | 0.12 | 0.10 | -0.13 | 0.02 | 0.02 | -0.21 | 0.09 | -0.09 | -0.13 | 0.18 | 0.29 | 0.15 | 0.04 | 0.19 | 0.23 | 0.27 |
| RZ Vel | -0.33 | 0.18 | 0.09 | 0.35 | 0.42 | 0.45 | 0.19 | 0.02 | 0.11 | 0.25 | -0.06 | 0.04 | -0.03 | 0.04 | 0.12 | 0.33 | 0.00 | -0.22 | 0.15 | 0.28 | 0.23 |
| ST Vel | -0.17 | 0.32 | 0.09 | 0.28 | -0.06 | 0.08 | 0.14 | 0.07 | 0.18 | 0.04 | -0.10 | 0.05 | 0.03 | -0.03 | 0.16 | 0.13 | 0.06 | -0.10 | 0.10 | 0.03 | 0.12 |
| SV Vel | -0.18 | 0.41 | 0.08 | 0.39 | 0.11 | 0.10 | 0.17 | 0.10 | 0.20 | 0.09 | -0.09 | 0.12 | 0.08 | -0.01 | 0.20 | 0.27 | 0.13 | -0.15 | 0.19 | 0.08 | 0.17 |
| SW Vel | -0.46 | 0.49 | 0.04 | 0.21 | 0.17 | 0.10 | 0.13 | 0.11 | 0.27 | -0.03 | -0.09 | 0.00 | 0.24 | -0.02 | 0.09 | 0.50 | 0.11 | 0.07 | 0.23 | 1.13 | 0.21 |
| SX Vel | -0.15 | 0.37 | 0.21 | 0.16 | -0.03 | 0.07 | 0.08 | -0.06 | 0.14 | -0.03 | -0.20 | 0.06 | -0.04 | -0.07 | 0.20 | 0.27 | 0.11 | -0.04 | 0.19 | 0.09 | 0.13 |
| XX Vel | -0.08 | 0.46 | 0.21 | 0.21 | 0.11 | 0.17 | 0.14 | 0.00 | 0.24 | 0.03 | -0.07 | 0.11 | 0.25 | 0.03 | 0.27 | 0.30 | 0.20 | -0.05 | 0.27 | 0.15 | 0.48 |
| AE Vel | -0.22 | 0.39 | 0.24 | 0.14 | 0.06 | 0.20 | 0.03 | -0.20 | -0.04 | -0.12 | -0.27 | 0.14 | -0.13 | -0.19 | 0.13 | 0.31 | 0.14 | -0.04 | 0.20 | -0.05 | 0.26 |
| AH Vel | -0.25 | 0.52 | 0.20 | 0.35 | 0.10 | 0.18 | 0.15 | 0.02 | 0.20 | 0.05 | -0.11 | 0.19 | 0.11 | -0.02 | 0.33 | 0.52 | 0.32 | 0.14 | 0.38 | 0.26 | 0.35 |

**Table 4**
Average [x/H] Ratios for Cepheids

| Cepheid | C | N | O | Na | Mg | Al | Si | Ca | Ti | Cr | Mn | Fe | Co | Ni | Y | La | Ce | Pr | Nd | Sm | Eu |
|---|---|---|---|---|---|---|---|---|---|---|---|---|---|---|---|---|---|---|---|---|---|
| BG Vel | -0.37 | 0.13 | 0.00 | 0.21 | 0.10 | 0.07 | 0.03 | -0.09 | 0.02 | -0.07 | -0.22 | 0.03 | -0.09 | -0.13 | 0.07 | 0.15 | 0.04 | -0.27 | 0.06 | 0.02 | -0.08 |
| CS Vel | -0.08 | 0.21 | 0.29 | 0.27 | 0.08 | 0.26 | 0.12 | -0.08 | 0.04 | 0.00 | -0.20 | 0.12 | -0.06 | -0.10 | 0.21 | 0.38 | 0.20 | 0.07 | 0.27 | -0.01 | 0.08 |
| CX Vel | -0.14 | 0.46 | 0.15 | 0.33 | 0.52 | 0.21 | 0.17 | 0.10 | 0.25 | 0.07 | -0.04 | 0.16 | 0.09 | 0.03 | 0.18 | 0.21 | 0.07 | -0.20 | 0.16 | 0.06 | ... |
| DK Vel | 0.01 | 0.48 | 0.29 | 0.21 | 0.10 | 0.11 | 0.15 | 0.02 | 0.23 | 0.07 | -0.07 | 0.18 | 0.07 | -0.02 | 0.30 | 0.39 | 0.19 | 0.16 | 0.27 | -0.06 | 0.22 |
| DR Vel | -0.12 | 0.20 | 0.22 | 0.36 | 0.13 | 0.24 | 0.18 | 0.08 | 0.12 | 0.13 | -0.09 | 0.18 | -0.03 | 0.02 | 0.24 | 0.28 | 0.26 | -0.02 | 0.31 | 0.20 | 0.26 |
| EX Vel | -0.25 | 0.39 | 0.12 | 0.15 | 0.07 | 0.15 | 0.08 | 0.01 | 0.11 | 0.00 | -0.21 | 0.07 | 0.05 | -0.07 | 0.14 | 0.28 | 0.13 | -0.06 | 0.21 | 0.13 | 0.23 |
| FG Vel | -0.30 | 0.11 | 0.16 | 0.17 | 0.01 | 0.15 | 0.02 | -0.18 | -0.06 | -0.09 | -0.32 | 0.02 | -0.20 | -0.20 | 0.05 | 0.27 | 0.02 | -0.15 | 0.14 | -0.23 | 0.05 |
| FN Vel | -0.21 | 0.19 | 0.09 | 0.14 | 0.23 | 0.15 | 0.14 | 0.09 | 0.14 | 0.05 | -0.12 | 0.15 | 0.03 | -0.02 | 0.11 | 0.27 | 0.11 | -0.12 | 0.17 | 0.13 | 0.27 |

**Table 5**
Abundance Gradients: Species = a* $R_G$ + b

| Species | Gradient a | Gradient b | Uncertainty a | Uncertainty b | σ | N |
|---:|---:|---:|---:|---:|---:|---:|
| [C/H]         | -0.080 |  0.469 | 0.004 | 0.039 | 0.166 | 313 |
| [N/H]         | -0.049 |  0.736 | 0.004 | 0.037 | 0.158 | 309 |
| [O/H]         | -0.056 |  0.613 | 0.003 | 0.028 | 0.118 | 313 |
| [(C+N )/H]    | -0.060 |  0.518 | 0.003 | 0.029 | 0.124 | 309 |
| [(C+N+O)/H]   | -0.057 |  0.578 | 0.003 | 0.025 | 0.106 | 309 |
| [C/Fe]        | -0.020 | -0.138 | 0.004 | 0.034 | 0.147 | 313 |
| [N/Fe]        |  0.011 |  0.137 | 0.004 | 0.035 | 0.147 | 309 |
| [O/Fe]        |  0.005 |  0.007 | 0.003 | 0.028 | 0.119 | 313 |
| [(C+N)/Fe]    |  0.000 | -0.081 | 0.003 | 0.024 | 0.102 | 309 |
| [[(C+N+O)/Fe] |  0.003 | -0.021 | 0.002 | 0.022 | 0.095 | 309 |
| C/O           | -0.014 |  0.391 | 0.003 | 0.025 | 0.106 | 313 |
| [Na/H]        | -0.047 |  0.647 | 0.003 | 0.031 | 0.131 | 312 |
| [Mg/H]        | -0.048 |  0.496 | 0.004 | 0.038 | 0.160 | 303 |
| [Al/H]        | -0.049 |  0.555 | 0.003 | 0.030 | 0.130 | 313 |
| [Si/H]        | -0.048 |  0.525 | 0.002 | 0.020 | 0.088 | 313 |
| [Ca/H]        | -0.041 |  0.354 | 0.003 | 0.026 | 0.114 | 313 |
| [Ti/H]        | -0.039 |  0.456 | 0.003 | 0.025 | 0.108 | 313 |
| [Cr/H]        | -0.048 |  0.427 | 0.003 | 0.024 | 0.104 | 313 |
| [Mn/H]        | -0.052 |  0.303 | 0.003 | 0.028 | 0.122 | 313 |
| [Fe/H]        | -0.061 |  0.607 | 0.003 | 0.024 | 0.104 | 313 |
| [Co/H]        | -0.034 |  0.305 | 0.003 | 0.030 | 0.127 | 313 |
| [Ni/H]        | -0.047 |  0.346 | 0.002 | 0.023 | 0.097 | 313 |
| [Y/H]         | -0.061 |  0.678 | 0.003 | 0.027 | 0.116 | 313 |
| [La/H]        | -0.031 |  0.560 | 0.004 | 0.034 | 0.144 | 312 |
| [Ce/H]        | -0.034 |  0.430 | 0.003 | 0.031 | 0.133 | 313 |
| [Pr/H]        | -0.040 |  0.242 | 0.004 | 0.033 | 0.143 | 311 |
| [Nd/H]        | -0.037 |  0.504 | 0.003 | 0.027 | 0.114 | 313 |
| [Sm/H]        | -0.035 |  0.408 | 0.005 | 0.043 | 0.185 | 306 |
| [Eu/H]        | -0.042 |  0.516 | 0.005 | 0.042 | 0.171 | 293 |

Note: σ is the standard deviation of the fit using N stars.

**Figures**

Figure 1: Top Panel - [Fe/H] versus galactocentric radius for the 403 Cepheid variables from this and preceding studies. Note the 5 discrepant stars from this analysis which are discussed in §4.3.1 and which are shown as (magenta) squares. Bottom Panel - The gradient in [Fe/H] for the 398 Cepheid variables from this and preceding studies. The blue (straight) line is a simple least squares fit to the entire dataset which yields a gradient $d[Fe/H]/dR_G$ = -0.062 dex kpc$^{-1}$. The confidence intervals shown are 95%. The red (not straight) line is a LOWESS fit which uses localized weighting. Note that the distance uncertainties (two representative examples shown) vary with $R_G$ with the error minimizing at $R_G$ = 7.9 kpc – the solar galactocentric radius.

Figure 2: The distribution of program Cepheids in the galactic plane. The key is the same as Figure 1.

Figure 3: Distance above or below the plane versus galactocentric radius. The stars at large distances from the plane do not perturb the derived gradients. The key is the same as Figure 1.

Figure 4: The azimuthal variation in [Fe/H] for several annuli at varying galactocentric radius. There is no discernible dependence on galactocentric angle.

Figure 5: A contour map of the [Fe/H] abundances versus spatial position. The salient feature of this plot is the smooth variation of the [Fe/H] ratio with position.

Figure 6: The [Fe/H], [C/Fe], [O/Fe], and [Na/Fe] ratios versus log(Period). There is no discernible dependence of the ratios on log(P) which can be interpreted as the degree of processing and mixing is comparable in all of these objects.

Figure 7: The gradient in [C/H] and [C/Fe]. Note that the gradient in [C/H] is steeper than that in [Fe/H] which manifests in the non-zero slope in [C/Fe]. For the [C/Fe] data we also show the 95% confidence interval.

Figure 8. The gradient in [N/H] and [N/Fe]. For the [N/Fe] data we also show the 95% confidence interval.

Figure 9: The gradient in [(C+N)/H] and [(C+N)/Fe]. For the [(C+N)/Fe] data we also show the 95% confidence interval. Note that the data beyond 12 kpc shows an increased scatter relative to the inner region.

Figure 10: [O/H] and [O/Fe] versus galactocentric distance. The scatter in oxygen is moderate but the gradient is comparable to that found in B stars. For the [O/Fe] data we also show the 95% confidence interval.

Figure 11: The gradient in [(C+N+O)/H] and [(C+N+O)/Fe]. For the [(C+N+O)/Fe] data we also show the 95% confidence interval. Note that the data beyond 12 kpc may show an increased scatter relative to the inner region and perhaps a somewhat larger mean abundance.

Figure 12: The gradient data for [Si/H], [Ca/H], [Y/H], [Nd/H], and [Eu/H]. All of these species show a significant gradient.

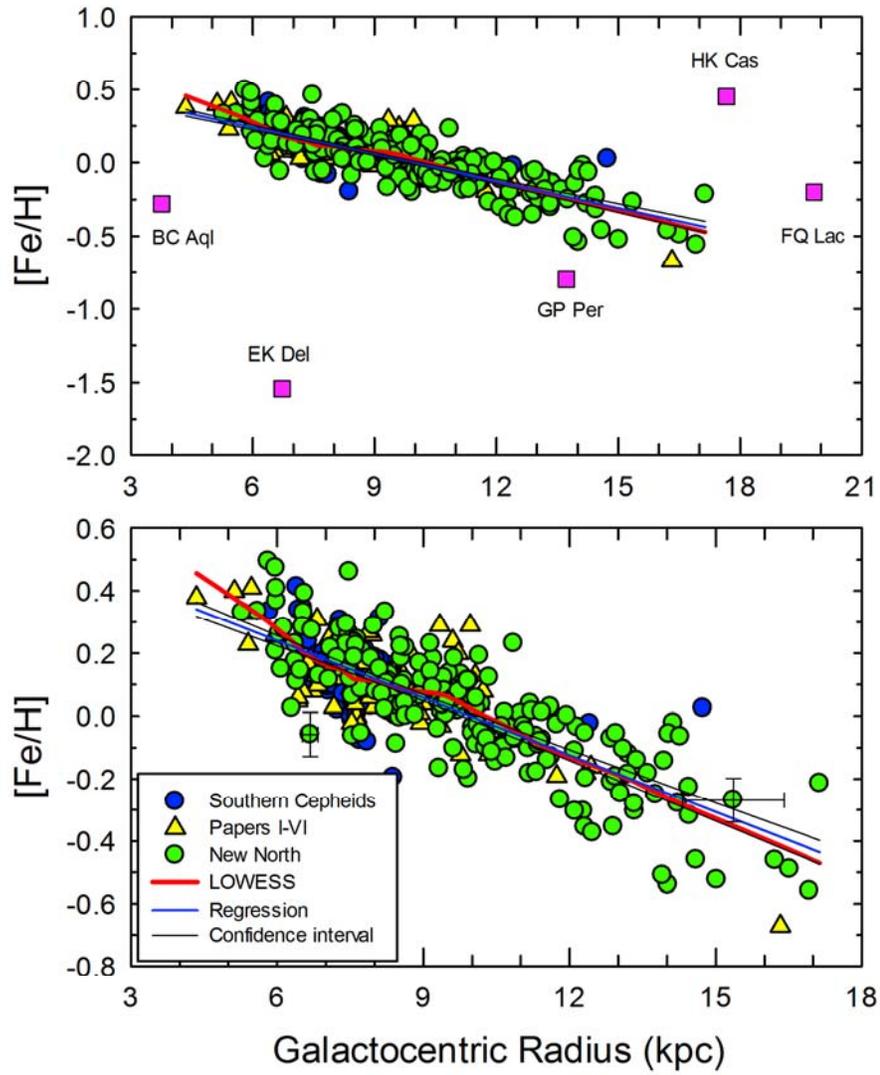

**Figure 1**

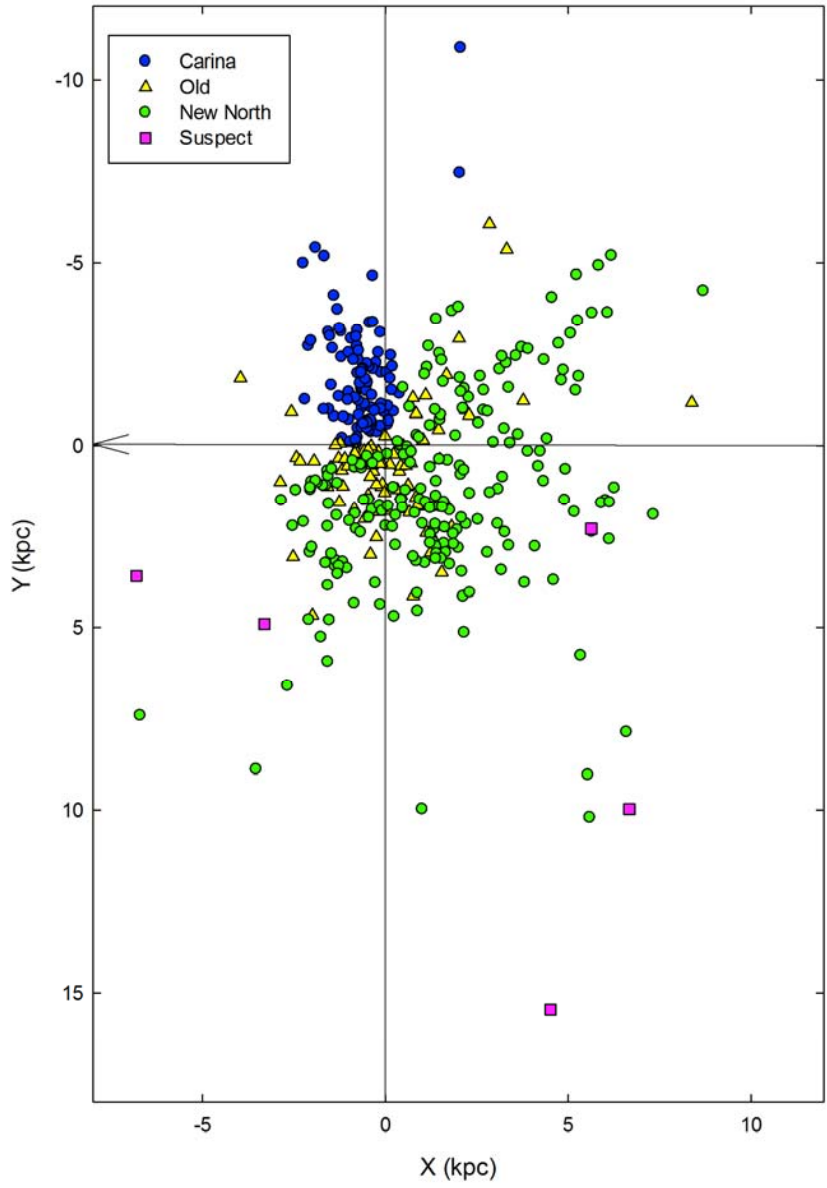

**Figure 2**

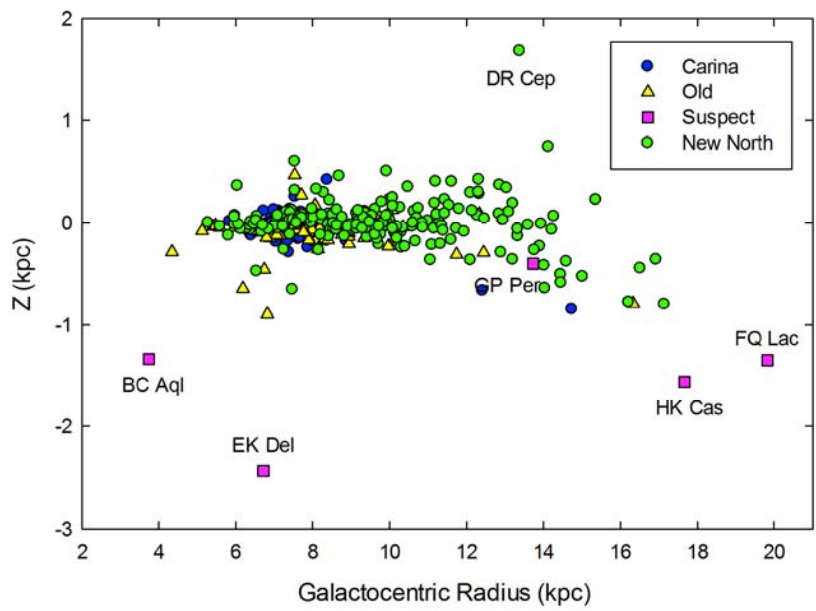

**Figure 3**

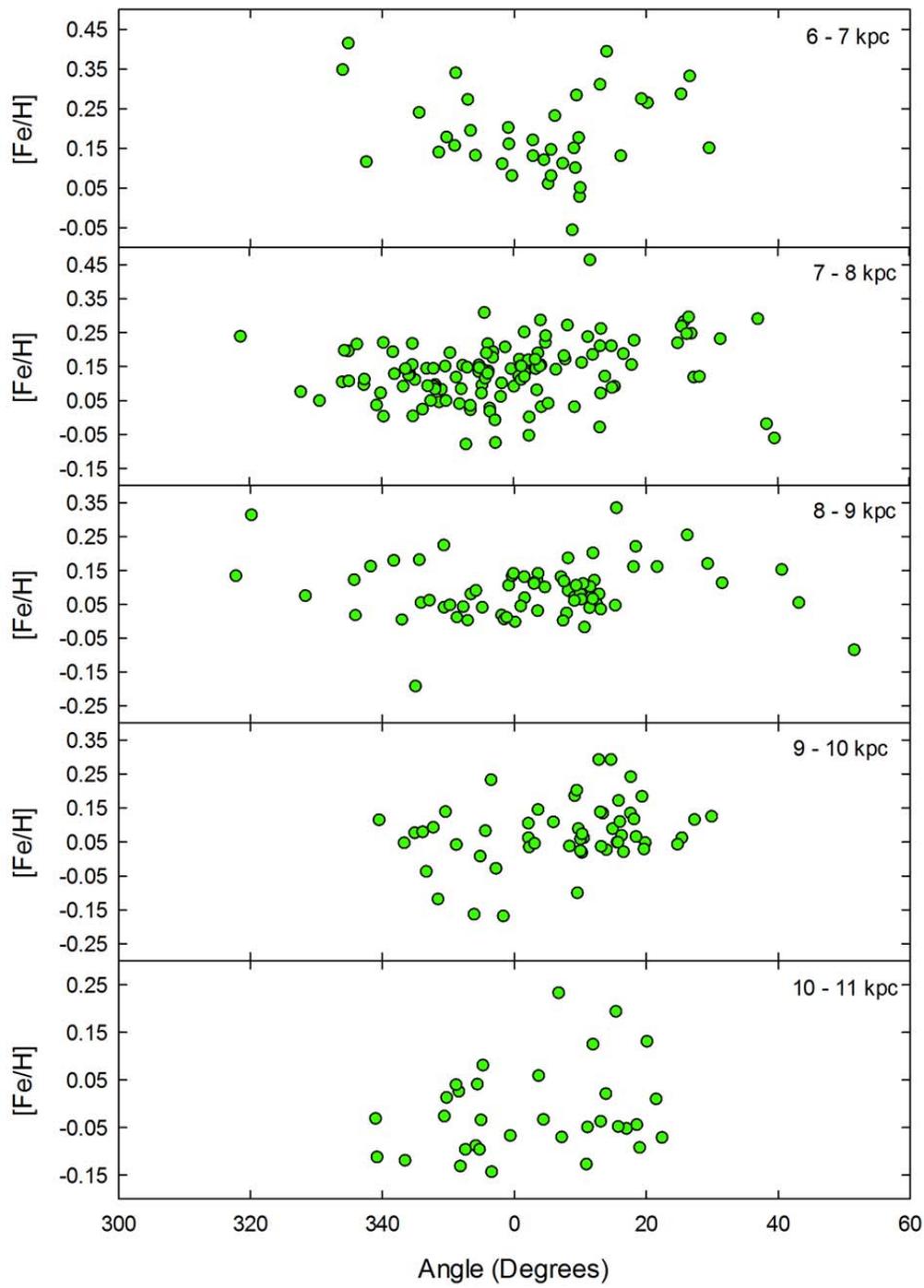

Figure 4

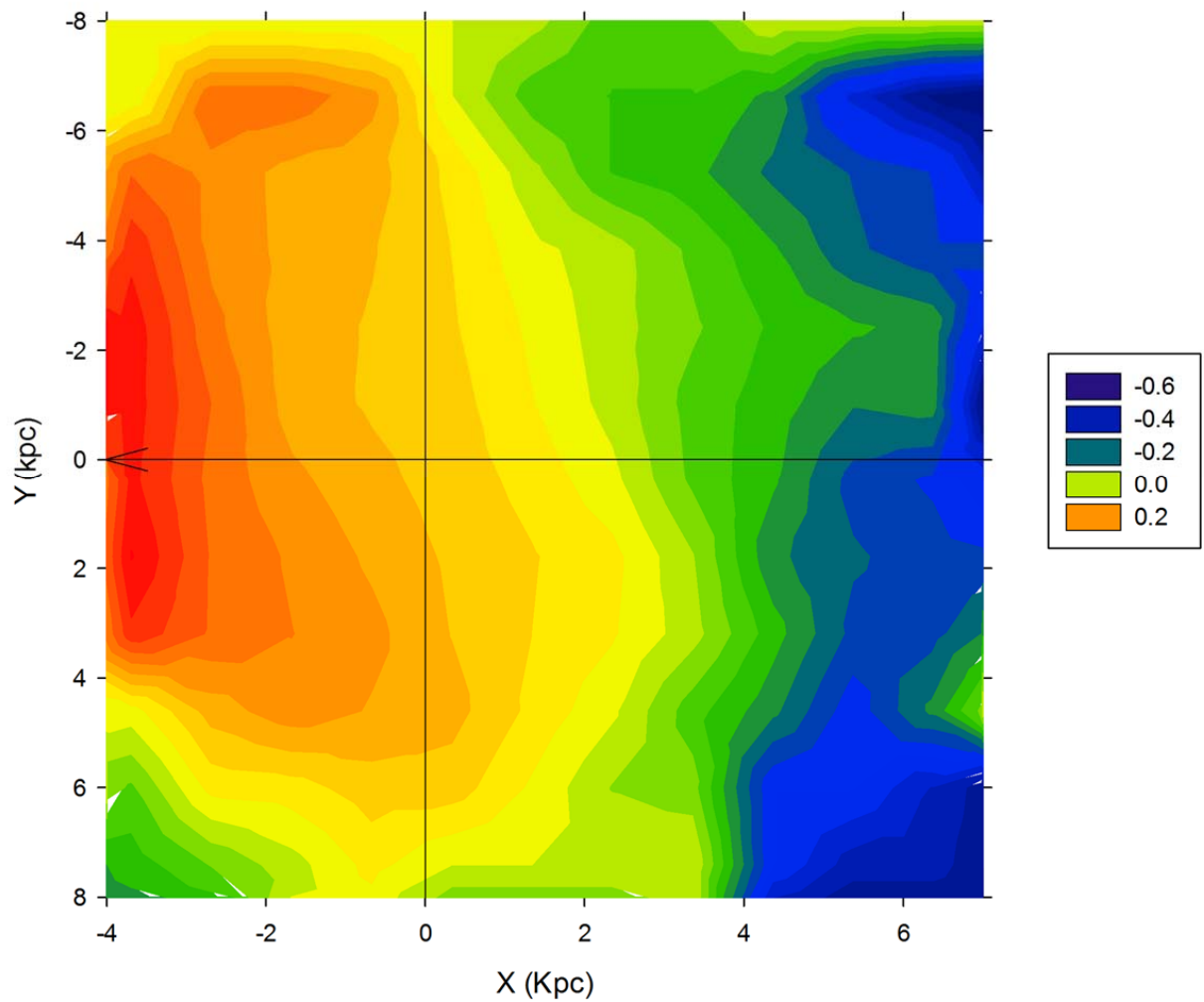

Figure 5

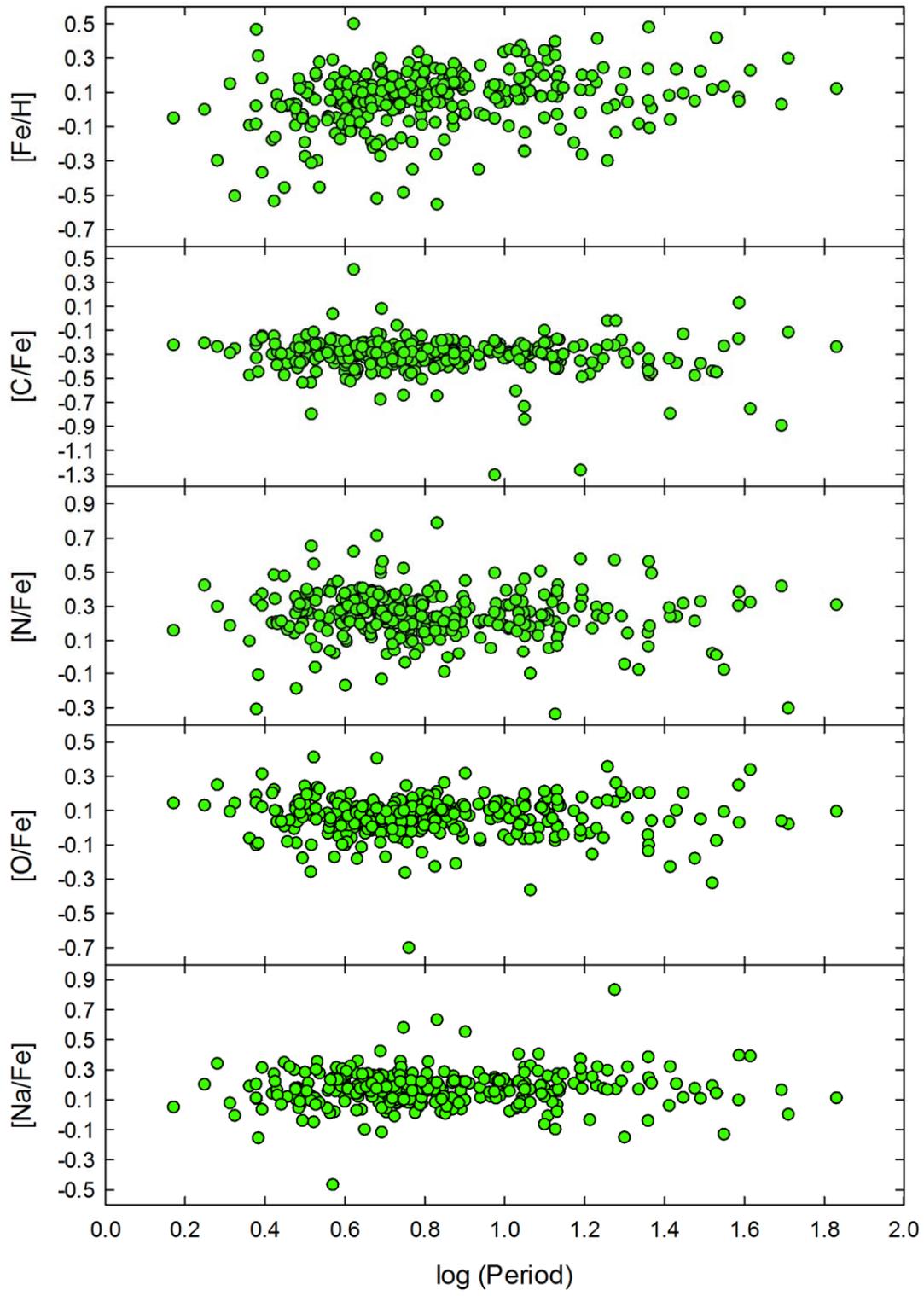

Figure 6

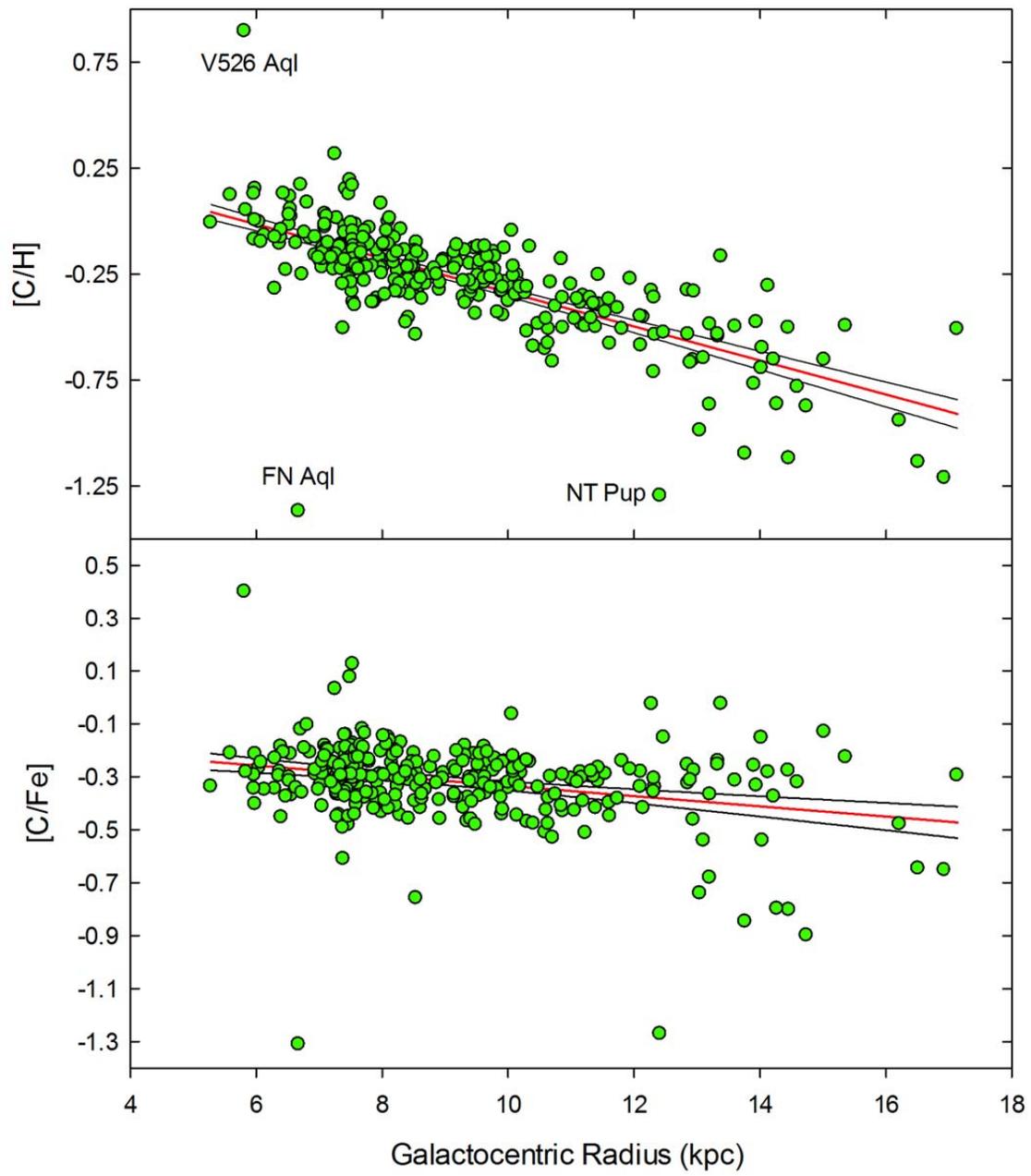

Figure 7

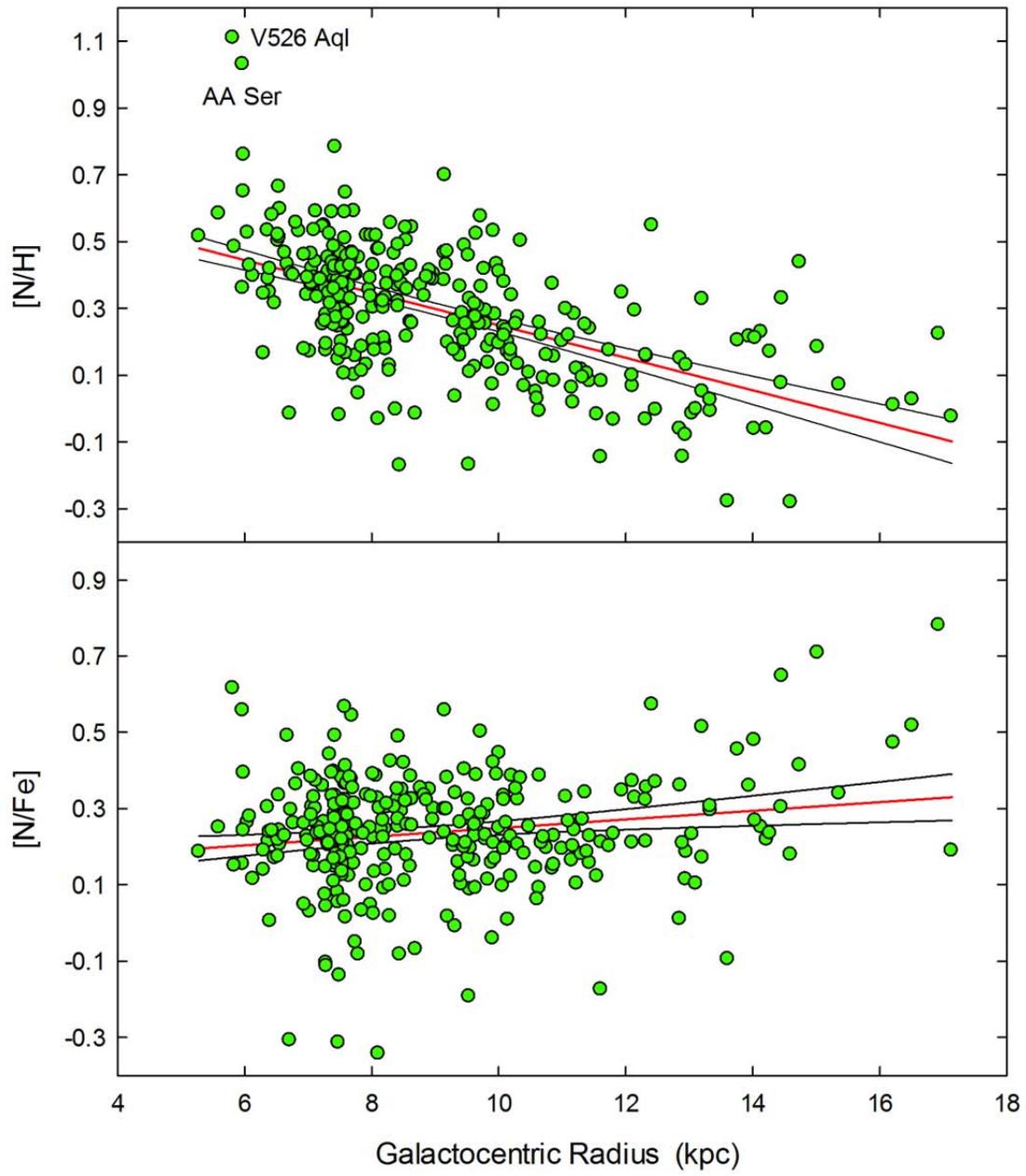

Figure 8

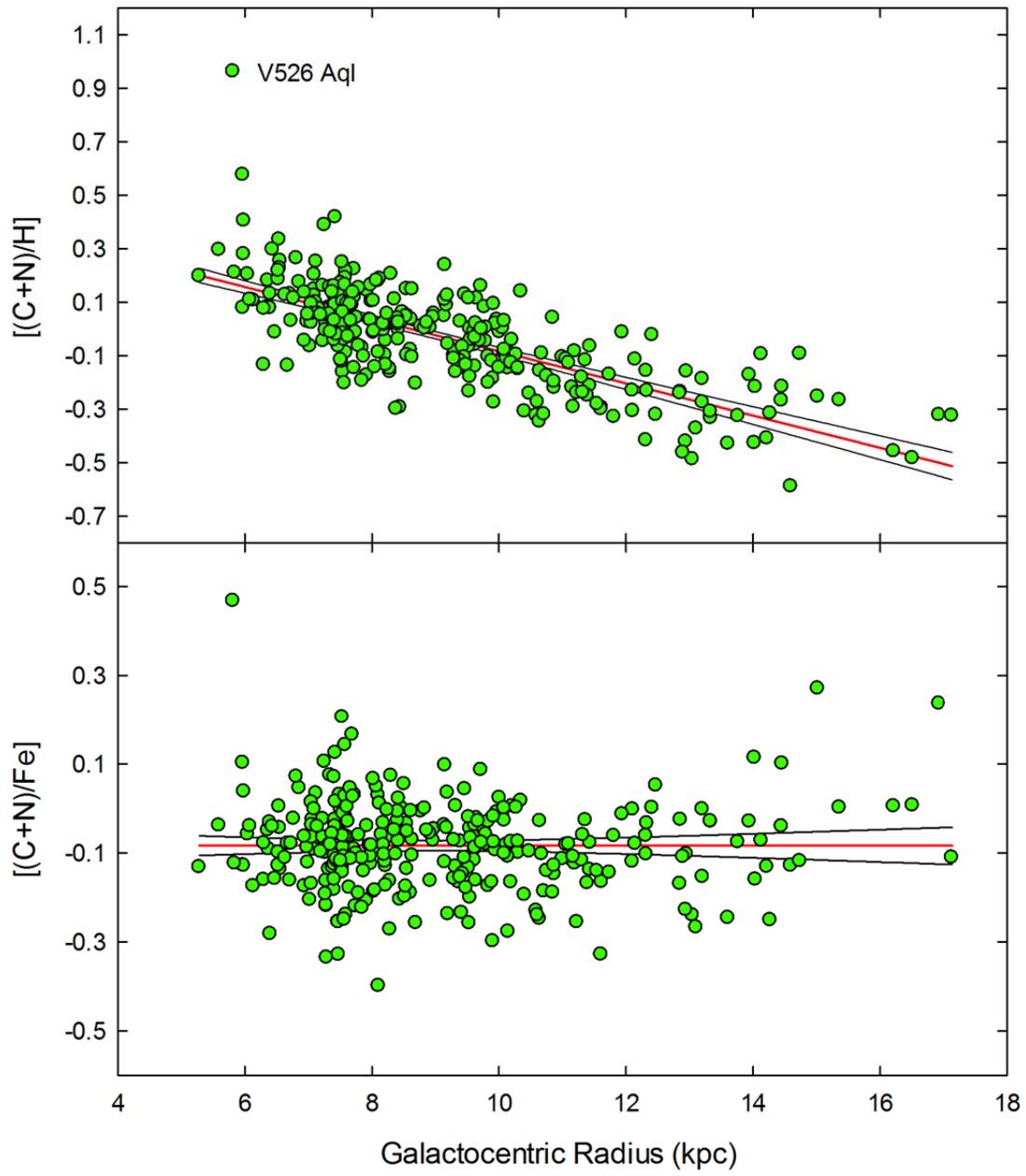

Figure 9

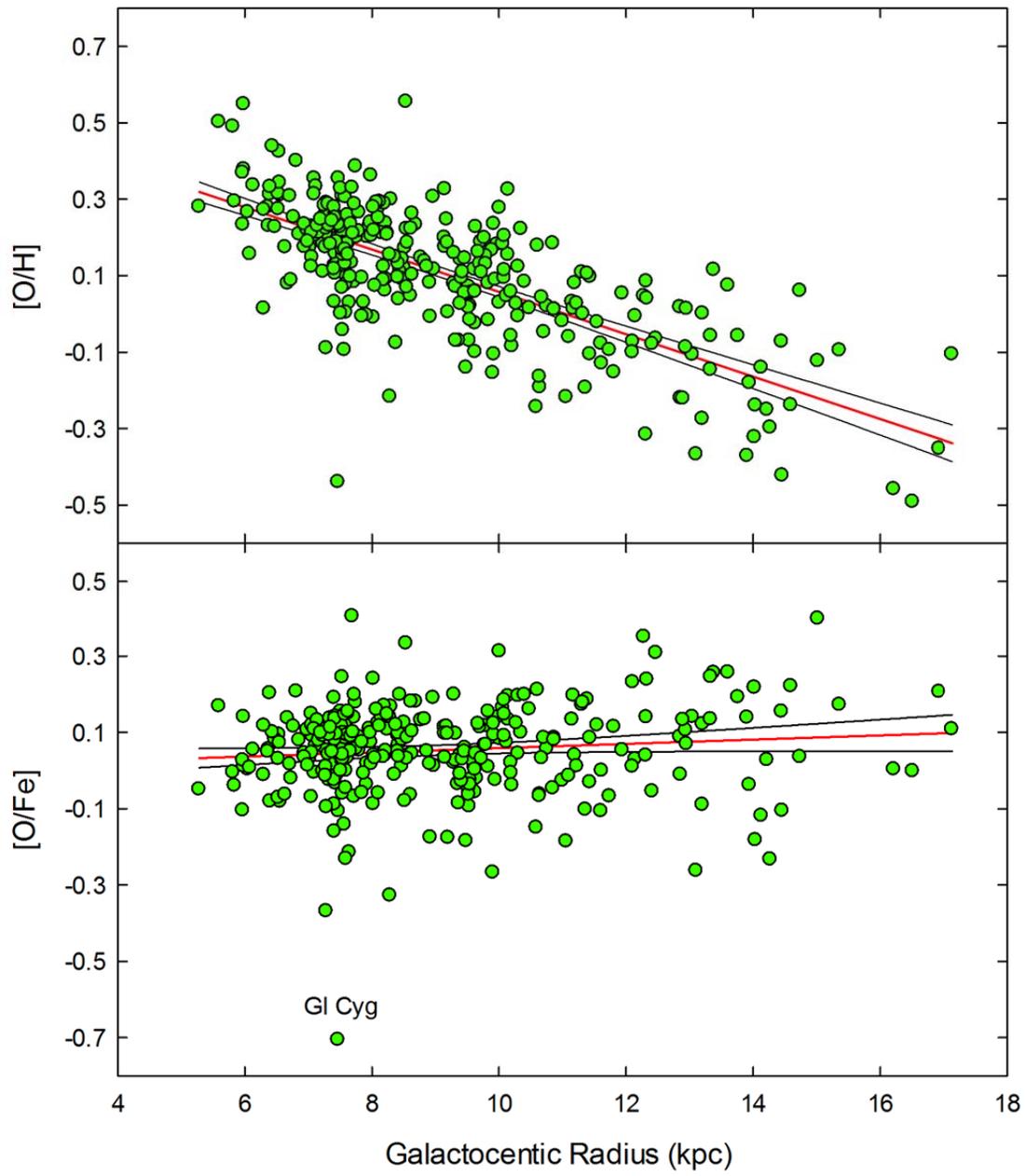

**Figure 10**

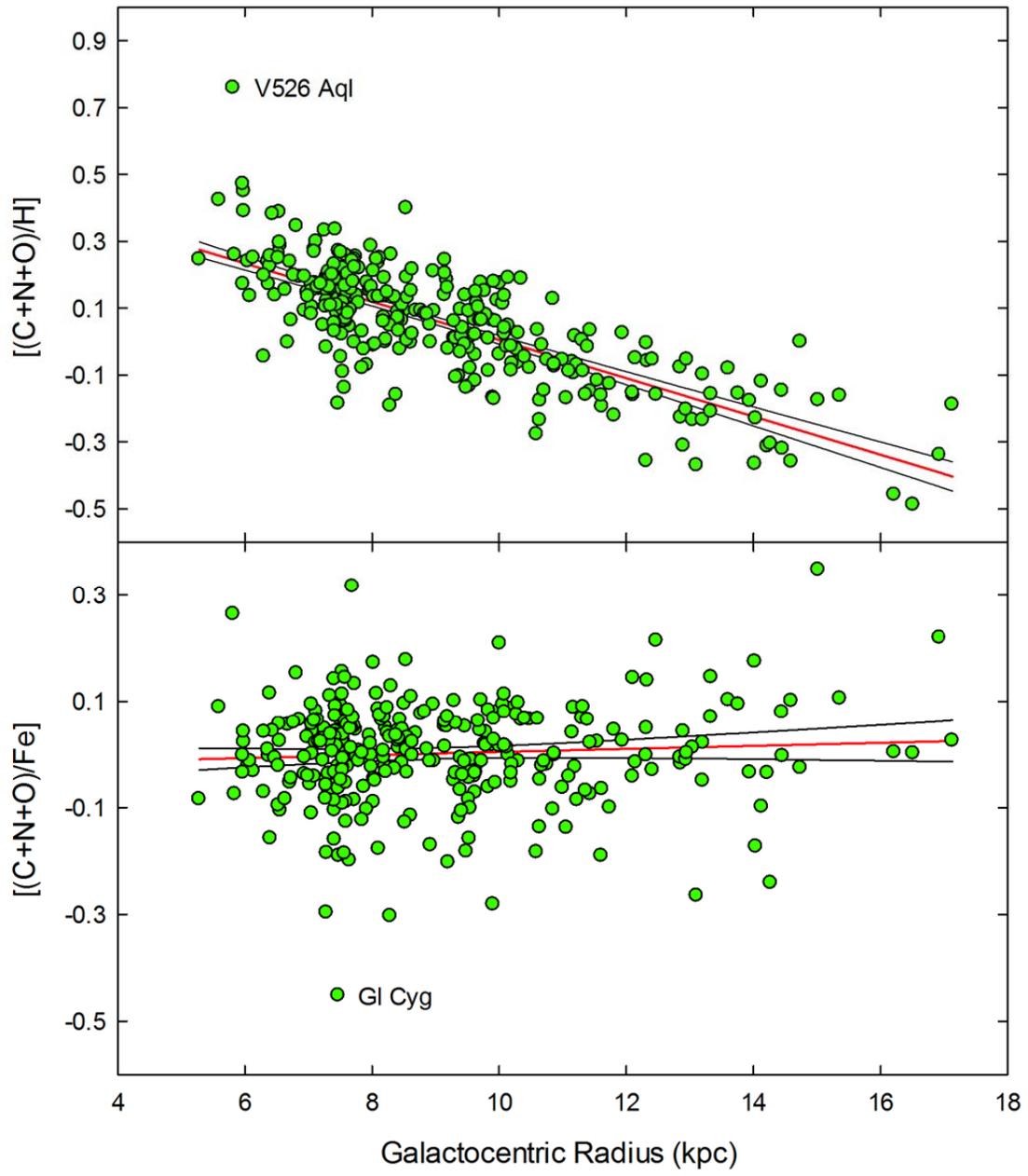

Figure 11

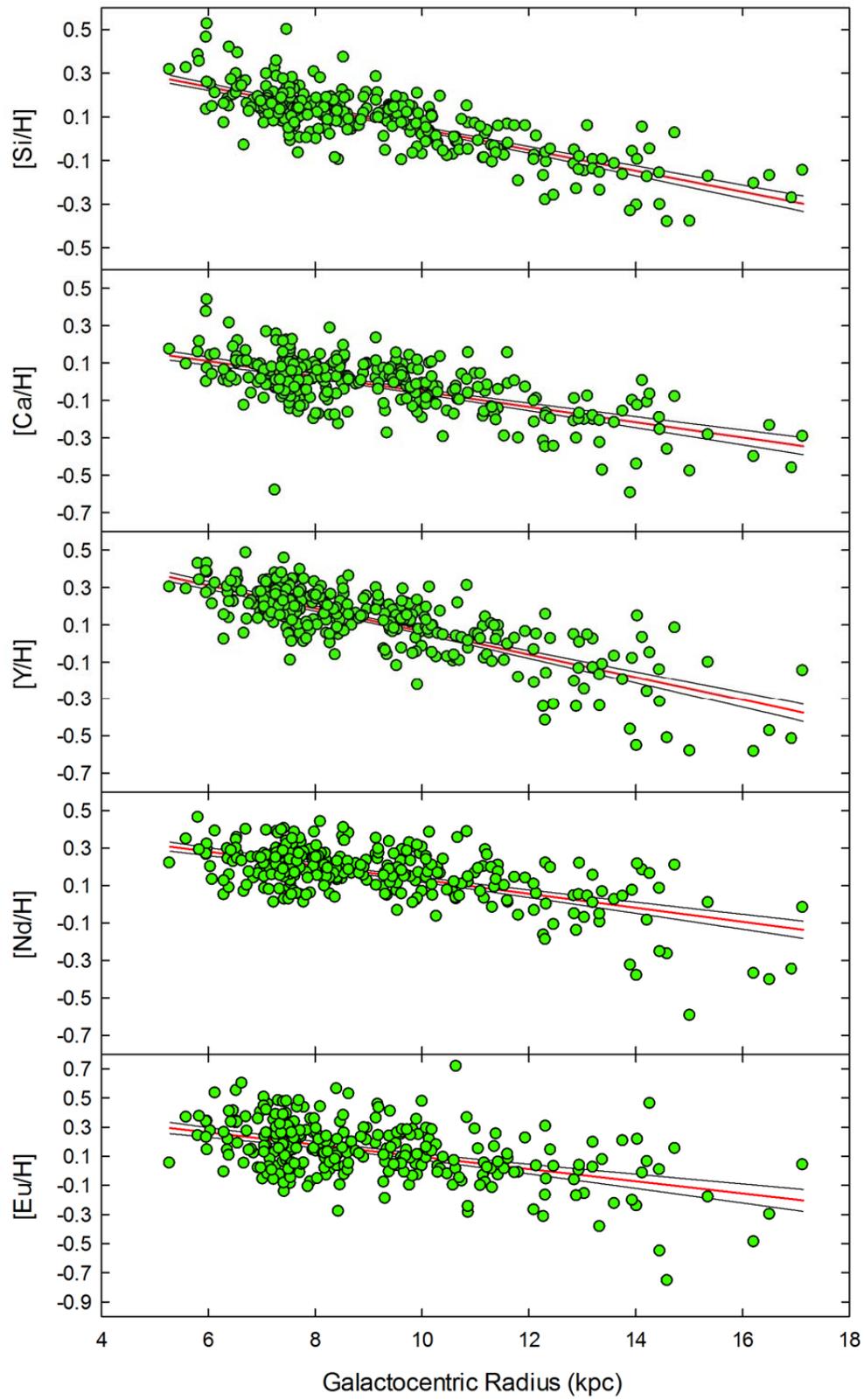

**Figure 12**